\documentclass[twocolumn]{aastex631}
\usepackage{CJK} % adding Chinese

\usepackage{amsmath}
\usepackage{amssymb}
\usepackage{epsfig}
\usepackage{apjfonts}
\usepackage{natbib}
\usepackage{hyperref}
\usepackage{epstopdf}
\usepackage{verbatim}
\usepackage{enumitem}
\usepackage{color}
\setlist[enumerate]{itemsep=-1mm}
\usepackage{soul, xcolor}
\setstcolor{blue}

\usepackage{textcase}

\usepackage{mathtools}

\usepackage{bigints}

\submitjournal{AJ (Accepted: May 31, 2025)}

\begin{document}

%%%%%%%%%%
%%% TEXT  %%%
%%%%%%%%%%

\begin{CJK*}{UTF8}{gbsn}

\title{ELemental abundances of Planets and brown dwarfs Imaged around Stars (ELPIS): \\II. The Jupiter-like Inhomogeneous Atmosphere of the First Directly Imaged Planetary-Mass Companion 2MASS~1207~b}

\author[0000-0002-3726-4881]{Zhoujian Zhang (张周健)} \thanks{NASA Sagan Fellow}
\affiliation{Department of Astronomy \& Astrophysics, University of California, Santa Cruz, CA 95064, USA}
\affiliation{Department of Physics \& Astronomy, University of Rochester, Rochester, NY 14627, USA}

\author[0000-0003-4096-7067]{Paul Molli\`{e}re}
\affiliation{Max-Planck-Institut f\"{u}r Astronomie, K\"{o}nigstuhl 17, 69117 Heidelberg, Germany}

\author[0000-0002-9843-4354]{Jonathan J. Fortney}
\affiliation{Department of Astronomy \& Astrophysics, University of California, Santa Cruz, CA 95064, USA}

\author[0000-0002-5251-2943]{Mark S. Marley}
\affiliation{Lunar \& Planetary Laboratory, University of Arizona, Tucson, AZ 85721, USA}

\begin{abstract}
2MASS~1207~b, the first directly imaged planetary-mass companion, has been instrumental in advancing our understanding of exoplanets and brown dwarfs over the past 20 years. We have performed extensive atmospheric retrieval analyses of 2MASS~1207~b's JWST/NIRSpec spectrum using \texttt{petitRADTRANS} and a new atmospheric inhomogeneity framework, which characterizes homogeneous atmospheres, patchy clouds, cloud-free hot spots, or the combination of patchy clouds and spots. Among 24 retrieval runs with various assumptions, the most statistically preferred model corresponds to the patchy cloud scheme, with $T_{\rm eff}=1174^{+4}_{-3}$~K, $\log{(g)}=3.62^{+0.03}_{-0.02}$~dex, and $R=1.399^{+0.008}_{-0.010}$~R$_{\rm Jup}$, along with near-solar atmospheric compositions of [M/H]$=-0.05\pm0.03$~dex and C/O$=0.440\pm0.012$. This model suggests $\sim9\%$ of 2MASS~1207~b's atmosphere is covered by thin iron and silicate clouds, producing L-dwarf-like spectra, while the remaining $91\%$ consists of thick iron and silicate clouds, emitting blackbody-like spectra. These thin-cloud patches and thick-cloud regions resemble Jupiter's belts and zones, respectively, and this scenario is consistently supported by other retrieval runs incorporating inhomogeneous atmospheres. We demonstrate that the weak CO absorption of 2MASS~1207~b can be explained by the veiling effects of patchy thick clouds; the absence of 3.3~$\mu$m CH$_{4}$ absorption is attributed to its hot thermal structure, which naturally leads to a CO-dominant, CH$_{4}$-deficient atmosphere. The retrieved atmospheric models also match the observed variability amplitudes of 2MASS~1207~b. Our analysis reveals that the inferred atmospheric properties show significant scatter in less statistically preferred retrieval runs but converge to consistent values among the preferred ones. This underscores the importance of exploring diverse assumptions in retrievals to avoid biased interpretations of atmospheric properties and formation pathways. 
\end{abstract}

\section{Introduction} 
\label{sec:introduction}

We are entering a new era in the atmospheric studies of exoplanets and brown dwarfs, marked by a rapidly growing number of spectroscopic analyses driven by unprecedented observational capabilities and continuous advancements in atmospheric modeling. For directly imaged exoplanets and brown dwarfs, thermal emission spectroscopy provides crucial insights into their atmospheric thermal structures, elemental abundances, surface gravities, radii, kinematics, spin rates, formation epoch, and atmospheric processes, such as cloud formation, vertical mixing, and planetesimal accretion \citep[e.g.,][]{2004A&A...425L..29C, 2008Sci...322.1348M, 2010Natur.468.1080M, 2011ApJ...729..128C, 2015Sci...350...64M, 2020A&A...640A.131M, 2020AJ....160..150W, 2021Natur.595..370Z, 2021AJ....162..148W, 2022AJ....163..189W, 2021AJ....161....5W, 2022ApJ...937...54X, 2023A&A...671L...5H, 2023ApJ...946L...6M, 2024arXiv241105917B, 2024A&A...688A.116D, 2024A&A...689A.212G, 2024ApJ...971....9H, 2024AJ....167..218I, 2024ApJ...971..121K, 2024A&A...682A..48L, 2024ApJ...963...73M, 2024A&A...687A.298N, 2024A&A...683A.214P, 2024ApJ...966L..11P, 2023AJ....166..198Z, 2025AJ....169....9Z}. In particular, measurements of these objects' atmospheric compositions --- including metallicity ([M/H]), carbon-to-oxygen ratio (C/O), and isotopologue ratios --- offer invaluable opportunities to compare these properties with those observed in the solar system and irradiated exoplanets \citep[e.g.,][]{2019ApJ...874L..31T, 2019ApJ...887L..20W, 2021Natur.595..370Z, 2023Natur.624..263B, 2023ApJ...957L..36G, 2023AJ....166...85H, 2024ApJ...970...71X, 2024arXiv241114541R, 2024ARA&A..62..333T}. Such comparisons enable population-level investigations into the origins and evolution of self-luminous worlds. 

Traditionally, the spectroscopic characterization of self-luminous exoplanets and brown dwarfs has relied on forward model grids, which have been foundational to the field \citep[see reviews by][]{1997ARA&A..35..137A, 2001RvMP...73..719B, 2015ARA&A..53..279M, 2018arXiv180408149F, 2020RAA....20...99Z, 2023ASPC..534..799C}. These pre-computed models use a limited number of free parameters to predict emergent spectra and thermal evolution properties under a set of assumptions, such as one-dimensional radiative-convective equilibrium, chemical (dis)equilibrium, and the presence and composition of clouds. By comparing observed spectrophotometry with synthetic atmospheric model spectra, these models provide self-consistent constraints on atmospheric properties. However, systematic offsets between observations and model predictions often arise, revealing the limitations of certain assumptions and motivating the development of next-generation forward models \citep[e.g.,][]{2021ApJ...916...53Z, 2021ApJ...921...95Z, 2024ApJ...961..121H, 2024ApJ...966L..11P}.

Over the past decade, the atmospheric retrieval technique has emerged as a major alternative for characterizing exoplanet and brown dwarf atmospheres \citep[e.g.,][]{2009ApJ...707...24M, 2012MNRAS.420..170L, 2013ApJ...778...97L, 2013ApJ...775..137L, 2015ApJ...807..183L, 2019A&A...627A..67M, 2023RNAAS...7...54M}. Unlike forward model grids, retrievals offer greater flexibility by treating atmospheric processes as free parameters rather than fixed assumptions, enabling constraints on a broader range of fundamental properties through a data-driven approach with significantly reduced data-model offsets. To date, retrievals have been applied to the near-infrared spectra of over ten directly imaged exoplanets \citep[e.g.,][]{2020A&A...633A.110G, 2023A&A...673A..98B, 2020A&A...640A.131M, 2020AJ....160..150W, 2021Natur.595..370Z, 2023AJ....166..203W, 2023AJ....166..198Z, 2024arXiv241105917B, 2024AJ....167..218I, 2024A&A...687A.298N} and over sixty brown dwarfs \citep[most of which are near the T/Y transition; e.g.,][]{2017ApJ...848...83L, 2019ApJ...877...24Z, 2022ApJ...936...44Z, 2021ApJ...923...19G, 2022ApJ...937...54X, 2024ApJ...970...71X, 2024A&A...688A.116D, 2024A&A...689A.212G, 2024ApJ...972..172P, 2024arXiv240508312H} across various spectral resolutions. Most retrieval studies, however, assume one-dimensional, homogeneous atmospheres --- a simplification that may be inadequate for some planetary-mass objects and brown dwarfs with detected atmospheric variabilities \citep[e.g.,][]{2009ApJ...701.1534A, 2014ApJ...793...75R, 2017Sci...357..683A, 2022ApJ...924...68V, 2022AJ....164..239Z, 2024MNRAS.532.2207B, 2024arXiv241116577M}. Such variability suggests top-of-atmosphere inhomogeneities, potentially caused by patchy clouds \citep[e.g.,][]{2001ApJ...556..872A, 2002ApJ...571L.151B, 2010ApJ...723L.117M, 2013ApJ...768..121A} and hot spots \citep[e.g.,][]{2014ApJ...785..158R, 2014ApJ...788L...6Z, 2019ApJ...883....4S, 2021MNRAS.502.2198T}, underscoring the need for more sophisticated atmospheric models.

Significant progress in retrieving inhomogeneous atmospheres has been demonstrated by \cite{2023ApJ...944..138V}, who analyzed the 1--15~$\mu$m spectra of two highly variable early-T brown dwarfs. To account for patchy clouds, \cite{2023ApJ...944..138V} employed a model combining two one-dimensional atmospheric columns with identical temperature-pressure (T-P) profiles but differing cloud properties. Their analysis suggested the presence of patchy forsterite clouds, which appear to be consistent with some of the observed photometric variability amplitudes of these brown dwarfs. This approach of combining two atmospheric columns has also been utilized in previous spectroscopic studies of variable brown dwarfs using self-consistent forward model grids \citep[e.g.,][]{2010ApJ...723L.117M, 2014ApJ...789L..14M, 2016ApJ...829L..32L}.

Here, we present a new generalized retrieval framework for characterizing inhomogeneous atmospheres of self-luminous exoplanets and brown dwarfs influenced by patchy clouds, hot spots, or a combination of both. We apply this framework to the JWST/NIRSpec spectrum of 2MASS~J12073346$-$3932539~b \citep[a.k.a. 2MASS~1207~b or TWA~27B;][]{2004A&A...425L..29C}, the first directly imaged planetary-mass companion, discovered 20 years ago. Orbiting a young M8 brown dwarf 2MASS~1207~A, 2MASS~1207~b has an L6 spectral type \citep{2024AJ....167..168M} with a variable atmosphere \citep{2016ApJ...818..176Z} and has been instrumental in advancing our understanding of self-luminous exoplanet and brown dwarf atmospheres.

The primary goal of this work is to characterize the potential inhomogeneous atmosphere of 2MASS~1207~b in the context of its detected photometric variabilities and to determine its atmospheric compositions as part of the Planets and brown dwarfs Imaged around Stars (ELPIS) program \citep{2023AJ....166..198Z}. We begin by reviewing the known properties of the 2MASS~1207~A+b system (Section~\ref{sec:2m1207}), followed by a description of the archival JWST/NIRSpec spectrum of the target 2MASS~1207~b (Section~\ref{sec:data}). We use thermal evolution models to contextualize the atmospheric properties of 2MASS~1207~b (Section~\ref{sec:evo}), and then perform retrieval analyses to explore various atmospheric scenarios: homogeneous atmospheres, patchy clouds, cloud-free hot spots, and a combination of patchy clouds and hot spots (Sections~\ref{sec:retrieval_framework} and \ref{sec:retrieval_planet}). Our analysis suggests that 2MASS~1207~b has near-solar atmospheric [M/H] and C/O ratios, along with a Jupiter-like, inhomogeneous atmosphere, with thin-cloud patches covering $9\%$ of its atmosphere and thick-cloud regions covering $91\%$. We discuss the implications of these results in Section~\ref{sec:discussion} and provide a summary in Section~\ref{sec:summary}.

\section{A Brief Review of 2MASS 1207 A+\lowercase{b}}
\label{sec:2m1207}

 In celebration of the 20th anniversary of the discovery of 2MASS~1207~A+b, we review the known properties of this remarkable system.

\subsection{2MASS~1207~A}
\label{subsec:2m1207a}

2MASS~1207~A (M8) was first identified as a candidate member of the TW Hya association \citep[TWA, $10 \pm 2$~Myr;][]{2023AJ....165..269L} by \cite{2002ApJ...575..484G}. The TWA membership of this object is supported by its youth, as suggested by the strong Balmer-series emission \citep[e.g.,][]{2002ApJ...575..484G, 2004ApJ...608L.113G, 2003ApJ...593L.109M, 2005ApJ...626..498M, 2005ApJ...629L..41S, 2006A&A...459..511B}, He~\textsc{i} emission \citep[$\lambda\lambda 4471$, $\lambda\lambda 5876$, $\lambda\lambda 6678$, $\lambda\lambda 7065$; e.g.,][]{2002ApJ...575..484G, 2003ApJ...593L.109M, 2005ApJ...626..498M}, Li~\textsc{i} $\lambda\lambda6708$ absorption \citep[e.g.,][]{2003ApJ...593L.109M, 2005ApJ...629L..41S}, weak alkali and FeH absorption \citep[e.g.,][]{2003ApJ...593L.109M, 2007ApJ...669L..97L, 2013ApJ...772...79A}, strong VO absorption \citep[e.g.,][]{2007ApJ...669L..97L, 2013ApJ...772...79A}, and a triangular-shaped $H$-band spectrum \citep[e.g.,][]{2007ApJ...669L..97L}. The parallax, proper motion, and radial velocity measurements have since placed 2MASS~1207~A as a confirmed kinematic member of TWA \citep[e.g.,][]{2005ApJ...634.1385M, 2007ApJ...669L..45G, 2007ApJ...669L..41B, 2008A&A...477L...1D, 2013ApJ...762...88M, 2016ApJ...833...95D}. 

The ongoing accretion of 2MASS~1207~A was suggested by multiple follow-up observations since its discovery. Its UV spectra resemble those of classical T Tauri stars, exhibiting numerous H$_{2}$ fluorescence lines and ion emission lines \citep[e.g.,][]{2005ApJ...630L..89G, 2010ApJ...715..596F}. Optical emission lines of Balmer series, He~\textsc{i}, Ca~\textsc{ii} H and K, Ca~\textsc{ii} infrared triplet, and O~\textsc{i} have been consistently detected \citep[e.g.,][]{2002ApJ...575..484G, 2003ApJ...593L.109M, 2005ApJ...626..498M, 2005ApJ...629L..41S, 2006ApJ...638.1056S, 2007ApJ...671..842S, 2009ApJ...696.1589H, 2019A&A...632A..46V}. The strongly variable shape and strength of H$\alpha$ suggest that the accretion rate of 2MASS~1207~A varies by more than one order of magnitude over months to years (e.g., \citealt{2005ApJ...626..498M, 2005ApJ...629L..41S, 2006ApJ...638.1056S, 2007ApJ...671..842S}; though see \citealt{2009ApJ...696.1589H}). No strong emission has been detected in the near-infrared wavelengths to date \citep[e.g.,][]{2019A&A...632A..46V, 2024AJ....167..168M}. The emission lines of 2MASS~1207~A likely result from its accretion and star-disk interactions, rather than potential chromospheric activities, especially given the absence of its X-ray \citep[][]{2004ApJ...608L.113G} and radio \citep[][]{2006ApJ...644L..67O} emission. Additionally, \cite{2009ApJ...697..373R} placed a $3\sigma$ upper limit of 1~kilogauss (with a maximum-likelihood value of 0) for the magnetic field of 2MASS~1207~A; this constraint is consistent with the weak magnetic field strength required to drive the observed accretion of this brown dwarf \citep[][]{2007ApJ...671..842S}. 

Our understanding of 2MASS~1207~A's circumstellar disk has greatly benefited from infrared and submillimeter observations. The infrared excess of this brown dwarf was only mildly detected in $L'$ band (3.8~$\mu$m) and Spitzer/IRAC [3.6] band \citep[][]{2003AJ....126.1515J, 2006ApJ...639L..79R}, significantly detected at longer wavelengths up to $\sim$70~$\mu$m \citep[][]{2004A&A...427..245S, 2006ApJ...639L..79R, 2008ApJ...681.1584R, 2012ApJ...744L...1H, 2024AJ....167..168M}, and not detected at 160~$\mu$m, 250~$\mu$m, 350~$\mu$m, 500~$\mu$m, and 850~$\mu$m \citep[based on Spitzer, Herschel, and JCMT;][]{2012ApJ...744L...1H, 2012MNRAS.422L...6R, 2012MNRAS.424L..74R, 2012A&A...548A..54R, 2013ApJ...773..168M}. The JWST/NIRSpec spectroscopy of this object also revealed the infrared excess at wavelengths beyond 2.5~$\mu$m \citep{2024AJ....167..168M}. Dust continuum emission at 0.89~mm and the CO emission from 2MASS~1207~A's disk were observed by ALMA \citep{2017AJ....154...24R}. 

Previous modeling efforts of 2MASS~1207~A's spectrophotometry suggested that its circumstellar disk does not have an inner hole \citep[e.g.,][]{2008ApJ...676L.143M} and has an outer radius of $\sim$10--50~au \citep[e.g.,][]{2012A&A...548A..54R, 2017AJ....154...24R}; this compact disk was likely tidally truncated by its planetary-mass companion, 2MASS~1207~b \citep[e.g.,][]{2012A&A...548A..54R, 2013ApJ...773..168M, 2017AJ....154...24R}. The disk inclination might be as high as 55$^{\circ}$--75$^{\circ}$ \citep[e.g.,][]{2004A&A...427..245S, 2007ApJ...661..354R, 2012A&A...548A..54R, 2011ApJ...732..107S}, which aligns with A's variable accretion emission lines \citep[e.g.,][]{2005ApJ...629L..41S, 2006ApJ...638.1056S} and the spectroastrometry of the observed bipolar and collimated outflows  \citep[e.g.,][]{2005ApJ...626..498M, 2007ApJ...659L..45W, 2012ApJ...761..120W}. A slightly lower disk inclination (${35^\circ}^{+20^{\circ}}_{-15^{\circ}}$) was derived by \cite{2017AJ....154...24R} based on ALMA observations, but as noted by these authors, such a low disk inclination would lead to an excessively high dynamical mass measurement for 2MASS~1207~A. The dust grains inside the disk may have grown to sizes larger than a few micrometers \citep[][]{2004A&A...427..245S, 2008ApJ...676L.143M, 2008ApJ...681.1584R}. The estimated dust reservoir surrounding 2MASS~1207~A's is $\sim 0.1$~M$_{\oplus}$ \citep[e.g.,][]{2013ApJ...773..168M, 2017AJ....154...24R}.

\begin{figure*}[t]
\begin{center}
\includegraphics[width=7.in]{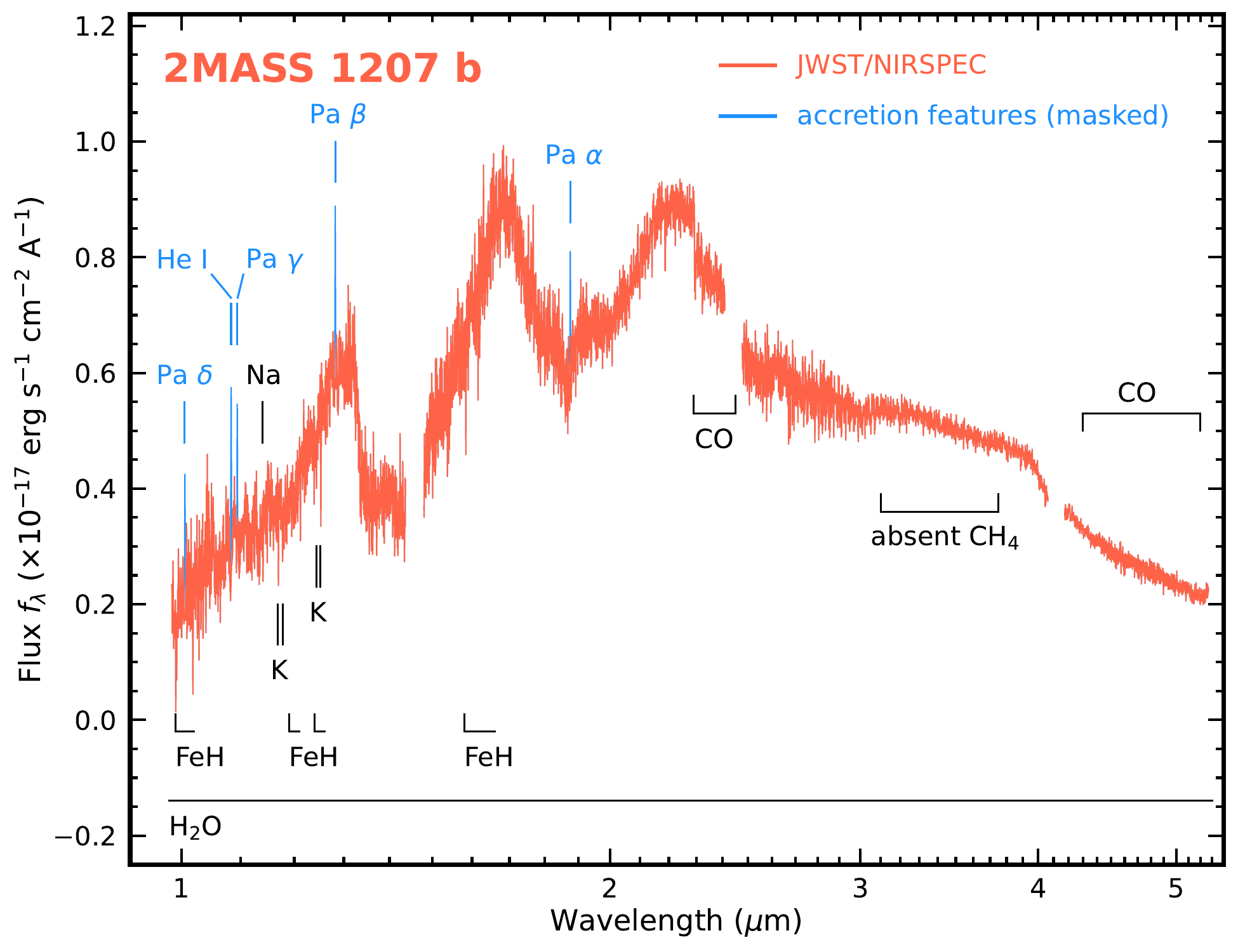}
\caption{The archival JWST/NIRSpec spectrum (red) of 2MASS~1207~b, with the x-axis shown in log scale. Key atomic and molecular features are labeled. Several prominent accretion emission lines, identified by \cite{2023ApJ...949L..36L} and with detection significance values above 3.5$\sigma$ \citep{2024ApJ...964...70M}, are highlighted in blue and are masked in our subsequent retrieval analysis.}
\label{fig:2m1207b_spec}
\end{center}
\end{figure*}

\subsection{2MASS~1207~b}
\label{subsec:2m1207b}

The planetary-mass companion, 2MASS~1207~b (L6), was first discovered by \cite{2004A&A...425L..29C} through high-contrast adaptive optics imaging, with its physical association to 2MASS~1207~A confirmed by \cite{2005A&A...438L..25C} and \cite{2006ApJ...652..724S}. Based on evolution models and 2MASS~1207~b's near-infrared absolute magnitudes, \cite{2004A&A...425L..29C} and \cite{2005A&A...438L..25C} estimated that this object has an effective temperature of $T_{\rm eff} = 1200 \pm 200$~K, a mass of $M = 5 \pm 2$~M$_{\rm Jup}$, a bolometric luminosity of $\log{(L/L_{\odot})} \approx -4.28$~dex, and a radius of $R \approx 1.5$~R$_{\rm Jup}$. As the first directly imaged planetary-mass companion, 2MASS~1207~b has received numerous follow-up observations focusing on its thermal emission spectrophotometry \citep[e.g.,][]{2004A&A...425L..29C, 2006ApJ...652..724S, 2007ApJ...657.1064M, 2010A&A...517A..76P, 2011ApJ...732..107S, 2014ApJ...792...17S, 2016ApJ...818..176Z, 2019AJ....157..128Z, 2023ApJ...949L..36L, 2024AJ....167..168M}. 

Spectroscopic studies of 2MASS~1207~b have profoundly advanced our understanding of the atmospheres of young exoplanets and brown dwarfs over the past two decades. Pioneering characterization efforts for 2MASS~1207~b encountered discrepancies among this companion's age, bolometric luminosity, and effective temperature. By comparing this object's near-infrared spectra with the available atmospheric models at the time, \cite{2007ApJ...657.1064M} inferred an effective temperature of $T_{\rm eff} \approx 1600$~K \citep[also see][]{2012A&A...540A..85P}. However, based on thermal evolution models and the system's young age, this high $T_{\rm eff}$ implied a bolometric luminosity $10\times$ brighter than what was measured from photometry \citep[][]{2005ApJ...634.1385M}. To explain the measured underluminosity, \cite{2007ApJ...657.1064M} thoroughly examined several scenarios and suggested a testable hypothesis that 2MASS~1207~b could be occulted by its potential gray edge-on disk. Around the same time, \cite{2007ApJ...668L.175M} proposed a novel alternative hypothesis that the companion might be an afterglow of a recent collision between two protoplanets and have a small radius and a hot temperature. 

Breakthroughs of the underluminosity problem for 2MASS~1207~b were achieved by \cite{2011ApJ...732..107S} and \cite{2011ApJ...735L..39B}, as sparked by the characterization of other newly discovered exoplanets and brown dwarfs at the time, such as HR~8799~bcde, HN~Peg~B, and HD~203030~B \citep[e.g.,][]{2006ApJ...651.1166M, 2007ApJ...654..570L, 2008Sci...322.1348M, 2010Natur.468.1080M, 2010ApJ...723..850B, 2011ApJ...733...65B, 2011ApJ...729..128C, 2011ApJ...737...34M}. By modeling the spectral energy distribution (SED) of 2MASS~1207~b, \cite{2011ApJ...732..107S} concluded that this companion is likely not influenced by an edge-on disk or an outer spherical dust shell. Instead, introducing thick clouds with disequilibrium chemistry processes into atmospheric models would lower the effective temperature to $T_{\rm eff} \approx 1000$~K \citep[e.g.,][]{2011ApJ...735L..39B, 2011ApJ...732..107S, 2014ApJ...792...17S}, resolving the underluminosity problem. Clouds and disequilibrium chemistry are now known as key processes in the atmospheres of exoplanets and brown dwarfs.

Inhomogeneities in the atmosphere of 2MASS1207b were indicated by time-series spectroscopic monitoring. Over an 8.7~hour baseline, \cite{2016ApJ...818..176Z} detected variability amplitudes of $1.36\% \pm 0.23\%$ and $0.78\% \pm 0.22\%$ at wavelengths near 1.25~$\mu$m and 1.54~$\mu$m, respectively, with a period of $10.7^{+1.2}_{-0.6}$~hr. 

Accretion emission lines, such as He~\textsc{i} and Paschen series, were recently revealed in 2MASS~1207~b's JWST/NIRSpec spectrum \citep{2023ApJ...949L..36L, 2024AJ....167..168M, 2024ApJ...964...70M}. The strengths of these lines suggest a mass accretion rate of several $\times 10^{-9}$~M$_{\rm Jup}$~yr$^{-1}$, with an estimated uncertainty of about an order of magnitude \citep{2024ApJ...964...70M, 2024AJ....168..155A}. These emission lines are suggested to originate from magnetospheric accretion rather than from active accretion through a potential circumplanetary disk \citep{2024ApJ...964...70M, 2024AJ....168..155A}. The presence of such a disk remains unconfirmed at the time of writing. Indeed, the W2 photometry (4.6~$\mu$m) of 2MASS~1207~b does not show strong excess, suggesting that the disk emission, if any, contributes to its SED beyond 5~$\mu$m \citep[][]{2023ApJ...949L..36L}. The mass of 2MASS~1207~b's dust disk is likely below 1 M$_{\rm moon}$ based on ALMA observations \citep{2017AJ....154...24R}.

\subsection{The 2MASS~1207~A+b Ecosystem}
\label{subsec:2m1207ab}

The mass ratio between 2MASS~1207~b and 2MASS~1207~A \citep[$M_{b} / M_{A} \approx 0.2$; e.g.,][]{2005A&A...438L..25C} suggests that this system is more analogous to stellar binaries than to typical planetary systems discovered by transits or radial velocity methods \citep[see also Figure~1 of][]{2025AJ....169....9Z}. Any additional companions with masses $>2$~$M_{\rm Jup}$ and orbital separation $>75$~au were ruled out by \cite{2019AJ....157..128Z}. For 2MASS~1207~b to have formed via disk instability, an unusually high initial disk mass for 2MASS~1207~A would be required \citep[e.g.,][]{2005ApJ...630L..89G}. Furthermore, the core accretion mechanism would likely be inefficient in assembling 2MASS~1207~b's mass at its current orbit ($\approx 50$~au) within its young age \citep[e.g.,][]{2005MNRAS.364L..91L}. However, outward migration could alleviate some of the tension regarding the formation timescale \citep[e.g.,][]{2007MNRAS.378.1589M}. Consequently, it is widely theorized that 2MASS~1207~A+b formed via gravitational instability during the collapse of a molecular cloud \citep[e.g.,][]{2005MNRAS.364L..91L, 2005ApJ...630L..89G, 2005A&A...438L..25C, 2006ApJ...637L.137B, 2009MNRAS.392..413S}. This scenario is further supported by the fact that the mass of 2MASS~1207~b is close to the estimated lowest mass threshold for star formation products \citep[e.g.,][]{2009A&A...493.1149Z, 2021ApJS..253....7K, 2022NatAs...6...89M, 2023arXiv231001231P, 2025arXiv250216349B}. 

The wide separation and relatively low total mass of the 2MASS~1207~A+b system suggest that it may not be stable over the long term \citep[e.g.,][]{2005AN....326..701M, 2006AJ....131.1007B}. Existing relative astrometry data of A and b only cover $0.06\%$ of the system's full orbit \citep[given their long orbital period of $\approx$1800~yr;][]{2017AJ....153..229B}, implying that the dynamical mass of 2MASS~1207~b will remain uncertain for the foreseeable future.

\section{Archival JWST/NIRSpec Spectroscopy of 2MASS~1207~\lowercase{b}}
\label{sec:data}

We collected archival JWST/NIRSpec spectra of 2MASS~1207~b from \cite{2023ApJ...949L..36L} and \cite{2024AJ....167..168M}, which cover 0.98--5.26~$\mu$m with moderate spectral resolution ($R \sim 2700$; Figure~\ref{fig:2m1207b_spec}). These data were taken using the G140H/F100LP, G235H/F170LP, and G395H/F290LP gratings, yielding an average signal-to-noise ratio (S/N) of 36, 70, 130, 240, and 128 per pixel near 1.3~$\mu$m, 1.7~$\mu$m, 2.2~$\mu$m, 3.7~$\mu$m, and 4.6~$\mu$m, respectively. 

For the subsequent retrieval analysis, we masked a few accretion emission lines with detection significance values exceeding $3.5\sigma$ (Figure~\ref{fig:2m1207b_spec}), as identified by \cite{2024ApJ...964...70M}. This includes Pa $\alpha$ (1.874--1.8765~$\mu$m), Pa $\beta$ (1.2812--1.2836~$\mu$m), Pa $\gamma$ (1.0935--1.0948~$\mu$m), Pa $\delta$ (1.004--1.0058~$\mu$m), and the He~\textsc{i} triplet (1.0824--1.0842~$\mu$m). Additionally, 2MASS~1207~b does not exhibit excess emission in the W2 band (4.6~$\mu$m), suggesting that its potential circumstellar disk does not significantly contribute to the JWST/NIRSpec spectroscopy (e.g., \citealt{2023ApJ...949L..36L}; see also Section~\ref{subsec:2m1207b}).

\begin{figure*}[t]
\begin{center}
\includegraphics[width=7.in]{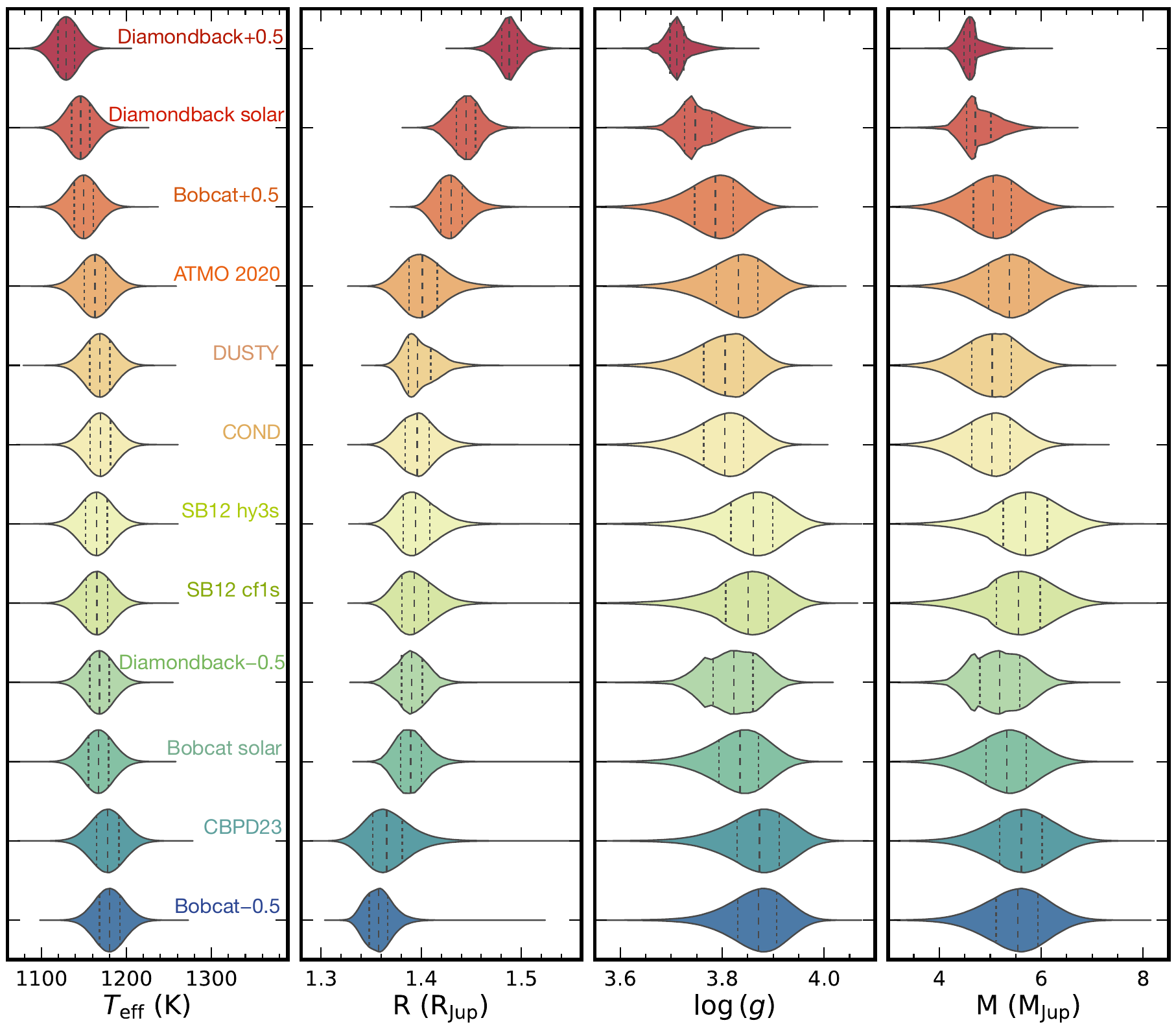}
\caption{Violin plot for the physical properties of 2MASS~1207~b inferred from its bolometric luminosity, age, and various evolution models (sorted by radii). The vertical dashed line inside each violin indicates the median value of the parameter, while the dotted lines represent the first and third quartiles of the distribution. Model labels and the parameter confidence intervals are summarized in Table~\ref{tab:evoparams}.  }
\label{fig:evo_companion}
\end{center}
\end{figure*}

{ 
\renewcommand{\arraystretch}{1.2} 
\begin{deluxetable*}{llcccc}
\setlength{\tabcolsep}{8pt} 
\tablecaption{Evolution-based properties of 2MASS 1207 b based on $\log{(L_{\rm bol}/L_{\odot})} = -4.467 \pm 0.026$~dex and an age of $10 \pm 2$~Myr } \label{tab:evoparams} 
\tablehead{ \multicolumn{1}{l}{Label\tablenotemark{\scriptsize a}}  & \multicolumn{1}{l}{Evolution Model} &  \multicolumn{1}{c}{$T_{\rm eff}$} &  \multicolumn{1}{c}{$R$} &  \multicolumn{1}{c}{$\log{g}$} &  \multicolumn{1}{c}{$M$} \\ 
\multicolumn{1}{l}{} &  \multicolumn{1}{l}{} &  \multicolumn{1}{c}{(K)} &  \multicolumn{1}{c}{($R_{\rm Jup}$)} &  \multicolumn{1}{c}{(dex)} &  \multicolumn{1}{c}{($M_{\rm Jup}$)} } 
\startdata 
\texttt{ATMO2020}  & \cite{2020AandA...637A..38P}: cloudless and [M/H]$=0$  &  $1163^{+18}_{-18}$  &  $1.401^{+0.023}_{-0.019}$  &  $3.83^{+0.05}_{-0.07}$  &  $5.4^{+0.6}_{-0.6}$  \\ 
\texttt{Bobcat-0.5}  & \cite{2021ApJ...920...85M}: cloudless and [M/H]$=-0.5$  &  $1180^{+17}_{-17}$  &  $1.357^{+0.014}_{-0.013}$  &  $3.87^{+0.05}_{-0.06}$  &  $5.5^{+0.6}_{-0.6}$  \\ 
\texttt{Bobcat solar}  & \cite{2021ApJ...920...85M}: cloudless and [M/H]$=0$  &  $1167^{+17}_{-17}$  &  $1.389^{+0.016}_{-0.014}$  &  $3.84^{+0.05}_{-0.06}$  &  $5.3^{+0.6}_{-0.6}$  \\ 
\texttt{Bobcat+0.5}  & \cite{2021ApJ...920...85M}: cloudless and [M/H]$=+0.5$  &  $1149^{+16}_{-16}$  &  $1.430^{+0.017}_{-0.015}$  &  $3.79^{+0.05}_{-0.06}$  &  $5.1^{+0.5}_{-0.6}$  \\ 
\texttt{CBPD23}  & \cite{2023AandA...671A.119C}: cloudless and [M/H]$=0$  &  $1178^{+19}_{-19}$  &  $1.365^{+0.024}_{-0.020}$  &  $3.87^{+0.06}_{-0.07}$  &  $5.6^{+0.6}_{-0.6}$  \\ 
\texttt{COND}  & \cite{2003AandA...402..701B}: cloudless and [M/H]$=0$  &  $1169^{+17}_{-17}$  &  $1.396^{+0.018}_{-0.018}$  &  $3.81^{+0.05}_{-0.06}$  &  $5.0^{+0.5}_{-0.6}$  \\ 
\texttt{Diamondback-0.5}  & \cite{2024ApJ...975...59M}: hybrid and [M/H]$=-0.5$  &  $1168^{+16}_{-16}$  &  $1.390^{+0.016}_{-0.015}$  &  $3.82^{+0.05}_{-0.06}$  &  $5.2^{+0.6}_{-0.5}$  \\ 
\texttt{Diamondback solar}  & \cite{2024ApJ...975...59M}: hybrid and [M/H]$=0$  &  $1146^{+15}_{-15}$  &  $1.445^{+0.014}_{-0.015}$  &  $3.75^{+0.05}_{-0.03}$  &  $4.7^{+0.5}_{-0.3}$  \\ 
\texttt{Diamondback+0.5}  & \cite{2024ApJ...975...59M}: hybrid and [M/H]$=+0.5$  &  $1129^{+14}_{-14}$  &  $1.488^{+0.012}_{-0.011}$  &  $3.71^{+0.03}_{-0.02}$  &  $4.6^{+0.2}_{-0.2}$  \\ 
\texttt{DUSTY}  & \cite{2000ApJ...542..464C}: cloudy and [M/H]$=0$  &  $1168^{+17}_{-17}$  &  $1.396^{+0.020}_{-0.013}$  &  $3.81^{+0.05}_{-0.06}$  &  $5.0^{+0.5}_{-0.6}$  \\ 
\texttt{SB12 cf1s}  & \cite{2012ApJ...745..174S}: cloudless and [M/H]$=0$  &  $1165^{+18}_{-18}$  &  $1.393^{+0.021}_{-0.017}$  &  $3.85^{+0.06}_{-0.07}$  &  $5.6^{+0.6}_{-0.7}$  \\ 
\texttt{SB12 hy3s}  & \cite{2012ApJ...745..174S}: hybrid and [M/H]$=+0.5$  &  $1164^{+18}_{-18}$  &  $1.394^{+0.022}_{-0.018}$  &  $3.86^{+0.06}_{-0.07}$  &  $5.7^{+0.6}_{-0.7}$  \\ 
\hline{\bf Adopted}\tablenotemark{\scriptsize b}  &   &  $1163^{+23}_{-21}$  &  $1.40^{+0.03}_{-0.04}$  &  $3.82^{+0.09}_{-0.07}$  &  $5.2^{+0.6}_{-0.7}$  \\ 
\enddata 
\tablenotetext{a}{Labels of the evolution model shown in \ref{fig:evo_companion}.}  
\tablenotetext{b}{The adopted parameters are determined from the concatenated chains of all these thermal evolution models (Section~\ref{subsec:evo}).}  
\end{deluxetable*} 
}

\section{Contextualizing the Properties of \\2MASS 1207 \lowercase{b} via Evolution Models}
\label{sec:evo}

We first derive the bolometric luminosity ($L_{\rm bol}$) of 2MASS~1207~b (Section~\ref{subsec:lbol}) by using its NIRSpec spectrum and a new bolometric correction approach we developed for self-luminous exoplanets, brown dwarfs, and low-mass stars (Appendix~\ref{app:bc}). By combining the $L_{\rm bol}$ and the age of 2MASS~1207~b, we use multiple thermal evolution models to derive its fundamental physical properties (Section~\ref{subsec:evo}), including effective temperatures ($T_{\rm eff}$), radii ($R$), surface gravities ($\log{(g)}$),\footnote{Throughout this manuscript, we use ``$\log{()}$'' and ``$\ln{()}$'' for 10-based and natural logarithm, respectively. Also, the surface gravity, $g$, is in cgs units of cm~s$^{-2}$.} and masses ($M$). These evolution-based constraints provide crucial context and physics-driven priors for our subsequent atmospheric retrieval analysis (see also Section~5.1 of \citealt{2023AJ....166..198Z} and references therein).

\subsection{Bolometric Luminosity}
\label{subsec:lbol}
We derive the bolometric luminosity of 2MASS~1207~b using its JWST/NIRSpec spectrum and the bolometric correction approach designed to mitigate modeling systematics, as described in Appendix~\ref{app:bc}. This method combines multiple grids of atmospheric models to estimate a spectroscopy-based bolometric correction term, \texttt{specBC$_{\lambda\lambda}$}, which converts the integrated spectral luminosity over a specific wavelength range $\lambda\lambda$ into the bolometric luminosity. 

For 2MASS~1207~b, we integrate its NIRSpec spectrum over the wavelength range $\lambda\lambda = 1-5$~$\mu$m. First, we estimate \texttt{specBC$_{\rm 1-5\mu m}$} for 2MASS~1207~b across a customized parameter space of $T_{\rm eff} =$1000--1400~K and $\log{(g)} < 5$~dex. By combining the bolometric correction terms from all model spectra within this parameter space, we find \texttt{specBC$_{\rm 1-5\mu m}$}$= 0.101 \pm 0.018$~dex based on the \texttt{Sonora Diamondback} grid and $0.112 \pm 0.025$~dex based on the \texttt{Exo-REM} grid. Other atmospheric model grids explored in Appendix~\ref{app:bc} are excluded either because their assumptions do not align with the cloudy atmospheres of 2MASS~1207~b (\texttt{ATMO2020}) or because their parameter spaces do not encompass the properties of 2MASS~1207~b (\texttt{BT-Settl} and \texttt{SPHINX}). To ensure conservative uncertainty estimates, we adopt \texttt{specBC$_{\rm 1-5\mu m}$}$= 0.112 \pm 0.025$~dex from \texttt{Exo-REM}.

Assuming 2MASS~1207~b shares the same distance as 2MASS~1207~A, $d = 64.47 \pm 0.47$~pc \citep{2021AJ....161..147B}, we derive a bolometric luminosity of $\log{(L_{\rm bol}/L_{\odot})} = -4.467 \pm 0.026$~dex for 2MASS~1207~b. This estimate is consistent with that reported by \cite{2023ApJ...949L..36L} but includes more conservative uncertainty estimates and mitigated modeling systematics.

\subsection{Evolution-based Physical Properties}
\label{subsec:evo}

We determine the physical properties of 2MASS~1207~b using the following hot-start thermal evolution models: 
\begin{enumerate}
\item[$\bullet$] \texttt{ATMO2020} models with solar metallicity \citep{2020AandA...637A..38P}.
\item[$\bullet$] \texttt{CBPD23} models with solar metallicity \citep{2023AandA...671A.119C}.
\item[$\bullet$] \texttt{COND} models with solar metallicity \citep{2003AandA...402..701B}.
\item[$\bullet$] \texttt{DUSTY} models with solar metallicity \citep{2000ApJ...542..464C}.
\item[$\bullet$] \texttt{Sonora Bobcat} models with [M/H]$=-0.5, 0, +0.5$~dex \citep{2021ApJ...920...85M}.
\item[$\bullet$] \texttt{Sonora Diamondback} hybrid models with [M/H]$=-0.5, 0, +0.5$~dex \citep{2024ApJ...975...59M}. These models transition from cloudy ($f_{\rm sed}=2$) to cloudless atmospheres at $T_{\rm eff} = 1300$~K. 
\item[$\bullet$] Models from \citep{2012ApJ...745..174S}, including cloudless atmospheres with solar metallicity (\texttt{SB12 cf1s}) and cloudy atmospheres with $3\times$ solar metallicity (\texttt{SB12 hy3s}).
\end{enumerate}

We infer the properties of 2MASS~1207~b in a Monte Carlo fashion. We draw $10^{6}$ random bolometric luminosities from a normal distribution $\mathcal{N}(\mu=-4.467\ {\rm dex}, \sigma=0.026\ {\rm dex})$, and the same number of random ages from a normal distribution $\mathcal{N}(\mu=10\ {\rm Myr}, \sigma=2\ {\rm Myr})$ truncated at zero \citep{2023AJ....165..269L}. Evolution models are interpolated linearly for $\log{(g)}$ and $\log{(L_{\rm bol}/L_{\odot})}$ and logarithmically for $T_{\rm eff}$, $R$, $M$, and age. 

Figures~\ref{fig:evo_companion} and Table~\ref{tab:evoparams} summarize the inferred properties of 2MASS~1207~b. The consistency across different evolution models allows us to concatenate the chains of each parameter derived by all models and compute their confidence intervals. We find that 2MASS~1207~b has $T_{\rm eff} = 1163^{+21}_{-23}$~K, $\log{(g)} = 3.82^{+0.07}_{-0.09}$~dex, $R = 1.40^{+0.04}_{-0.03}$~R$_{\rm Jup}$, and $M = 5.2^{+0.7}_{-0.6}$~M$_{\rm Jup}$. 

The $3\sigma$ confidence intervals across all evolution models span 1.30--1.55~R$_{\rm Jup}$ in $R$, 3.54--4.02~dex in $\log{(g)}$, and 3.0--7.5~M$_{\rm Jup}$ in $M$. These evolution-based ranges will serve as uniform parameter priors in some of our subsequent retrieval runs (Section~\ref{sec:retrieval_framework}).

\section{Atmospheric Retrieval Framework} 
\label{sec:retrieval_framework}

We perform atmospheric retrievals for the JWST/NIRSpec spectrum of 2MASS~1207~b using \texttt{petitRADTRANS} \citep{2019A&A...627A..67M, 2024JOSS....9.5875N}. The thermal structure, chemistry, and cloud models employed in the retrievals are described in Sections~\ref{subsec:temperature_model}--\ref{subsec:cloud_model}. 

In Section~\ref{subsec:inhomo_framework}, we introduce an ``atmospheric inhomogeneity'' framework, which models the potentially inhomogeneous atmospheres of exoplanets, brown dwarfs, and low-mass stars. This framework uses multiple atmospheric columns, each with distinct cloud properties and/or temperature-pressure (T-P) profiles. Unless otherwise specified, each atmospheric column adheres to the same modeling setup described in Sections~\ref{subsec:temperature_model}--\ref{subsec:cloud_model}.

\begin{figure*}[t]
\begin{center}
\includegraphics[width=6.8in]{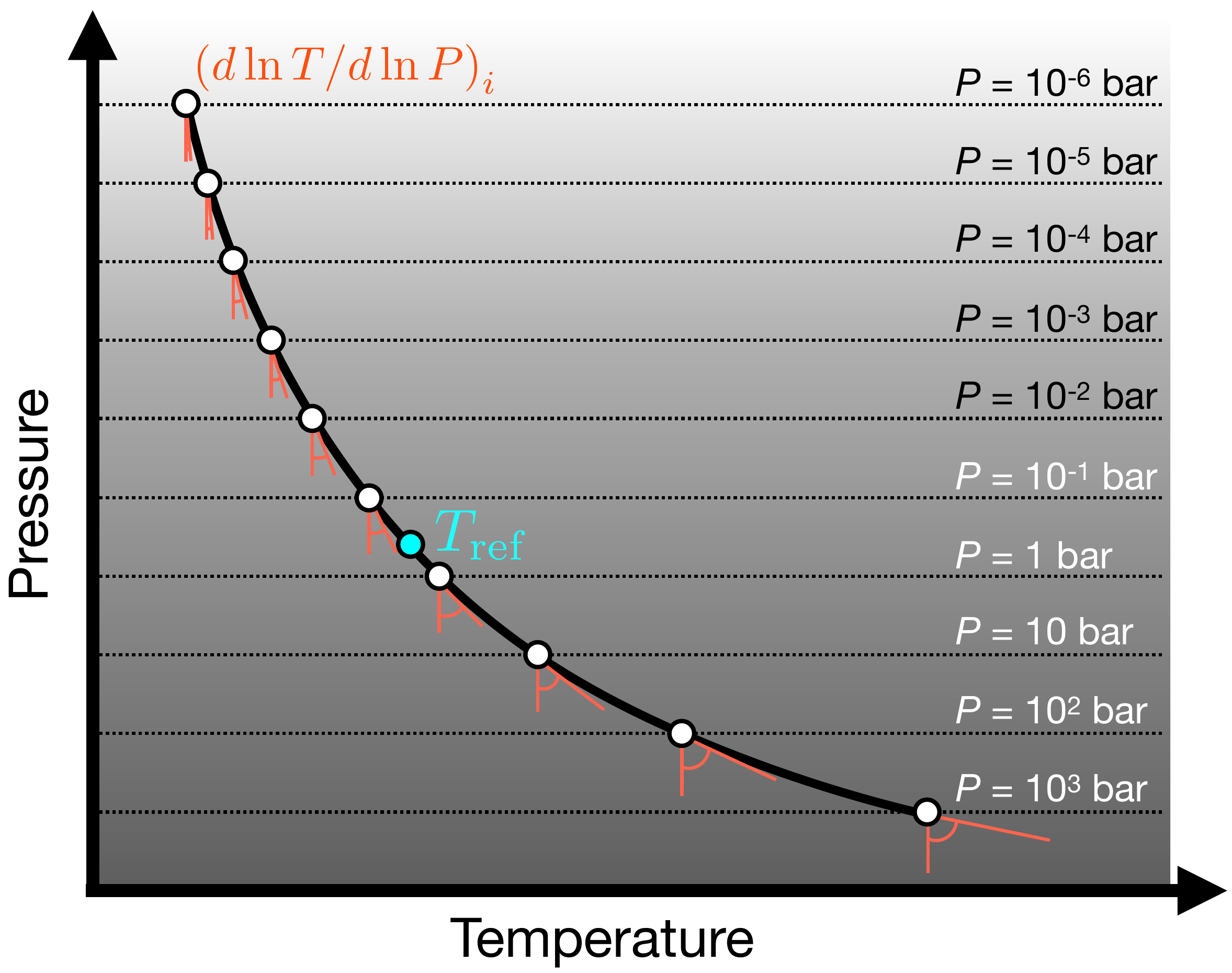}
\caption{Sketch of the temperature model described in Section~\ref{subsec:temperature_model}.}
\label{fig:tp}
\end{center}
\end{figure*}

\begin{figure*}[t]
\begin{center}
\includegraphics[width=7in]{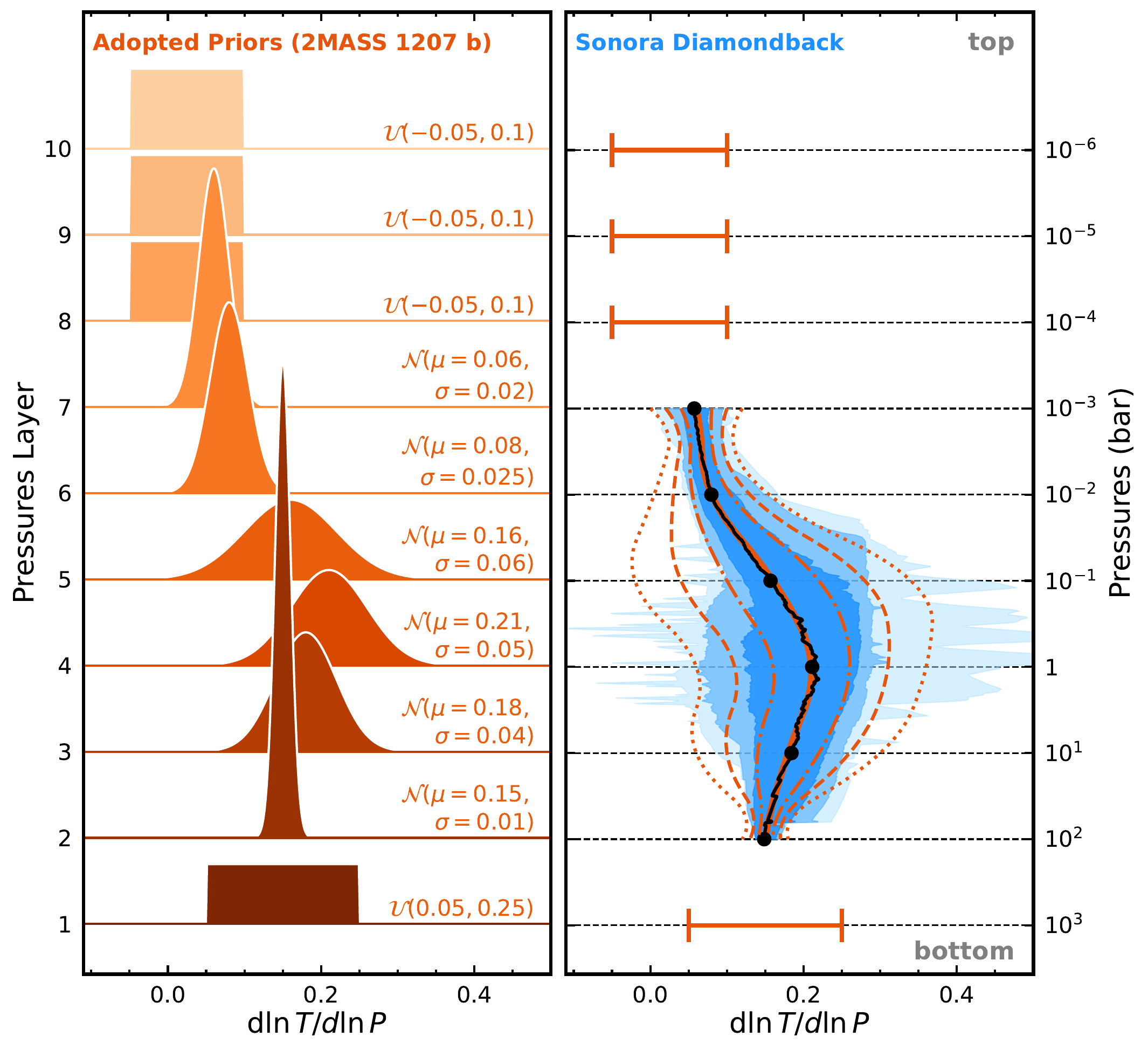}
\caption{{\it Left}: Our adopted priors for the temperature gradient, $(d\ln{T}/d\ln{P})$, at each of the 10 pressure layers evenly spaced in $\log{P}$ from $10^{3}$~bar (layer 1) to $10^{-6}$~bar (layer 10).  {\it Right}: The $1\sigma/2\sigma/3\sigma$ (from dark to light blues) confidence intervals of the $(d\ln{T}/d\ln{P})_{\rm RCE}$ profiles computed from the \texttt{Sonora Diamondback} models. The black solid line represents the median $(d\ln{T}/d\ln{P})_{\rm RCE}$ profile, with black circles corresponding to the temperature gradient values at the 10 pre-defined layers. For pressures between $10^{2}$~bar and $10^{-3}$~bar, we use orange lines to highlight the median values (solid), as well as the $1\sigma$ (dash-dotted), $2\sigma$ (dashed), and $3\sigma$ boundaries (dotted) of our adopted Gaussian priors. For pressure layers with $P < 10^{-3}$~bar or $P > 10^{2}$~bar, the adopted uniform prior is indicated by horizontal error bars.}
\label{fig:rce_dlnt_dlnp_companion}
\end{center}
\end{figure*}

\subsection{Temperature Model Coupled with Radiative-Convective Equilibrium} 
\label{subsec:temperature_model}

Our temperature model builds on the parameterization of \cite{2023AJ....166..198Z}. The atmosphere is divided into 10 pressure layers, spaced logarithmically from $10^{3}$ to $10^{-6}$ bar (Figure~\ref{fig:tp}). The thermal structure is characterized by 11 parameters, (1) the temperature gradient at each of the 10 layers, $\left(d\ln{T}/d\ln{P}\right)_{i}$, where $i=1,2,\dots,10$, and (2) the reference temperature, $T_{\rm ref}$, at a pre-defined pressure $P_{\rm ref}$.\footnote{If a user-defined $P_{\rm ref}$ is not one of the 10 layers, its corresponding temperature gradient $\left(d\ln{T}/d\ln{P}\right)_{\rm ref}$ is computed using quadratic interpolation of $\left(d\ln{T}/d\ln{P}\right)_{i}$.} With $T_{\rm ref}$ and the quadratically interpolated $\left(d\ln{T}/d\ln{P}\right)_{i}$, we construct the T-P profile over a finer grid of 1000 levels (for cloudy models) or 100 levels (for cloudless models) that are evenly spaced in $\log{P}$. The temperature at each pressure grid level (indexed by $j$) is computed as:
\begin{equation} \label{eq:tp}
\begin{aligned} 
T_{j \star} =& \exp{\left[ \ln{T_{\rm ref}} + \left( \ln{P_{j \star}} - \ln{P_{\rm ref}} \right) \times \left( \frac{d\ln{T}}{d\ln{P}} \right)_{\rm ref} \right]}, \\ 
& {\rm if\ P_{\rm ref} \not\in \{10^{3}, 10^{2}, \dots, 10^{-6}}\ {\rm bar} \} \\
T_{j \pm 1} =& \exp{\left[ \ln{T_{j}} + \left( \ln{P_{j \pm 1}} - \ln{P_{j}} \right) \times \left( \frac{d\ln{T}}{d\ln{P}} \right)_{\rm j} \right]} \\
\end{aligned} 
\end{equation}
Here, a larger $j$ corresponds to a higher altitude or lower pressure, and $j\star$ refers to the closest pressure grid level to the reference pressure if $P_{\rm ref}$ is not a direct pressure grid level. 

As demonstrated by \cite{2023AJ....166..198Z}, modeling the thermal structure via temperature gradients (Equation~\ref{eq:tp}) enables the incorporation of radiative-convective equilibrium (RCE) as parameterized priors for the T-P profiles. These priors can be derived from the temperature gradients of pre-computed atmospheric model grids, typically generated under RCE assumptions. Specifically, for a given atmospheric model grid, the T-P profile of each model spectrum is used to compute the corresponding temperature gradient as a function of pressure, i.e., $(d\ln{T}/d\ln{P})_{\rm RCE}$. Stacking the temperature gradient profiles of individual grid points offers insights into the distribution of $(d\ln{T}/d\ln{P})_{\rm RCE}$ across pressure layers. These distributions help define the priors for the gradients in our temperature model. Using RCE priors mitigates the degeneracies between cloud properties and T-P profiles, preventing retrievals from converging on cloudless solutions for cloudy atmospheres, thus providing stronger constraints on cloud properties \citep[see Sections~6.2 and 8.3 of][]{2023AJ....166..198Z}. Additionally, while these priors naturally {\it favour} the T-P profiles with similar shapes of the RCE forward models, they do not {\it fix} the temperature gradients during the retrievals and thereby flexibly allow the T-P profiles to deviate from the RCE as in the scenarios proposed by \cite{2016ApJ...817L..19T}.

For the retrievals of 2MASS~1207~b, we set RCE priors for $\left(d\ln{T}/d\ln{P}\right)$ following Section~6.2 of \cite{2023AJ....166..198Z}. We use the \texttt{Sonora Diamondback} models with $T_{\rm eff} \in [900,1500]$~K, $\log{(g)} \in [3.5, 4.5]$~dex, [M/H]$\in [-0.5, +0.5]$~dex, and $f_{\rm sed} \in [1,4]$. This leads to a total of 252 grid points. The majority of these models extend from $10^{-3}$~bar to $10^{2}$~bar, so we adopt Gaussian priors for the $\left(d\ln{T}/d\ln{P}\right)_{i}$ in this pressure range. Some Gaussian priors are further narrowed at a few pressure layers (Table~\ref{tab:retrieval_params}) based on the skewed distributions of $(d\ln{T}/d\ln{P})_{\rm RCE}$ from the \texttt{Sonroa Diamondback} models. We use broad uniform priors for pressure layers with $P < 10^{-3}$~bar or $P > 10^{2}$~bar. These uniform priors are designed such that the potential upper-atmosphere thermal inversion (i.e., $d\ln{T}/d\ln{P} < 0$) is allowed.

\subsection{Chemistry Model} 
\label{subsec:chemistry_model}

\subsubsection{Opacities}
\label{subsubsec:opacities}

The atomic and molecular line species included in our retrievals are Na \citep{2019A&A...628A.120A}, K \citep{2016A&A...589A..21A}, H$_{2}$O \citep{2018MNRAS.480.2597P}, CH$_{4}$ \citep{2020ApJS..247...55H}, CO \citep{2010JQSRT.111.2139R}, CO$_{2}$ \citep{2020MNRAS.496.5282Y}, NH$_{3}$ \citep{2019MNRAS.490.4638C}, PH$_{3}$ \citep{2015MNRAS.446.2337S}, H$_{2}$S \citep{2016MNRAS.460.4063A}, FeH \citep{2003ApJ...594..651D, 2020JQSRT.24006687B}, VO \citep{2016MNRAS.463..771M}, and TiO \citep{2019MNRAS.488.2836M}. We also account for Rayleigh scattering by H$_{2}$ \citep{1962ApJ...136..690D} and He \citep{1965PPS....85..227C}, and include the collision-induced absorption by H$_{2}$--H$_{2}$ and H$_{2}$--He \citep{1988ApJ...326..509B, 1989ApJ...336..495B, 1989ApJ...341..549B, 2001JQSRT..68..235B, 2002A&A...390..779B, 2012JQSRT.113.1276R}.

\subsubsection{An Expanded Grid of Chemical Equilibrium Abundances}
\label{subsubsec:ceq_grid}
\texttt{petitRADTRANS} includes a pre-computed grid of chemical-equilibrium abundances for key opacity sources as functions of [M/H] and C/O \citep{2015ApJ...813...47M, 2017A&A...600A..10M, 2019A&A...627A..67M}. This grid provides a convenient proxy for atmospheric equilibrium chemistry in retrievals and has been widely used \citep[e.g.,][]{2020A&A...640A.131M, 2021A&A...652A..57K, 2022ApJ...937...54X, 2023A&A...673A..98B, 2023AJ....166..198Z, 2024arXiv241105917B, 2024AJ....167..218I, 2024arXiv240508312H, 2024A&A...687A.298N}. 

In this work, we expand the \texttt{petitRADTRANS} chemical equilibrium abundance grid over an even broader parameter space and include more opacity sources. This new grid is computed using \texttt{easyCHEM}\footnote{\url{https://easychem.readthedocs.io/en/latest/}} \citep{2017A&A...600A..10M, 2024arXiv241021364L}. Our grid spans [40~K, 6000~K] in temperature ($T$), [$10^{-8}$~bar, $10^{3}$~bar] in pressure ($P$), [$-2$ dex, $+3$ dex] in [M/H], and [$0.1$, $1.6$] in C/O, with 120, 100, 40, and 20 grid points, respectively. These grid points are evenly spaced logarithmically for $T$ and $P$ and linearly for [M/H] and C/O. Compared to the grid from \cite{2019A&A...627A..67M}, this new grid extends to higher temperatures and includes additional species, such as AlH, C$_{2}$, CN, OH, Al, Ca, Fe, Mg, Si, and Ti. This grid is available via the Zenodo repository, \url{https://doi.org/10.5281/zenodo.15654694}.

\subsubsection{Chemical Abundance Profiles}
\label{subsubsec:abund_profiles}

Our retrievals assume chemical equilibrium and account for disequilibrium effects of carbon chemistry. We compute abundance profiles of atomic and molecular species based on equilibrium chemistry for a given [M/H] and C/O (Section~\ref{subsubsec:ceq_grid}). An additional free parameter, the carbon-chemistry quench pressure ($P_{\rm quench}$), accounts for disequilibrium chemistry. At pressures smaller than $P_{\rm quench}$, the abundances of H$_{2}$O, CO, and CH$_{4}$ are reset to their values at $P_{\rm quench}$ \citep[e.g.,][]{2014ApJ...797...41Z}.

\begin{figure*}[t]
\begin{center}
\includegraphics[width=7.in]{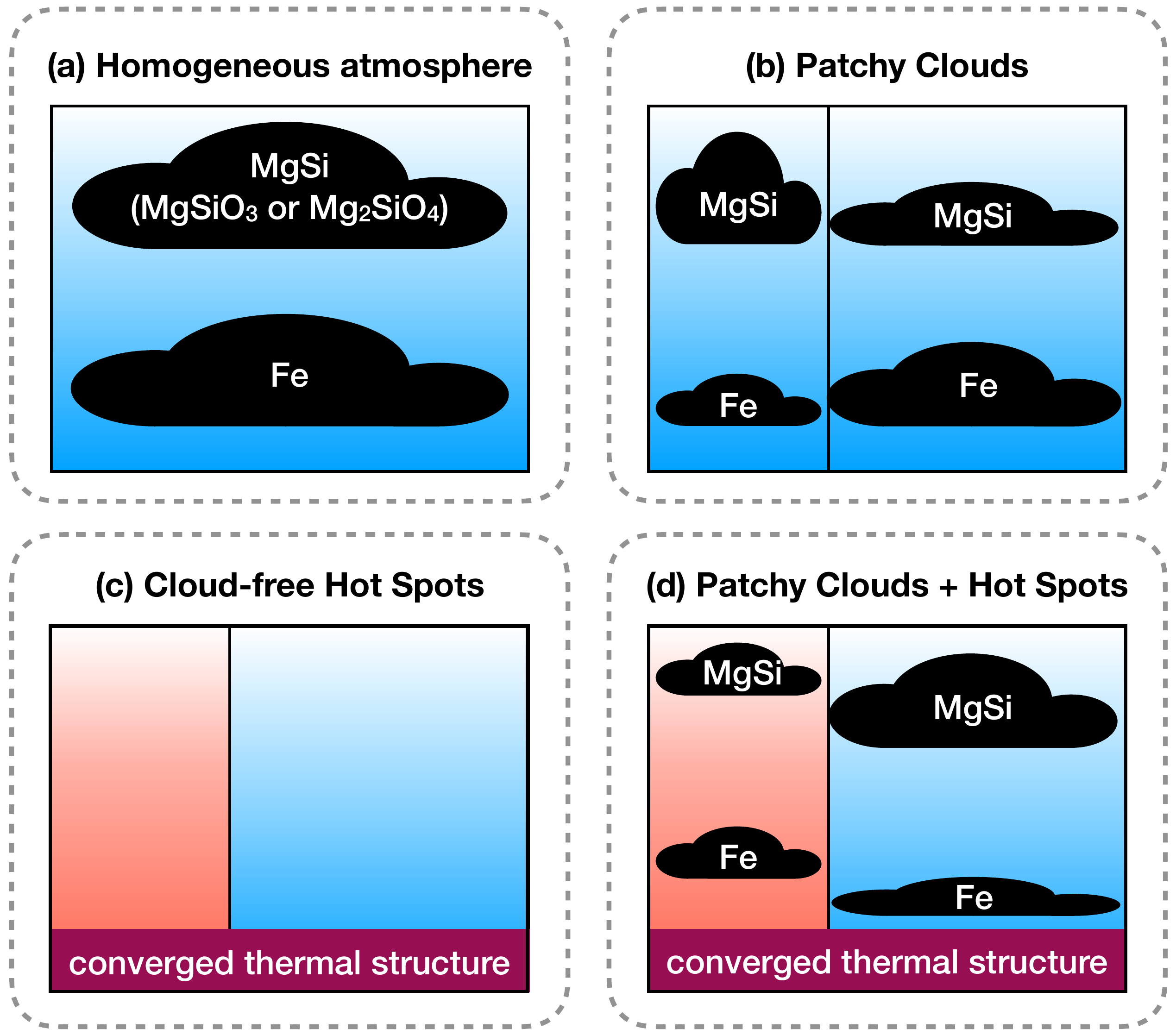}
\caption{Four schemes in our atmospheric inhomogeneity framework (Section~\ref{subsec:inhomo_framework}). }
\label{fig:inhomo}
\end{center}
\end{figure*}

\subsection{Cloud Model} 
\label{subsec:cloud_model}

Our cloud model is based on the prescriptions by \cite{2001ApJ...556..872A}. The free parameters include (1) $\sigma_{g}$, the standard deviation of the log-normal cloud particle size distribution; (2) $K_{\rm zz}$, the eddy diffusion coefficient; (3) $f_{\rm sed}$, the condensate sedimentation efficiency; and (4) $X_{0}^{(c)}$, the cloud mass fraction at the cloud base pressure. When multiple cloud species or multiple atmospheric columns (under the inhomogeneous scenario; Section~\ref{subsec:inhomo_framework}) are modeled, we assume that $\sigma_{g}$ and $K_{\rm zz}$ are shared by all cloud species within a given column. However, $f_{\rm sed}$ and $X_{0}^{(c)}$ are allowed to vary between different cloud species within the same column or for the same cloud species across different columns. 

The base pressure for each cloud species is computed by comparing the T-P profile to its condensation curve derived from equilibrium chemistry. However, in some retrieval runs, the cloud base pressure for a given cloud species, $P_{\rm base}^{(c)}$, is treated as a free parameter.

Our retrievals are performed using both cloud-free and cloudy scenarios. For the cloudy setups, we test various combinations of cloud species, including amorphous silicates\footnote{We focus on Mg$_{2}$SiO$_{4}$ and MgSiO$_{3}$ clouds in this work but note that the upcoming MIRI spectra will better constrain the specific type of silicate clouds in 2MASS~1207~b's atmosphere (GTO program 1270), as well as its potential circumsecondary disk.} of forsterite (Mg$_{2}$SiO$_{4}$) and enstatite (MgSiO$_{3}$) stoichiometry \citep{2003JQSRT..79..765J} and iron \citep[Fe;][]{1994ApJ...421..615P} with Mie scattering particle shapes. Additionally, in some cloudy retrieval runs, we incorporate a gray cloud layer with a varying pressure $P_{\rm gray}$; the opacity for deeper pressure layers with $P > P_{\rm gray}$, $\kappa_{\rm gray}$, is introduced as a free parameter, assumed to remain constant as a function of pressures.

\subsection{Atmospheric Inhomogeneity Framework} 
\label{subsec:inhomo_framework}

Traditional atmospheric retrieval analyses often assume one-dimensional, homogeneous atmospheres (Section~\ref{sec:introduction}). However, this simplification may not suffice for certain exoplanets and brown dwarfs exhibiting top-of-atmosphere inhomogeneities indicated by variability monitoring. Patchy clouds and hot spots have been suggested as two primary mechanisms driving such variabilities, highlighting the need for more sophisticated atmospheric models to accurately retrieve the properties of variable objects, such as 2MASS~1207~b.

To address this, we introduce a new retrieval framework comprising four schemes (Figure~\ref{fig:inhomo}), designed to characterize both homogeneous and inhomogeneous atmospheres of exoplanets, brown dwarfs, and low-mass stars.\footnote{We note that our retrieval framework can also be flexibly adapted to model the blended spectra of unresolved binaries or multiple systems.}

\subsubsection{Scheme (a): homogeneous Atmosphere}
\label{subsubsec:scheme_a}
This scheme represents the most widely used retrieval approach for brown dwarfs and directly imaged exoplanets \citep[e.g.,][]{2015ApJ...807..183L, 2017ApJ...848...83L, 2017MNRAS.470.1177B, 2020ApJ...905...46G, 2020A&A...640A.131M, 2021Natur.595..370Z, 2022ApJ...937...54X, 2023AJ....166..198Z, 2024AJ....167..237L}. In this work, we explore scenarios of (1) a cloud-free atmosphere, (2) a combination of Fe and MgSiO$_{3}$ clouds, and (3) a combination of Fe and Mg$_{2}$SiO$_{4}$ clouds. The T-P profile is modeled using the temperature gradients at ten pressure layers (as described in Section~\ref{subsec:temperature_model}) and a reference pressure of 0.1~bar, i.e., $T_{\rm ref} = T_{\rm 0.1\ bar}$.

{ 
\renewcommand{\arraystretch}{1.1} 
\begin{deluxetable*}{lll}
\setlength{\tabcolsep}{15pt} 
\tablecaption{Free Parameters of Retrievals} \label{tab:retrieval_params} 
\tablehead{ \multicolumn{1}{l}{Parameter} &  \multicolumn{1}{l}{Prior} &  \multicolumn{1}{l}{Description} } 
\startdata 
\hline 
\multicolumn{3}{c}{Temperature Model (Sections~\ref{subsec:temperature_model})}   \\ 
\hline 
$T_{\rm 0.1}$  &  $\mathcal{U}(500\ {\rm K}, 4000\ {\rm K})$  &   Temperature at $P=0.1$ bar in schemes (a--b) (Section~\ref{subsec:inhomo_framework})   \\ 
$T_{\rm 10}$  &  $\mathcal{U}(500\ {\rm K}, 7000\ {\rm K})$  &   Temperature at $P=10$ bar in schemes (c--d) (Section~\ref{subsec:inhomo_framework})   \\ 
$\left(d\ln{T}/d\ln{P}\right)_{1}$  &  $\mathcal{U}(0.05, 0.25)$  &   Temperature gradient at $P = 10^{3}$ bar.    \\ 
$\left(d\ln{T}/d\ln{P}\right)_{2}$  &  $\mathcal{N}(\mu=0.15, \sigma=0.01)$, within $[-2\sigma,+2\sigma]$  &   Temperature gradient at $P = 10^{2}$ bar.    \\ 
$\left(d\ln{T}/d\ln{P}\right)_{3}$  &  $\mathcal{N}(\mu=0.18, \sigma=0.04)$, within $[-2\sigma,+2\sigma]$  &   Temperature gradient at $P = 10^{1}$ bar.    \\ 
$\left(d\ln{T}/d\ln{P}\right)_{4}$  &  $\mathcal{N}(\mu=0.21, \sigma=0.05)$, within $[-2\sigma,+2\sigma]$  &   Temperature gradient at $P = 10^{0}$ bar.    \\ 
$\left(d\ln{T}/d\ln{P}\right)_{5}$  &  $\mathcal{N}(\mu=0.16, \sigma=0.06)$, within $[-1\sigma,+2\sigma]$  &   Temperature gradient at $P = 10^{-1}$ bar.    \\ 
$\left(d\ln{T}/d\ln{P}\right)_{6}$  &  $\mathcal{N}(\mu=0.08, \sigma=0.025)$, within $[-1\sigma,+2\sigma]$  &   Temperature gradient at $P = 10^{-2}$ bar.    \\ 
$\left(d\ln{T}/d\ln{P}\right)_{7}$  &  $\mathcal{N}(\mu=0.06, \sigma=0.02)$, within $[-2\sigma,+2\sigma]$  &   Temperature gradient at $P = 10^{-3}$ bar.    \\ 
$\left(d\ln{T}/d\ln{P}\right)_{8}$  &  $\mathcal{U}(-0.05, 0.10)$  &   Temperature gradient at $P = 10^{-4}$ bar.    \\ 
$\left(d\ln{T}/d\ln{P}\right)_{9}$  &  $\mathcal{U}(-0.05, 0.10)$  &   Temperature gradient at $P = 10^{-5}$ bar.    \\ 
$\left(d\ln{T}/d\ln{P}\right)_{10}$  &  $\mathcal{U}(-0.05, 0.10)$  &   Temperature gradient at $P = 10^{-6}$ bar.    \\ 
\hline  
\multicolumn{3}{c}{Chemistry Model (Section~\ref{subsec:chemistry_model})}   \\ 
\hline  
[M/H]  &  $\mathcal{U}(-1\ {\rm dex}, 2\ {\rm dex})$  &   Atmospheric metallicity.    \\ 
C/O  &  $\mathcal{U}(0.1, 1.6)$  &   Atmospheric carbon-to-oxygen ratio.    \\ 
$\log{(P_{\rm quench})}$  &  $\mathcal{U}(-6\ {\rm dex}, 3\ {\rm dex})$  &   Quench pressure of H$_{2}$O, CH$_{4}$, and CO (unit: bar).   \\  
\hline  
\multicolumn{3}{c}{Cloud Model (Section~\ref{subsec:cloud_model})\tablenotemark{\scriptsize a}}   \\ 
\hline   
$\sigma_{g}$  &  $\mathcal{U}(1.02, 3)$  &   Width of the log-normal cloud particle size distribution.     \\ 
$\log{(K_{\rm zz})}$  &  $\mathcal{U}(5\ {\rm dex}, 13\ {\rm dex})$  &   Vertical eddy diffusion coefficient of clouds.   \\   
$f_{\rm sed}$  &  $\mathcal{U}(0, 10)$  &   Sedimentation efficiency of a given cloud species.\tablenotemark{\scriptsize b}   \\  
$\log{(X_{0}^{(c)})}$  &  $\mathcal{U}(-10, 0)$  &   Mass fraction of a given cloud species at its base pressure.\tablenotemark{\scriptsize b}   \\  
$\log{(P_{\rm base}^{(c)})}$  &  $\mathcal{U}(-6, 3)$  &   Base pressure of a given cloud species (unit: bar).\tablenotemark{\scriptsize b}   \\  
$\log{(P_{\rm gray})}$  &  $\mathcal{U}(-6, -1)$  &   Pressure of the gray cloud layer (unit: bar).   \\  
$\log{(\kappa_{\rm gray})}$  &  $\mathcal{U}(-6, 9)$  &   Opacity below the gray cloud layer (unit: cm$^{2}/$g).   \\  
\hline  
\multicolumn{3}{c}{Atmospheric Inhomogeneity (Section~\ref{subsec:inhomo_framework})}   \\ 
\hline  
$\xi_{\rm thin}$  &  $\mathcal{U}(0, 1)$  &   Fraction of one atmospheric column in schemes (b--d) (Section~\ref{subsec:inhomo_framework}).    \\  
\hline   
\multicolumn{3}{c}{Other Parameters (Section~\ref{subsec:params_priors})\tablenotemark{\scriptsize c}}   \\ 
\hline   
$\log{(g)}$  &  $\mathcal{U}(2.5\ {\rm dex}, 5.5\ {\rm dex})$  &   Surface gravity (unit: cm$/$s$^{2}$).    \\  
  &  or\ $\mathcal{U}(3.54\ {\rm dex}, 4.02\ {\rm dex})$ &   A narrower prior constrained by thermal evolution (Section~\ref{subsec:evo})    \\  
$R$  &  $\mathcal{U}(0.5\ {R_{\rm Jup}}, 2.5\ {R_{\rm Jup}})$  &   Radius.    \\  
  &  or\ $\mathcal{U}(1.30\ {R_{\rm Jup}}, 1.55\ {R_{\rm Jup}})$ &   A narrower prior constrained by thermal evolution (Section~\ref{subsec:evo})    \\  
$\log{(b_{\rm chunk})}$  &  $\mathcal{U}(\log{(\min{(\sigma_{\rm chunk}))}}, \log{( 10^{3} \times \max{(\sigma_{\rm chunk})})} $  &   Additive flux error of each wavelength chunk of JWST/NIRSPEC.    \\  
\enddata 
\tablenotetext{a}{For most of our retrieval runs, the base pressure of a given cloud species is self-consistently computed following equilibrium chemistry, but several retrieval runs treat the cloud base pressures, $\log{(P_{\rm base}^{(c)})}$, as free parameters.   }  
\tablenotetext{b}{The cloud species in our retrievals include Fe, MgSiO$_{3}$, and Mg$_{2}$SiO$_{4}$. The $f_{\rm sed}$, $\log{(X_{0}^{(c)})}$, and $\log{(P_{\rm base}^{(c)})}$ values vary across different cloud species within the same atmospheric column or the same cloud species from different columns (Section~\ref{subsec:cloud_model}).    }  
\tablenotetext{c}{Some of our retrieval runs adopt the physically motivated narrower priors for $\log{(g)}$, radius, and mass as determined by thermal evolution models (Section~\ref{subsec:evo}). When the evolution-based priors for $\log{(g)}$ and $R$ are used in retrievals, we assume a uniform prior for $M$ of $\mathcal{U}(3.0\ {M_{\rm Jup}}, 7.5\ {M_{\rm Jup}})$.  }  
\end{deluxetable*} 
}

\subsubsection{Scheme (b): patchy clouds}
\label{subsubsec:scheme_b}
Patchy clouds, or spatial inhomogeneities in cloud cover and thickness, are observed in the atmospheres of Earth, Jupiter, and Saturn and are suggested for self-luminous exoplanets and brown dwarfs \citep[e.g.,][]{2001ApJ...556..872A, 2002ApJ...571L.151B, 2010ApJ...723L.117M, 2013ApJ...768..121A}. In this scheme, we model patchy clouds by combining two atmospheric columns that share the same T-P profile but differ in cloud properties \citep[see also][]{2010ApJ...723L.117M, 2014ApJ...789L..14M, 2016ApJ...829L..32L, 2023ApJ...944..138V}. For each column, we use the setup from the scheme (a) and explore combinations of Fe and MgSiO$_{3}$ clouds or Fe and Mg$_{2}$SiO$_{4}$ clouds. The same cloud species are used for both columns, but their properties --- $\sigma_{g}$, $K_{\rm zz}$, $f_{\rm sed}$, and $X_{0}^{(c)}$ --- are allowed to vary (even for the same species). 

The emergent spectrum is computed as:
\begin{equation} \label{eq:patchy}
F_{\lambda} = \xi_{\rm thin} \times F_{\rm \lambda, thin} + (1 - \xi_{\rm thin}) \times F_{\rm \lambda, thick}
\end{equation}
Here, $F_{\rm \lambda, thin}$ represents the spectrum from the atmospheric column with less cloud content (thin clouds), and $F_{\rm \lambda, thick}$ represents the spectrum from the column with thicker clouds. The free parameter $\xi_{\rm thin}$ describes the fraction of the thin-cloud column (i.e., column 1). 

This scheme accommodates an extreme scenario, such as cloud decks with completely depleted cloud holes, and an intermediate case, such as a combination of thin and thick clouds.

\subsubsection{Scheme (c): cloud-free hot spots}
\label{subsubsec:scheme_c}

Inhomogeneities in atmospheric thermal structures, or hot spots, may also explain the observed variabilities of exoplanets and brown dwarfs \citep[e.g.,][]{2014ApJ...788L...6Z, 2014ApJ...785..158R, 2019ApJ...874..111T, 2017Sci...357..683A, 2019ApJ...883....4S, 2021MNRAS.502.2198T, 2023MNRAS.523.4477L}. These temperature variations may be induced by dynamic processes such as moist convection, cloud radiative feedback, thermal perturbations, or planetary-scale waves \citep[see also][]{2020RAA....20...99Z}. Although clouds are often invoked in relevant dynamic modeling efforts, certain processes, such as thermocompositional diabatic convection, are theorized to occur in the atmospheres without any clouds \citep{2016ApJ...817L..19T, 2019ApJ...876..144T, 2020A&A...643A..23T}.

We simplify this scenario by combining two cloud-free atmospheric columns with non-identical T-P profiles. Each T-P profile is modeled following Section~\ref{subsec:temperature_model}, but unlike schemes (a) and (b), we use a reference pressure of 10~bar, i.e., $T_{\rm ref} = T_{\rm 10\ bar}$. To ensure that the thermal structures of the two atmospheric columns converge in the deep atmosphere or interior, we require both two T-P profiles to share the same $T_{\rm 10\ bar}$ and temperature gradients at 10~bar, $10^{2}$~bar, and $10^{3}$~bar. The choice of a 10~bar reference pressure aligns with the typical location of surface boundary conditions used when coupling atmospheric models with thermal evolution models \citep[e.g.,][]{1996ApJ...460..993S, 1997ApJ...491..856B, 2008ApJ...689.1327S}. For the upper atmospheres with $P < 10$~bar, we allow the two T-P profiles to have different temperature gradients and temperatures, reflecting variations in the balance between radiative cooling and dynamical timescales in these two columns.

The emergent spectrum is computed as:
\begin{equation} \label{eq:hotspots}
F_{\lambda} = \xi_{\rm thin} \times F_{\rm \lambda, hot} + (1 - \xi_{\rm thin}) \times F_{\rm \lambda, cold}
\end{equation}
Here, $F_{\rm \lambda, hot}$ and $F_{\rm \lambda, cold}$ represent the spectra from the hotter and colder regions, respectively. We keep the notation $\xi_{\rm thin}$ in Equation~\ref{eq:patchy} to describe the surface fraction of hot spots.

\subsubsection{Scheme (d): the combination of patchy clouds and hot spots.}
\label{subsubsec:scheme_d}

This scheme combines schemes (b) and (c), enabling the coexistence of patchy clouds and temperature variations within the atmosphere. We configure the cloud properties following scheme (b) and model the T-P profiles following scheme (c).

\subsection{Nested Sampling and Free Parameters} 
\label{subsec:params_priors}

Our retrievals are performed using nested sampling algorithm \texttt{MultiNest} \citep{2008MNRAS.384..449F, 2009MNRAS.398.1601F, 2019OJAp....2E..10F} via its Python wrapper \texttt{PyMultiNest} \citep{2014A&A...564A.125B}. This algorithm has been extensively applied in various fields, including cosmology \citep[e.g.,][]{2014A&A...571A..22P}, particle physics \citep[e.g.,][]{2008JHEP...12..024T}, and X-ray astronomy \citep[e.g.,][]{2014A&A...564A.125B}, and is among the most widely used tools for atmospheric retrieval analyses of exoplanets and brown dwarfs \citep[see a list of retrieval packages summarized by][]{2023RNAAS...7...54M}. Recently, \cite{2024OJAp....7E..79D} suggested that combining \texttt{PyMultiNest} with complementary algorithms, such as the Markov-chain Monte Carlo \citep[e.g.,][]{2013PASP..125..306F}, could help address potential limitations of individual techniques and improve the robustness of parameter inference. This practice remains rare in the field considering the computational demands, particularly when multiple retrieval runs are required, but it represents an important direction for future studies. For the scope of this work, we focus on using \texttt{PyMultiNest} and set 2000 live points and a constant sampling efficiency of 0.05. The spectral resolution of the JWST/NIRSpec spectrum is downgraded from $R \sim 2700$ to $R=1000$ to manage the computational cost of the multiple retrieval runs conducted in this work, especially since the cloud scattering included in our retrieval runs are numerically expensive. 

In addition to the parameters described in previous sections, we introduce a hyper-parameter, $\log{(b)}$, for each of the four segments of the JWST/NIRSpec spectrum, separated by the wavelength gaps of the G140H, G235H, and G395H gratings. These hyper-parameters are added to the measured flux uncertainties in quadrature when evaluating the log-likelihood, addressing (1) any underestimation of flux uncertainties in the observed NIRSpec spectra and (2) systematic differences between models and data. The surface gravity ($\log{(g)}$) and radius ($R$) of 2MASS~1207~b are also constrained by our retrievals. The free parameters and priors in our analysis are summarized in Table~\ref{tab:retrieval_params}.

{ 
\renewcommand{\arraystretch}{1.3} 
\begin{deluxetable*}{cccccccccccccc}
\setlength{\tabcolsep}{2pt} 
\tablecaption{Retrieved properties of 2MASS 1207 b (Quenching + Equilibrium Chemistry): Fundamental Parameters} \label{tab:retrieved_qeq_key} 
\tablehead{\multicolumn{1}{l}{}  & \multicolumn{1}{c}{}  & \multicolumn{1}{c}{}   &     \multicolumn{4}{c}{Atmospheric Setup}  & \multicolumn{1}{c}{}   & \multicolumn{1}{c}{}   & \multicolumn{1}{c}{}   & \multicolumn{1}{c}{}    & \multicolumn{1}{c}{}    & \multicolumn{1}{c}{}    & \multicolumn{1}{c}{}    \\ 
\cline{4-7}  
\multicolumn{1}{l}{Label\tablenotemark{\scriptsize a} }  &  \multicolumn{1}{c}{$\Delta$BIC\tablenotemark{\scriptsize b}}  & \multicolumn{1}{c}{}  &   \multicolumn{1}{c}{Column 1}   &   \multicolumn{1}{c}{Column 2}   &   \multicolumn{1}{c}{Cloud Base\tablenotemark{\scriptsize c}}   &   \multicolumn{1}{c}{Priors\tablenotemark{\scriptsize d}}  &    &   \multicolumn{1}{c}{$T_{\rm eff}$}   & \multicolumn{1}{c}{$\log{(g)}$}   & \multicolumn{1}{c}{$R$}    & \multicolumn{1}{c}{[M/H]}    & \multicolumn{1}{c}{C/O}    & \multicolumn{1}{c}{$\log{(P_{\rm quench})}$}   \\ 
\multicolumn{1}{l}{}    &   \multicolumn{1}{c}{}  & \multicolumn{1}{c}{}    &   \multicolumn{1}{c}{}    &   \multicolumn{1}{c}{}    &   \multicolumn{1}{c}{}    &   \multicolumn{1}{c}{}    &   \multicolumn{1}{c}{}    &   \multicolumn{1}{c}{(K)}    &   \multicolumn{1}{c}{(dex)}    &   \multicolumn{1}{c}{($R_{\rm Jup}$)}    &   \multicolumn{1}{c}{(dex)}    &   \multicolumn{1}{c}{}    &   \multicolumn{1}{c}{(bar)}     }  
\startdata 
\multicolumn{14}{c}{Scheme (b): Patchy Clouds (Figure~\ref{fig:inhomo}) }   \\ 
\hline  
\textbf{QEQ-1}  &  \textbf{0}  &   &    \textbf{ Forsterite$\mathbf{^{(1)}}$+Iron$\mathbf{^{(1)}}$}  &  \textbf{Forsterite$\mathbf{^{(2)}}$+Iron$\mathbf{^{(2)}}$}  &  \textbf{CEQ}   &   \textbf{Evo.}   &   &   $\mathbf{1174^{+4}_{-3}}$   &   $\mathbf{3.62^{+0.03}_{-0.02}}$   &   $\mathbf{1.399^{+0.008}_{-0.010}}$   &   $\mathbf{-0.05^{+0.03}_{-0.03}}$   &   $\mathbf{0.440^{+0.012}_{-0.012}}$   &   $\mathbf{-3.3^{+1.2}_{-1.2}}$   \\ 
QEQ-2  &  4  &  &    Forsterite$^{(1)}$+Iron$^{(1)}$  &  Forsterite$^{(2)}$+Iron$^{(2)}$  &  Flex. MgSi   &   Evo.   &   &  $1136^{+6}_{-6}$   &   $3.58^{+0.03}_{-0.02}$   &   $1.492^{+0.018}_{-0.016}$   &   $-0.12^{+0.04}_{-0.04}$   &   $0.459^{+0.014}_{-0.016}$   &   $-3.7^{+1.2}_{-1.2}$   \\ 
QEQ-3  &  42  &  &    Enstatite$^{(1)}$+Iron$^{(1)}$  &  Enstatite$^{(2)}$+Iron$^{(2)}$  &  CEQ   &   Evo.   &   &  $1199^{+5}_{-5}$   &   $3.65^{+0.02}_{-0.02}$   &   $1.334^{+0.012}_{-0.012}$   &   $-0.01^{+0.03}_{-0.03}$   &   $0.451^{+0.011}_{-0.012}$   &   $-3.7^{+1.2}_{-1.2}$    \\ 
QEQ-4  &  89  &  &    Forsterite$^{(1)}$+Iron$^{(1)}$  &  Forsterite$^{(2)}$+Iron$^{(2)}$+Gray  &  CEQ   &   Evo.   &   &    $1203^{+3}_{-3}$   &   $3.67^{+0.04}_{-0.02}$   &   $1.322^{+0.008}_{-0.007}$   &   $0.09^{+0.03}_{-0.03}$   &   $0.486^{+0.013}_{-0.013}$   &   $-3.8^{+1.1}_{-1.1}$   \\ 
QEQ-5  &  92  &  &     Enstatite$^{(1)}$+Iron$^{(1)}$  &  Enstatite$^{(2)}$+Iron$^{(2)}$+Gray  &  CEQ   &   Evo.   &   &    $1206^{+2}_{-2}$   &   $3.67^{+0.03}_{-0.02}$   &   $1.315^{+0.005}_{-0.005}$   &   $0.10^{+0.03}_{-0.03}$   &   $0.506^{+0.011}_{-0.011}$   &   $-3.8^{+1.1}_{-1.1}$   \\ 
QEQ-10  &  333  &  &    Enstatite$^{(1)}$+Iron$^{(1)}$  &  Enstatite$^{(2)}$+Iron$^{(2)}$  &  Flex. MgSi   &   Evo.   &   &    $1192^{+3}_{-3}$   &   $3.63^{+0.01}_{-0.01}$   &   $1.353^{+0.007}_{-0.007}$   &   $0.15^{+0.02}_{-0.03}$   &   $0.358^{+0.012}_{-0.014}$   &   $-3.5^{+1.0}_{-1.0}$   \\ 
\hline  
\multicolumn{14}{c}{Scheme (d): Patchy Clouds + Hot Spots (Figure~\ref{fig:inhomo}) }   \\ 
\hline  
QEQ-6  &  109  &  &    Enstatite$^{(1)}$+Iron$^{(1)}$  &  Enstatite$^{(2)}$+Iron$^{(2)}$  &  CEQ   &   Broad   &   &     $1221^{+6}_{-7}$   &   $3.40^{+0.07}_{-0.07}$   &   $1.286^{+0.014}_{-0.014}$   &   $0.02^{+0.03}_{-0.03}$   &   $0.522^{+0.014}_{-0.012}$   &   $-3.9^{+1.0}_{-1.0}$   \\ 
QEQ-7  &  123  &  &    Forsterite$^{(1)}$+Iron$^{(1)}$  &  Forsterite$^{(2)}$+Iron$^{(2)}$  &  CEQ   &   Broad   &   &     $1235^{+6}_{-6}$   &   $3.62^{+0.08}_{-0.06}$   &   $1.256^{+0.012}_{-0.012}$   &   $0.05^{+0.03}_{-0.02}$   &   $0.468^{+0.013}_{-0.015}$   &   $-3.8^{+1.1}_{-1.1}$   \\ 
QEQ-8 &  165  &  &    Forsterite$^{(1)}$+Iron$^{(1)}$  &  Forsterite$^{(2)}$+Iron$^{(2)}$+Gray  &  CEQ   &   Broad   &   &    $1392^{+3}_{-2}$   &   $3.34^{+0.06}_{-0.05}$   &   $0.987^{+0.003}_{-0.004}$   &   $0.07^{+0.03}_{-0.03}$   &   $0.531^{+0.010}_{-0.013}$   &   $-4.1^{+0.9}_{-0.9}$     \\ 
QEQ-9  &  169  &  &    Forsterite$^{(1)}$+Iron$^{(1)}$  &  Forsterite$^{(2)}$+Iron$^{(2)}$+Gray  &  CEQ   &   Evo.   &   &   $1209^{+1}_{-1}$   &   $3.70^{+0.04}_{-0.03}$   &   $1.309^{+0.002}_{-0.003}$   &   $0.18^{+0.02}_{-0.02}$   &   $0.522^{+0.009}_{-0.009}$   &   $-3.4^{+0.8}_{-0.9}$   \\ 
QEQ-11 &  532  &  &    Enstatite$^{(1)}$+Iron$^{(1)}$  &  Enstatite$^{(2)}$+Iron$^{(2)}$  &  CEQ   &   Evo.   &   &    $1191^{+1}_{-1}$   &   $3.61^{+0.01}_{-0.00}$   &   $1.364^{+0.003}_{-0.003}$   &   $0.34^{+0.02}_{-0.02}$   &   $0.322^{+0.008}_{-0.008}$   &   $-3.5^{+1.2}_{-1.2}$    \\ 
QEQ-12 &  586  &  &    Enstatite$^{(1)}$+Iron$^{(1)}$  &  Enstatite$^{(2)}$+Iron$^{(2)}$+Gray  &  CEQ   &   Broad   &   &    $1381^{+1}_{-1}$   &   $2.54^{+0.02}_{-0.02}$   &   $1.002^{+0.002}_{-0.002}$   &   $-0.20^{+0.02}_{-0.03}$   &   $0.483^{+0.010}_{-0.010}$   &   $-3.8^{+0.8}_{-0.8}$     \\ 
QEQ-13 &  589  &  &    Forsterite$^{(1)}$+Iron$^{(1)}$  &  Forsterite$^{(2)}$+Iron$^{(2)}$  &  CEQ   &   Evo.   &   &    $1186^{+2}_{-2}$   &   $3.60^{+0.01}_{-0.01}$   &   $1.374^{+0.006}_{-0.006}$   &   $-0.17^{+0.03}_{-0.02}$   &   $0.199^{+0.010}_{-0.010}$   &   $-3.6^{+1.5}_{-1.3}$   \\ 
QEQ-18 &  907  & &     Enstatite$^{(1)}$+Iron$^{(1)}$  &  Enstatite$^{(2)}$+Iron$^{(2)}$+Gray  &  CEQ   &   Evo.   &   &    $1143^{+4}_{-4}$   &   $3.58^{+0.02}_{-0.02}$   &   $1.469^{+0.011}_{-0.011}$   &   $0.86^{+0.03}_{-0.03}$   &   $0.115^{+0.010}_{-0.007}$   &   $-3.4^{+1.2}_{-1.2}$     \\ 
\hline  
\multicolumn{14}{c}{Scheme (a): Homogeneous Atmosphere (Figure~\ref{fig:inhomo}) }   \\ 
\hline  
QEQ-14 &  610  &  &    Forsterite$^{(1)}$+Iron$^{(1)}$  &  N/A  &  CEQ   &   Broad   &   &    $1239^{+2}_{-2}$   &   $2.51^{+0.01}_{-0.01}$   &   $1.248^{+0.004}_{-0.003}$   &   $0.13^{+0.02}_{-0.02}$   &   $0.593^{+0.010}_{-0.010}$   &   $2.0^{+0.0}_{-0.0}$     \\ 
QEQ-15 &  622  &  &    Enstatite$^{(1)}$+Iron$^{(1)}$  &  N/A  &  Flex. Both   &   Evo.   &   &    $1122^{+2}_{-2}$   &   $3.57^{+0.02}_{-0.02}$   &   $1.530^{+0.005}_{-0.005}$   &   $0.94^{+0.02}_{-0.02}$   &   $0.848^{+0.004}_{-0.004}$   &   $1.9^{+0.0}_{-0.0}$    \\ 
QEQ-16 &  693  &  &    Enstatite$^{(1)}$+Iron$^{(1)}$  &  N/A  &  CEQ   &   Broad   &   &    $1215^{+1}_{-1}$   &   $2.50^{+0.00}_{-0.00}$   &   $1.295^{+0.003}_{-0.003}$   &   $0.33^{+0.03}_{-0.03}$   &   $0.660^{+0.002}_{-0.002}$   &   $1.6^{+0.1}_{-0.1}$    \\ 
QEQ-17 &  753  &  &    Forsterite$^{(1)}$+Iron$^{(1)}$  &  N/A  &  Flex. Both   &   Evo.   &   &    $1136^{+2}_{-2}$   &   $3.58^{+0.02}_{-0.02}$   &   $1.488^{+0.005}_{-0.005}$   &   $0.83^{+0.02}_{-0.01}$   &   $0.837^{+0.003}_{-0.003}$   &   $1.7^{+0.0}_{-0.0}$    \\ 
QEQ-19 &  963  &  &    Forsterite$^{(1)}$+Iron$^{(1)}$  &  N/A  &  CEQ   &   Evo.   &   &    $1214^{+1}_{-1}$   &   $3.65^{+0.01}_{-0.00}$   &   $1.302^{+0.002}_{-0.002}$   &   $0.22^{+0.04}_{-0.04}$   &   $0.169^{+0.031}_{-0.026}$   &   $-0.4^{+0.5}_{-3.6}$     \\ 
QEQ-20 &  984  &  &    Enstatite$^{(1)}$+Iron$^{(1)}$  &  N/A  &  CEQ   &   Evo.   &   &    $1214^{+1}_{-1}$   &   $3.65^{+0.01}_{-0.00}$   &   $1.302^{+0.002}_{-0.001}$   &   $0.19^{+0.04}_{-0.04}$   &   $0.154^{+0.015}_{-0.018}$   &   $-2.5^{+2.3}_{-2.0}$    \\ 
QEQ-22 &  2468  &  &    Cloud-free  &  N/A  &  CEQ   &   Broad   &   &    $1302^{+1}_{-2}$   &   $2.50^{+0.00}_{-0.00}$   &   $1.111^{+0.003}_{-0.003}$   &   $0.85^{+0.03}_{-0.03}$   &   $0.101^{+0.001}_{-0.001}$   &   $-0.4^{+0.1}_{-0.1}$     \\ 
QEQ-24 &  4753  &  &    Cloud-free  &  N/A  &  CEQ   &   Evo.   &   &     $1210^{+1}_{-1}$   &   $3.64^{+0.00}_{-0.00}$   &   $1.300^{+0.000}_{-0.000}$   &   $1.98^{+0.01}_{-0.01}$   &   $0.100^{+0.001}_{-0.000}$   &   $0.6^{+0.1}_{-0.2}$     \\ 
\hline  
\multicolumn{14}{c}{Scheme (c): Hot Spots (Figure~\ref{fig:inhomo}) }   \\ 
\hline  
QEQ-21 &  1601  &  &    Cloud-free  &  Cloud-free  &  CEQ   &   Broad   &   &     $1152^{+3}_{-3}$   &   $2.50^{+0.00}_{-0.00}$   &   $1.462^{+0.007}_{-0.007}$   &   $1.80^{+0.03}_{-0.04}$   &   $0.101^{+0.001}_{-0.001}$   &   $-0.2^{+0.1}_{-0.1}$     \\ 
QEQ-23 &  3766  &  &    Cloud-free  &  Cloud-free  &  CEQ   &   Evo.   &   &     $1141^{+4}_{-5}$   &   $3.54^{+0.00}_{-0.00}$   &   $1.479^{+0.014}_{-0.010}$   &   $1.74^{+0.03}_{-0.02}$   &   $0.100^{+0.000}_{-0.000}$   &   $-3.0^{+0.8}_{-1.2}$     \\ 
\enddata 
\tablenotetext{a}{Label numbers indicate the BIC ranks of retrieval runs. ``QEQ-1'' corresponds to the most preferred setup among all retrieval runs that incorporate quenching and equilibrium chemistry.  }  
\tablenotetext{b}{The $\Delta$BIC is computed by comparing the BIC of each retrieval run with that of QEQ-1.  }  
\tablenotetext{c}{``CEQ'' indicates that the base pressures all cloud species are self-consistently computed based on chemical equilibrium; ``Flex. MgSi'' indicates that the base pressures of iron clouds are computed based on chemical equilibrium, while those of the magnesium silicate clouds are free parameters; ``Flex. Both'' indicates that the base pressures of all cloud species are free parameters.  }  
\tablenotetext{d}{``Broad'' indicates that broad priors are adopted for $\log{(g)}$, $R$, and $M$; ``Evo.'' indicates that the evolution-based priors for these parameters are adopted.   }  
\end{deluxetable*} 
}

{ 
\renewcommand{\arraystretch}{1.3} 
\begin{deluxetable*}{lcccccccccccccc}
\setlength{\tabcolsep}{3pt} 
\tablecaption{Retrieved properties of 2MASS 1207 b (Quenching + Equilibrium Chemistry): Cloud Parameters} \label{tab:retrieved_qeq_cloud} 
\tablehead{\multicolumn{1}{l}{}  & \multicolumn{1}{c}{}  & \multicolumn{1}{c}{}   &     \multicolumn{5}{c}{Column 1 (Thin-cloud patches or hot spots)\tablenotemark{\scriptsize a}}  & \multicolumn{1}{c}{}   & \multicolumn{6}{c}{Column 2 (Thick-cloud regions or cold spots)\tablenotemark{\scriptsize a}}    \\ 
\cline{4-8}  \cline{10-15}  
\multicolumn{1}{c}{Label}  &  \multicolumn{1}{c}{$\Delta$BIC}  & \multicolumn{1}{c}{}  &   \multicolumn{1}{c}{$f_{\rm sed,MgSi}$}   &   \multicolumn{1}{c}{$\log{(X_{\rm 0,MgSi})}$}   &   \multicolumn{1}{c}{$f_{\rm sed,Fe}$}   &   \multicolumn{1}{c}{$\log{(X_{\rm 0,Fe})}$}   &   \multicolumn{1}{c}{$\xi_{\rm thin}$\tablenotemark{\scriptsize a}}  &    &   \multicolumn{1}{c}{$f_{\rm sed,MgSi}$}   &   \multicolumn{1}{c}{$\log{(X_{\rm 0,MgSi})}$}   &   \multicolumn{1}{c}{$f_{\rm sed,Fe}$}   &   \multicolumn{1}{c}{$\log{(X_{\rm 0,Fe})}$}  &  \multicolumn{1}{c}{$\log{(P_{\rm gray})}$}   & \multicolumn{1}{c}{$\log{(\kappa_{\rm gray})}$}   \\ 
\multicolumn{1}{c}{}  &  \multicolumn{1}{c}{}  & \multicolumn{1}{c}{}  &   \multicolumn{1}{c}{}   &   \multicolumn{1}{c}{(dex)}   &   \multicolumn{1}{c}{}   &   \multicolumn{1}{c}{(dex)}   &   \multicolumn{1}{c}{}  &    &   \multicolumn{1}{c}{}   &   \multicolumn{1}{c}{(dex)}   &   \multicolumn{1}{c}{}   &   \multicolumn{1}{c}{(dex)}  &  \multicolumn{1}{c}{(dex)}   & \multicolumn{1}{c}{(dex)}   }  
\startdata 
\multicolumn{15}{c}{Scheme (b): Patchy Clouds (Figure~\ref{fig:inhomo}) }   \\ 
\hline  
\textbf{QEQ-1}  &  0  &   &   $\mathbf{4.54^{+2.45}_{-2.34}}$   &   $\mathbf{-7.77^{+1.10}_{-1.06}}$   &   $\mathbf{4.53^{+0.74}_{-0.55}}$   &   $\mathbf{-4.53^{+0.11}_{-0.07}}$    &  $\mathbf{8.8^{+0.2}_{-0.2}}$ \textbf{\%}  &   &   $\mathbf{1.20^{+0.07}_{-0.06}}$   &   $\mathbf{-0.67^{+0.13}_{-0.12}}$   &   $\mathbf{5.59^{+2.19}_{-2.26}}$   &   $\mathbf{-4.84^{+2.38}_{-2.45}}$   &   \textbf{N/A}   &   \textbf{N/A}     \\ 
QEQ-2  &  4  &   &   $4.82^{+2.61}_{-2.55}$   &   $-6.58^{+2.27}_{-1.93}$   &   $3.14^{+0.49}_{-0.40}$   &   $-4.67^{+0.07}_{-0.07}$    &  $9.5^{+0.3}_{-0.4}$ \%  &   &   $5.24^{+2.48}_{-2.54}$   &   $-5.11^{+2.50}_{-2.52}$   &   $1.48^{+0.06}_{-0.08}$   &   $-0.22^{+0.12}_{-0.16}$   &   N/A   &   N/A     \\ 
QEQ-3  &  42  &   &   $4.71^{+2.58}_{-2.36}$   &   $-7.68^{+1.17}_{-1.19}$   &   $2.39^{+0.31}_{-0.26}$   &   $-4.23^{+0.05}_{-0.05}$    &  $9.1^{+0.2}_{-0.2}$ \%  &   &   $1.50^{+0.09}_{-0.07}$   &   $-0.20^{+0.10}_{-0.12}$   &   $5.59^{+2.25}_{-2.38}$   &   $-5.00^{+2.43}_{-2.53}$   &   N/A   &   N/A    \\ 
QEQ-4  &  89  &   &   $4.80^{+2.56}_{-2.52}$   &   $-7.69^{+1.23}_{-1.20}$   &   $1.79^{+0.52}_{-0.42}$   &   $-4.25^{+0.10}_{-0.09}$    &  $11.0^{+0.2}_{-0.3}$ \%  &   &   $5.40^{+2.47}_{-2.62}$   &   $-5.16^{+2.59}_{-2.57}$   &   $5.07^{+2.52}_{-2.48}$   &   $-5.03^{+2.53}_{-2.53}$   &   $-5.5^{+0.3}_{-0.3}$   &   $0.90^{+0.05}_{-0.05}$    \\ 
QEQ-5  &  92  &   &   $5.02^{+2.42}_{-2.42}$   &   $-7.40^{+1.31}_{-1.27}$   &   $1.32^{+0.26}_{-0.30}$   &   $-4.32^{+0.06}_{-0.08}$    &  $11.4^{+0.2}_{-0.2}$ \%  &   &   $4.82^{+2.42}_{-2.32}$   &   $-4.85^{+2.33}_{-2.40}$   &   $5.14^{+2.37}_{-2.34}$   &   $-4.96^{+2.36}_{-2.39}$   &   $-5.3^{+0.3}_{-0.3}$   &   $0.88^{+0.06}_{-0.05}$    \\ 
QEQ-10  &  333  &   &   $2.76^{+0.72}_{-0.57}$   &   $-2.85^{+0.23}_{-0.20}$   &   $9.93^{+0.03}_{-0.04}$   &   $-0.45^{+0.13}_{-0.15}$    &  $12.4^{+0.3}_{-0.3}$ \%  &   &   $5.12^{+2.09}_{-2.17}$   &   $-4.42^{+2.15}_{-2.30}$   &   $1.52^{+0.03}_{-0.03}$   &   $-0.13^{+0.06}_{-0.09}$   &   N/A   &   N/A    \\ 
\hline  
\multicolumn{15}{c}{Scheme (d): Patchy Clouds + Hot Spots (Figure~\ref{fig:inhomo}) }   \\ 
\hline  
QEQ-6  &  109  &   &   $4.50^{+2.46}_{-2.24}$   &   $-7.41^{+1.35}_{-1.31}$   &   $2.90^{+0.63}_{-0.54}$   &   $-4.30^{+0.12}_{-0.12}$    &  $12.8^{+0.7}_{-0.7}$ \%  &   &   $4.95^{+2.37}_{-2.30}$   &   $-4.80^{+2.30}_{-2.32}$   &   $0.11^{+0.05}_{-0.03}$   &   $-0.95^{+0.13}_{-0.10}$   &   N/A   &   N/A    \\ 
QEQ-7  &  123  &   &   $5.16^{+2.20}_{-2.26}$   &   $-7.34^{+1.16}_{-1.18}$   &   $7.95^{+0.94}_{-1.07}$   &   $-4.17^{+0.06}_{-0.07}$    &  $11.6^{+0.4}_{-0.4}$ \%  &   &   $4.47^{+2.26}_{-2.05}$   &   $-5.25^{+2.21}_{-2.14}$   &   $0.13^{+0.04}_{-0.03}$   &   $-0.69^{+0.16}_{-0.10}$   &   N/A   &   N/A    \\ 
QEQ-8  &  165  &   &   $3.92^{+2.27}_{-2.01}$   &   $-7.52^{+1.08}_{-1.05}$   &   $1.39^{+0.32}_{-0.21}$   &   $-4.55^{+0.09}_{-0.06}$    &  $23.9^{+0.5}_{-0.6}$ \%  &   &   $5.13^{+2.24}_{-2.27}$   &   $-4.86^{+2.26}_{-2.29}$   &   $4.59^{+2.23}_{-2.18}$   &   $-5.70^{+2.22}_{-2.09}$   &   $-4.7^{+0.1}_{-0.1}$   &   $7.25^{+0.81}_{-0.81}$    \\ 
QEQ-9  &  169  &   &   $3.74^{+2.07}_{-1.90}$   &   $-7.35^{+1.09}_{-1.12}$   &   $1.70^{+0.34}_{-0.20}$   &   $-4.19^{+0.07}_{-0.05}$    &  $11.1^{+0.2}_{-0.3}$ \%  &   &   $5.09^{+2.22}_{-2.23}$   &   $-5.55^{+2.12}_{-2.06}$   &   $5.26^{+2.15}_{-2.11}$   &   $-4.69^{+2.17}_{-2.33}$   &   $-4.9^{+0.4}_{-0.4}$   &   $0.95^{+0.19}_{-0.17}$    \\ 
QEQ-11  &  532  &   &   $9.55^{+0.20}_{-0.21}$   &   $-0.03^{+0.01}_{-0.02}$   &   $4.58^{+1.99}_{-1.99}$   &   $-5.24^{+1.87}_{-1.94}$    &  $28.3^{+0.4}_{-0.4}$ \%  &   &   $1.38^{+0.07}_{-0.06}$   &   $-0.70^{+0.10}_{-0.09}$   &   $6.15^{+1.86}_{-2.10}$   &   $-3.85^{+1.80}_{-1.93}$   &   N/A   &   N/A    \\ 
QEQ-12  &  586  &   &   $1.80^{+0.02}_{-0.02}$   &   $-5.43^{+0.06}_{-0.06}$   &   $9.89^{+0.05}_{-0.07}$   &   $-0.66^{+0.11}_{-0.10}$    &  $41.0^{+0.5}_{-0.5}$ \%  &   &   $4.98^{+2.18}_{-2.15}$   &   $-5.83^{+2.14}_{-1.91}$   &   $5.44^{+2.05}_{-2.08}$   &   $-5.49^{+2.16}_{-2.12}$   &   $-4.6^{+0.0}_{-0.0}$   &   $7.42^{+0.80}_{-0.98}$    \\ 
QEQ-13  &  589  &   &   $0.99^{+0.09}_{-0.06}$   &   $-0.89^{+0.08}_{-0.07}$   &   $5.13^{+2.18}_{-2.17}$   &   $-4.80^{+2.23}_{-2.26}$    &  $19.7^{+0.7}_{-0.5}$ \%  &   &   $9.43^{+0.31}_{-0.41}$   &   $-1.12^{+0.08}_{-0.07}$   &   $9.22^{+0.40}_{-0.52}$   &   $-0.50^{+0.14}_{-0.14}$   &   N/A   &   N/A    \\ 
QEQ-18  &  907  &   &   $4.50^{+0.34}_{-0.36}$   &   $-0.07^{+0.03}_{-0.04}$   &   $6.36^{+1.79}_{-2.17}$   &   $-5.15^{+2.21}_{-2.17}$    &  $35.9^{+0.5}_{-0.5}$ \%  &   &   $5.23^{+2.20}_{-2.23}$   &   $-5.15^{+2.33}_{-2.25}$   &   $0.35^{+0.13}_{-0.11}$   &   $-0.92^{+0.43}_{-0.48}$   &   $-4.9^{+0.1}_{-0.1}$   &   $3.55^{+0.15}_{-0.15}$    \\ 
\hline  
\multicolumn{15}{c}{Scheme (a): Homogeneous Atmosphere (Figure~\ref{fig:inhomo}) }   \\ 
\hline  
QEQ-14  &  610  &   &   $0.08^{+0.00}_{-0.00}$   &   $-4.48^{+0.01}_{-0.01}$   &   $6.26^{+0.11}_{-0.11}$   &   $-0.10^{+0.04}_{-0.05}$    &  $100$ \%  &   &   N/A   &   N/A   &   N/A   &   N/A   &   N/A   &   N/A     \\ 
QEQ-15  &  622  &   &   $1.30^{+0.11}_{-0.14}$   &   $-0.10^{+0.04}_{-0.05}$   &   $0.88^{+0.05}_{-0.05}$   &   $-0.36^{+0.06}_{-0.05}$    &  $100$ \%  &   &   N/A   &   N/A   &   N/A   &   N/A   &   N/A   &   N/A     \\ 
QEQ-16  &  693  &   &   $0.14^{+0.00}_{-0.00}$   &   $-2.47^{+0.01}_{-0.01}$   &   $5.78^{+2.44}_{-2.81}$   &   $-6.06^{+2.27}_{-2.21}$    &  $100$ \%  &   &   N/A   &   N/A   &   N/A   &   N/A   &   N/A   &   N/A     \\ 
QEQ-17  &  753  &   &   $4.99^{+0.58}_{-0.46}$   &   $-1.82^{+0.09}_{-0.07}$   &   $1.57^{+0.04}_{-0.04}$   &   $-2.13^{+0.03}_{-0.03}$    &  $100$ \%  &   &   N/A   &   N/A   &   N/A   &   N/A   &   N/A   &   N/A     \\ 
QEQ-19  &  963  &   &   $0.09^{+0.00}_{-0.00}$   &   $-3.03^{+0.02}_{-0.01}$   &   $4.26^{+0.12}_{-0.12}$   &   $-0.44^{+0.21}_{-0.21}$    &  $100$ \%  &   &   N/A   &   N/A   &   N/A   &   N/A   &   N/A   &   N/A     \\ 
QEQ-20  &  984  &   &   $0.04^{+0.00}_{-0.00}$   &   $-2.90^{+0.01}_{-0.01}$   &   $4.14^{+0.15}_{-0.29}$   &   $-0.11^{+0.06}_{-0.08}$    &  $100$ \%  &   &   N/A   &   N/A   &   N/A   &   N/A   &   N/A   &   N/A     \\ 
QEQ-22  &  2468  &   &   N/A   &   N/A   &   N/A   &   N/A    &  $100$ \%  &   &   N/A   &   N/A   &   N/A   &   N/A   &   N/A   &   N/A     \\ 
QEQ-24  &  4753  &   &   N/A   &   N/A   &   N/A   &   N/A    &  $100$ \%  &   &   N/A   &   N/A   &   N/A   &   N/A   &   N/A   &   N/A     \\ 
\hline  
\multicolumn{15}{c}{Scheme (c): Hot Spots (Figure~\ref{fig:inhomo}) }   \\ 
\hline  
QEQ-21  &  1601  &   &   N/A   &   N/A   &   N/A   &   N/A    &  $28.5^{+0.4}_{-0.4}$ \%  &   &   N/A   &   N/A   &   N/A   &   N/A   &   N/A   &   N/A     \\ 
QEQ-23  &  3766  &   &   N/A   &   N/A   &   N/A   &   N/A    &  $42.0^{+16.5}_{-1.4}$ \%  &   &   N/A   &   N/A   &   N/A   &   N/A   &   N/A   &   N/A     \\ 
\enddata 
\tablenotetext{a}{For schemes with patchy clouds or a combination of patchy clouds and hot spots, the column 1 represents the thin-cloud patches and the column 2 represents the thick-cloud regions; the parameter $\xi_{\rm thin}$ describes the coverage of thin-cloud patches. In the hot-spot scheme, column 1 represents the atmospheric column with an overall hotter thermal structure (i.e., hot spots), while the column 2 represents the cold spots; the parameter $\xi_{\rm thin}$ describes the coverage of hot spots.}  
\end{deluxetable*} 
}

\section{Retrieving the JWST/NIRSpec spectrum of 2MASS 1207 \lowercase{b}} 
\label{sec:retrieval_planet}

We have performed multiple retrieval analyses on the JWST/NIRSpec spectrum of 2MASS~1207~b using an extensive array of setups. These setups explore the effects of variations in the following properties: 
\begin{enumerate}
\item[$\bullet$] Atmospheric inhomogeity. We investigate all four atmospheric schemes described in Section~\ref{subsec:inhomo_framework} and Figure~\ref{fig:inhomo}, including homogeneous atmospheres, patchy clouds, cloud-free hot spots, and the combination of patchy clouds and hot spots.   

\item[$\bullet$] Priors for surface gravity, radius, and mass. Both broad and narrow priors are explored for these parameters (Table~\ref{tab:retrieval_params}), with the narrow priors informed by thermal evolution models using the bolometric luminosity and age of 2MASS~1207~b (Section~\ref{sec:evo}).

\item[$\bullet$] Cloud composition. We consider several combinations of cloud species, including (1) Mg$_{2}$SiO$_{4}$ and Fe clouds, (2) Mg$_{2}$SiO$_{4}$, Fe, and gray clouds, (3) MgSiO$_{3}$ and Fe clouds, (4) MgSiO$_{3}$, Fe, and gray clouds, and (5) cloud-free atmospheres. For inhomogeneous atmospheres, we assume the same cloud composition for both atmospheric columns, but the cloud properties are allowed to vary between columns. When included, gray clouds are confined to the second atmospheric column. 

\item[$\bullet$] Cloud base pressure. Most of our retrievals compute cloud base pressures assuming chemical equilibrium. However, we treat the base pressures of magnesium silicate clouds and/or iron clouds as additional free parameters in certain retrieval runs. 
\end{enumerate}
We have conducted 24 retrieval runs with variations in the setups described above. These runs incorporate chemical equilibrium and quench pressures for carbon chemistry (CO, CH$_{4}$, H$_{2}$O) and are labeled as \texttt{QEQ} throughout this work. We further derive the posterior distribution of bolometric luminosities for each retrieval run by integrating the spectra computed from the retrieved parameters. Using the Stefan-Boltzmann law, we calculate the posterior distributions of effective temperatures based on $L_{\rm bol}$ and the retrieved $R$. Median values and confidence intervals of key retrieved parameters are summarized in Tables~\ref{tab:retrieved_qeq_key}--\ref{tab:retrieved_qeq_cloud}.

\begin{figure*}[t]
\begin{center}
\includegraphics[width=7in]{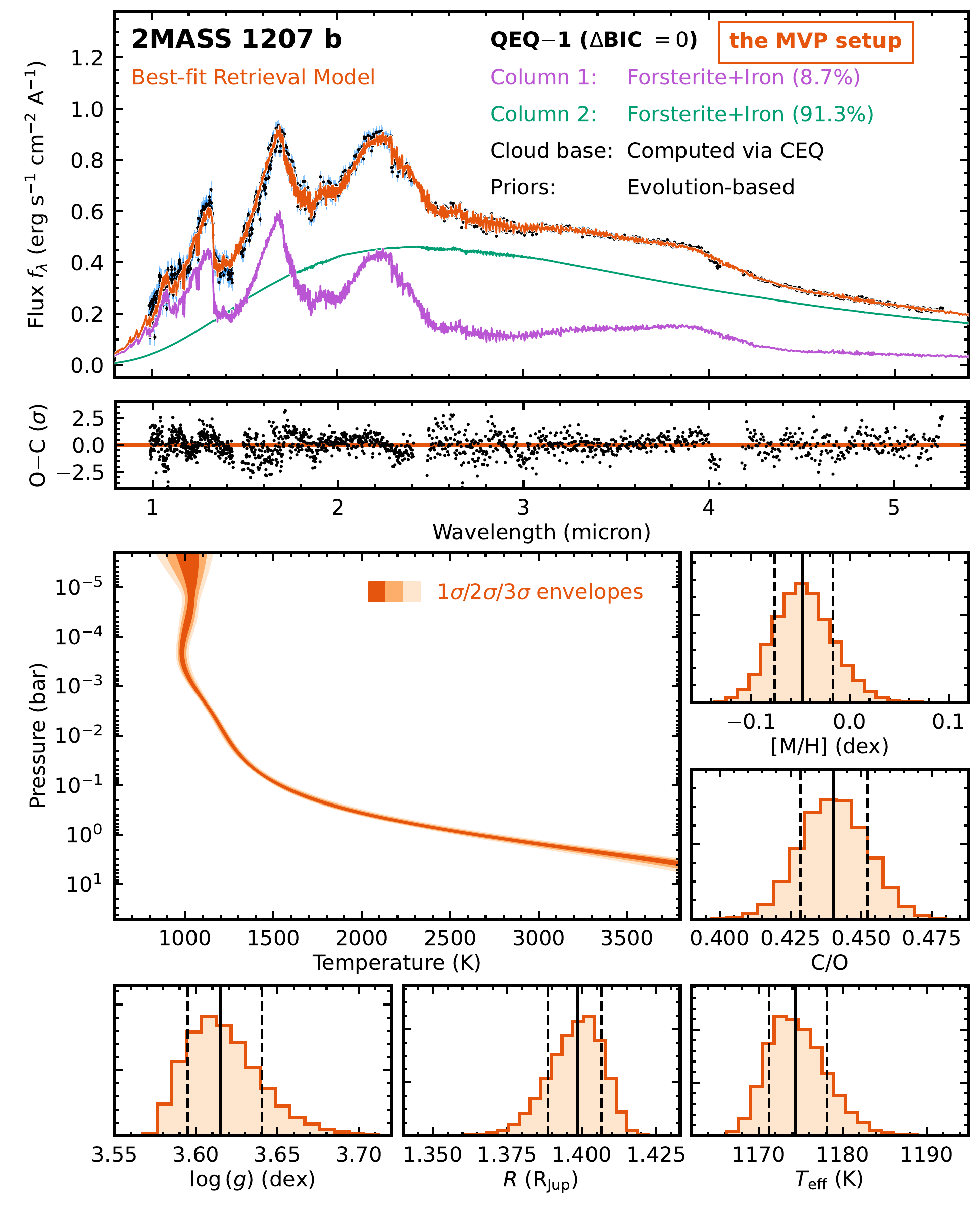}
\caption{Results of our most preferred retrieval setup, \texttt{QEQ-1} (see Tables~\ref{tab:retrieved_qeq_key}--\ref{tab:retrieved_qeq_cloud} and discussions in Section~\ref{subsec:inhomo}), which has the lowest BIC value. The top panel compares the observed JWST/NIRSpec spectrum of 2MASS~1207~b (black circles) with the best-fit spectrum from \texttt{QEQ-1} (orange). The emergent spectra from each atmospheric column are plotted in purple and green, with fluxes scaled by their corresponding coverage fractions. The blue error bars ($\sigma$) represent the total error budget, where $\sigma = \sqrt{\sigma_{\rm obs}^{2} + b_{\rm chunk}^{2}}$, with $\sigma_{\rm obs}$ as the observed flux uncertainty and $b_{\rm chunk}$ as the fitted hyper-parameter (explained in Section~\ref{subsec:params_priors}). The second panel from the top shows the residual spectrum between the observed and best-fit spectra, scaled by $\sigma$. The middle-left panel presents the confidence intervals for the retrieved T-P profiles. The remaining panels present the posteriors of key parameters, with median values and confidence intervals summarized in Table~\ref{tab:retrieved_qeq_key}. }
\label{fig:patchy_fesifor_evo}
\end{center}
\end{figure*}

\begin{figure*}[t]
\begin{center}
\includegraphics[width=7in]{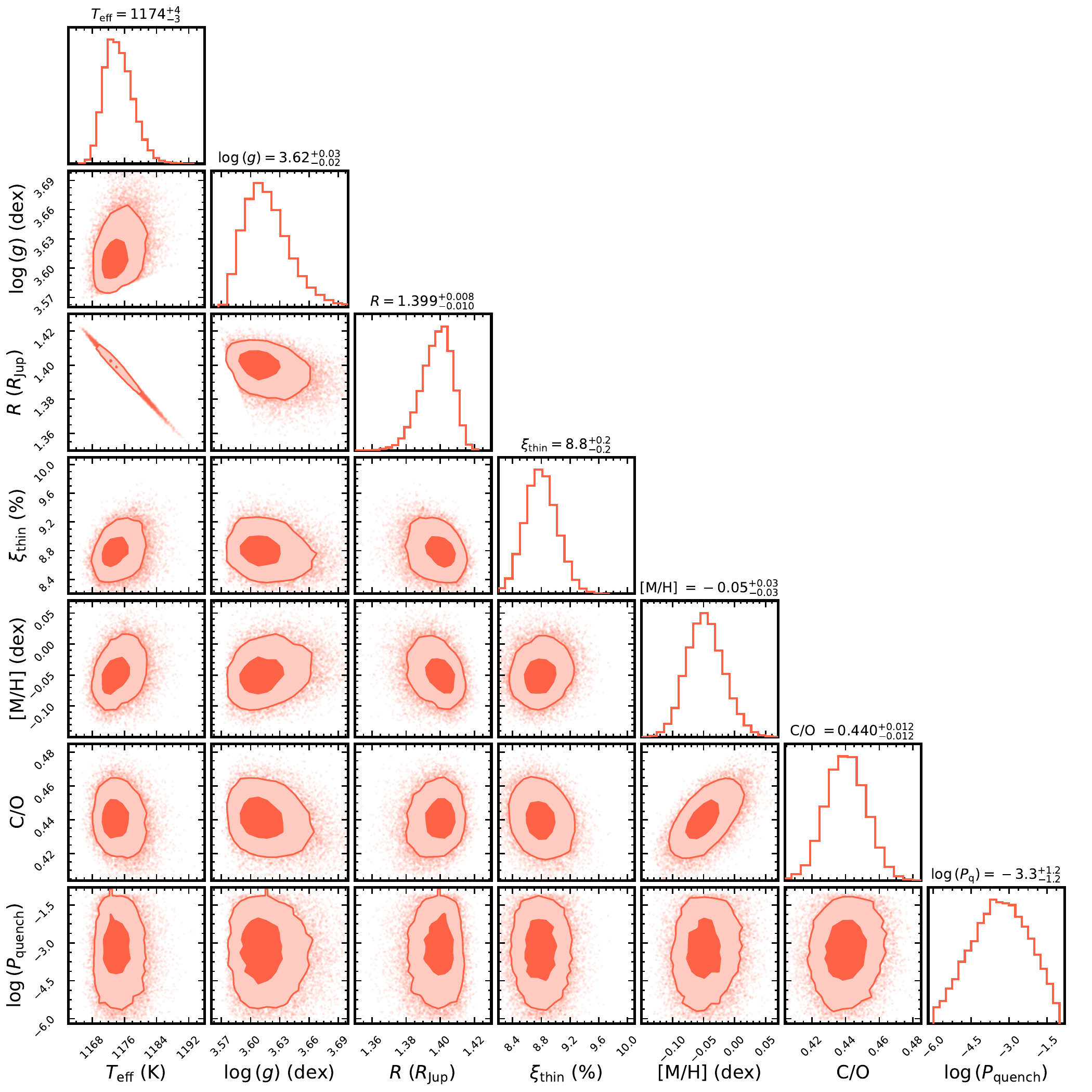}
\caption{Corner plot for the inferred parameters of 2MASS~1207~b based on the most preferred retrieval setup, \texttt{QEQ-1}. The $1\sigma$ and $2\sigma$ confidence intervals are highlighted for each parameter. }
\label{fig:corner}
\end{center}
\end{figure*}

To evaluate the performance of each retrieval run, we calculate the Bayesian Information Criterion \citep[BIC;][]{1978AnSta...6..461S, 2007MNRAS.377L..74L}:
\begin{equation}
{\rm BIC} \equiv - 2\ln{\mathcal{L}_{\rm max}} + k\ln{N} 
\end{equation}
where $N$ is the number of NIRSpec wavelength elements, $k$ is the number of free parameters, and $\mathcal{L}_{\rm max}$ is the maximum likelihood. The retrieval run with the lowest BIC value is considered the most preferred model for explaining the observations. The significance of this preference over another run is assessed by the difference in their BIC values, i.e., $\Delta$BIC. According to \cite{Kass1995}, such preference is absent if $0<\Delta$BIC$<2$, indicated if $2<\Delta$BIC$<6$, strongly indicated if $6<\Delta$BIC$<10$, and very strongly indicated if $\Delta$BIC$>10$.

In our analysis, the lowest BIC corresponds to a patchy cloud scenario with inhomogeneous forsterite and iron clouds, identified as the scheme (b) in our atmospheric inhomogeneity framework (Section~\ref{subsec:inhomo_framework}). This run outperforms other runs with $\Delta$BIC values ranging from 4 to nearly 5000, making it the most favored model for interpreting the JWST NIRSpec spectrum of 2MASS~1207~b. Throughout this manuscript, we refer to this most preferred model as \texttt{QEQ-1} and label other, less favored models as \texttt{QEQ-x}, where \texttt{x} denotes their ranks by BIC values. A larger \texttt{x} corresponds to a higher $\Delta$BIC and a less preferred model. We compare the observations with the fitted model spectra of \texttt{QEQ-1} in Figure~\ref{fig:patchy_fesifor_evo} and present the posterior distributions of key retrieved parameters in Figure~\ref{fig:corner}.\footnote{The best-fit model spectra and the fitted T-P profiles of all individual retrieval runs, as well as the \texttt{QEQ-1} parameter posteriors, are accessible through this Zenodo repository, \url{https://doi.org/10.5281/zenodo.15654303}} 

In the next section (Section~\ref{sec:discussion}), we provide a detailed discussion of these retrieval results, with a particular focus on the most preferred retrieval model, \texttt{QEQ-1}. In Section~\ref{subsec:inhomo}, we discuss the inhomogeneous atmosphere of 2MASS~1207~b, as consistently indicated by \texttt{QEQ-1} and several other moderately preferred retrieval runs, and compare these findings with atmospheric properties of Jupiter. In Section~\ref{subsec:weakCO_noCH4}, we explore unique spectral features of 2MASS~1207~b that have not been fully explained in previous studies. These features include weak CO absorption over 4.4--5.2~$\mu$m and the absence of 3.3~$\mu$m CH$_{4}$ absorption, which we demonstrate can be understood within the context of 2MASS~1207~b's inhomogeneous atmosphere, as revealed by our retrieval analyses. Section~\ref{subsec:unconstrained_quench} discusses that the presence of disequilibrium chemistry in the atmosphere of 2MASS~1207~b cannot be constrained via CH$_{4}$ quenching as commonly performed for colder L, T, and Y dwarfs, given this object's hot and CO-dominant atmosphere. We further place the results of \texttt{QEQ-1} in the context of 2MASS~1207~b's observed photometric variabilities in Section~\ref{subsec:variability}, and discuss the atmospheric [M/H] and C/O of this planetary-mass companion based on all retrieval runs in Section~\ref{subsec:mhco}. Finally, we comment on the results of retrievals with various atmospheric assumptions in Section~\ref{subsec:impact},

\begin{figure*}[t]
\begin{center}
\includegraphics[width=7in]{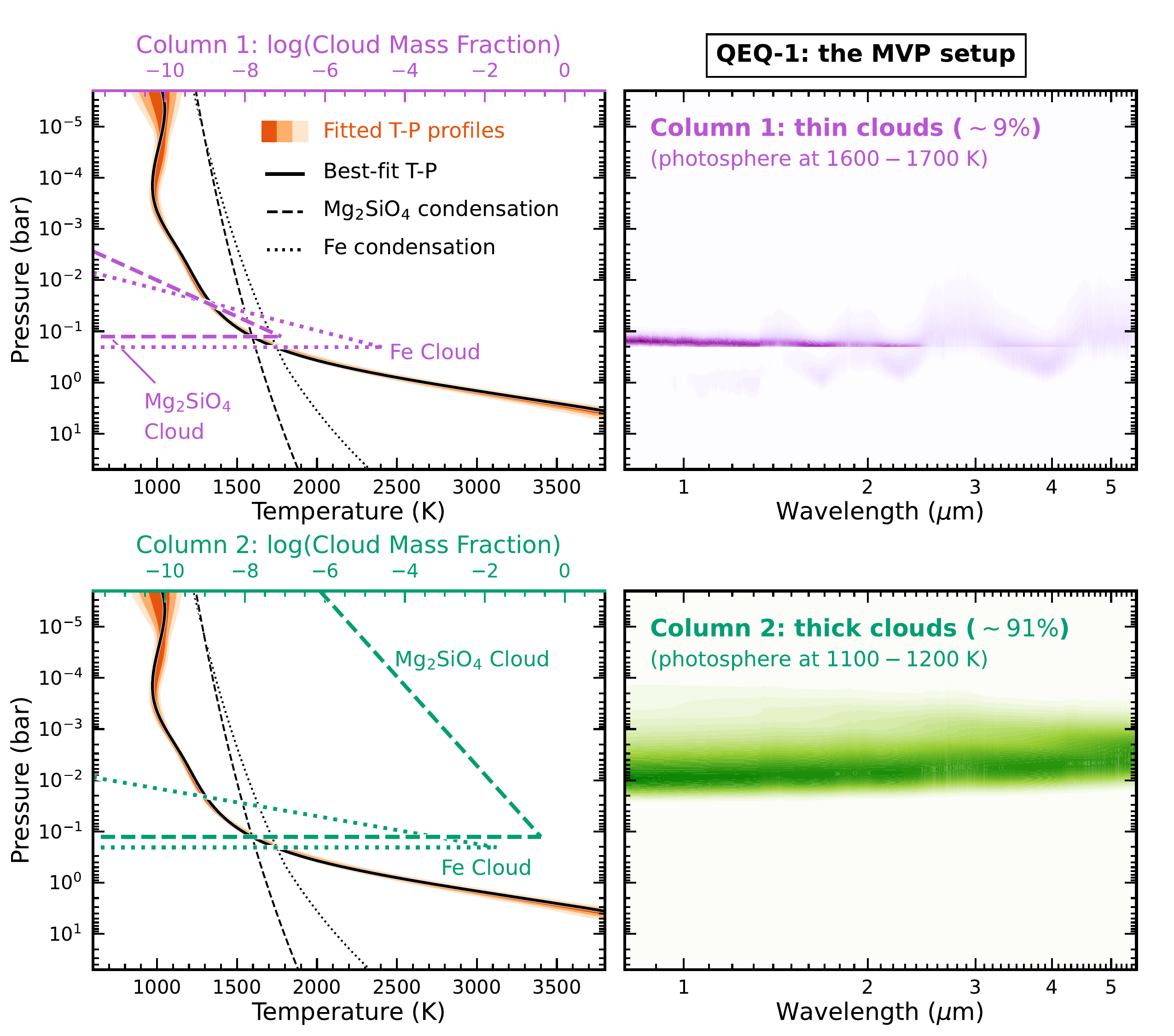}
\caption{{\it Top Left}: Retrieved T-P profiles from the \texttt{QEQ-1} setup, with $1\sigma/2\sigma/3\sigma$ confidence intervals shown in orange. The solid black line represents the best-fit T-P profile. Condensation curves of Mg$_{2}$SiO$_{4}$ and Fe, calculated using the maximum-likelihood [M/H] and C/O values, are shown by dashed and dotted lines, respectively. The best-fit cloud mass fraction profiles for the first column (thin-cloud patches covering about $9\%$ of 2MASS~1207~b's atmosphere) are plotted in purple, with values indicated on the upper x-axis.  {\it Top right}: Maximum-likelihood contribution function of the thin-cloud patches. {\it Bottom}: These two panels show the best-fit cloud content and contribution function for the second column (thick-cloud regions covering about $91\%$ of the atmosphere), following the same format as the top two panels.  }
\label{fig:qeq1_properties}
\end{center}
\end{figure*}

\section{Discussion} 
\label{sec:discussion}

\subsection{The Jupiter-like Inhomogeneous Atmosphere of 2MASS~1207~\lowercase{b}: $91\%$ thick clouds with $9\%$ thin-cloud patches}
\label{subsec:inhomo}

Figure~\ref{fig:patchy_fesifor_evo} presents the fitted model spectra of \texttt{QEQ-1}, the most favored model for interpreting the JWST/NIRSpec spectrum of 2MASS~1207~b in our retrieval analyses. As summarized in Table~\ref{tab:retrieved_qeq_key}, this model corresponds to the patchy cloud scheme illustrated in Figure~\ref{fig:inhomo} (b), incorporated with evolution-based priors for surface gravity, radius, and mass. \texttt{QEQ-1} yields an effective temperature of $T_{\rm eff} = 1174^{+4}_{-3}$~K, a surface gravity of $\log{(g)} = 3.62^{+0.03}_{-0.02}$~dex, a radius of $R = 1.399^{+0.008}_{-0.010}$~R$_{\rm Jup}$, a solar-like atmospheric metallicity of [M/H]$=-0.05 \pm 0.03$~dex, and an atmospheric C/O ratio of $0.440 \pm 0.012$ (see Section~\ref{subsec:mhco} for further discussions of atmospheric compositions).

Notably, \texttt{QEQ-1} suggests that the atmosphere of 2MASS~1207~b contains patchy forsterite and iron clouds with a significant contrast in cloud coverage between two atmospheric columns (Tables~\ref{tab:retrieved_qeq_key}--\ref{tab:retrieved_qeq_cloud}) that share the same thermal structure, detailed as below:
\begin{enumerate}
\item[$\bullet$] {\bf Thin-cloud patches ($\sim$9\% coverage).} About $9\%$ of the atmosphere has thin iron clouds with a mass fraction of $\log{(X_{\rm 0, Fe})} = -4.53^{+0.14}_{-0.08}$ at the base pressure and an intermediate sedimentation efficiency of $f_{\rm sed} = 4.5^{+0.7}_{-0.6}$. The forsterite cloud content is significantly lower, with a mass fraction of $\log{(X_{\rm 0, Mg_{2}SiO_{4}})} = -7.8 \pm 1.1$ at the base pressure. As shown in Figure~\ref{fig:patchy_fesifor_evo}, the emergent spectrum from these thin-cloud patches resembles that of a young L dwarf with a triangular $H$-band shape \citep[e.g.,][]{2007ApJ...669L..97L, 2013ApJ...772...79A} which aligns with the retrieved low surface gravity of $\log{(g)} = 3.62^{+0.02}_{-0.02}$~dex. As shown in Figure~\ref{fig:qeq1_properties}, the contribution function of this atmospheric column spans pressures from $10^{-2}$~bar to 1~bar, peaking at around 0.2~bar, meaning that the corresponding photosphere has a brightness temperature of 1600--1700~K.

\item[$\bullet$] {\bf Thick-cloud regions ($\sim$91\% coverage).} The remaining $91\%$ atmosphere is dominated by vertically extended Mg$_{2}$SiO$_{4}$ and iron clouds, with high mass fractions at the base pressures and low sedimentation efficiencies. The significant cloud opacity results in a contribution function distributed in the upper atmosphere, spanning pressures from $2\times10^{-4}$~bar to $2\times10^{-2}$~bar, meaning that the corresponding photosphere has a brightness temperature of 1100--1200~K (Figure~\ref{fig:qeq1_properties}). This leads to a blackbody-like emergent spectrum with flux peaking at $\sim 2.5$~$\mu$m (Figure~\ref{fig:patchy_fesifor_evo}). 
\end{enumerate}

Interestingly, the combination of thin-cloud patches with L-dwarf-like spectra and thick-cloud regions with blackbody-like spectra, as suggested by \texttt{QEQ-1}, is consistently supported by nearly all \texttt{QEQ} retrievals that include clouds and inhomogeneous atmospheres. In particular, when a gray cloud layer is incorporated into one of the atmospheric columns, the retrieved models indicate that the thick-cloud regions are characterized by a high-altitude gray cloud layer at low pressures around $10^{-5}$~bar (\texttt{QEQ-4, 5, 8, 9, 12, 18}). Additionally, retrieval runs with the seven lowest BIC values (\texttt{QEQ-1} to \texttt{QEQ-7}) indicate scenarios where $87\%-91\%$ of 2MASS~1207~b's atmosphere is covered by thick clouds, with $9\%-13\%$ covered by thin-cloud patches (Tables~\ref{tab:retrieved_qeq_key}--\ref{tab:retrieved_qeq_cloud}). These findings align with the results of \cite{2013ApJ...768..121A}, who suggested that the variability of brown dwarfs near the L/T transition is better explained by inhomogeneous atmospheres with a mixture of thin-cloud and thick-cloud patches rather than by a simple combination of clouds and purely cloud-free holes. 

The inhomogeneous atmosphere of 2MASS~1207~b is analogous to that of Jupiter, which exhibits a banded structure of zones and belts \citep[e.g.,][]{1981Natur.292..677P, 1996Sci...274..377B, 1999gpoj.book.....H, 2020SSRv..216...30F, 2021JGRE..12606858F}. Zones are cloudy regions that strongly reflect sunlight, while belts are thin-cloud regions that appear darker in reflected light. Although these dark belts occupy only about $1\%$ of Jupiter's atmosphere \citep[e.g.,][]{1969ApJ...157L..63W, 1996Sci...272..839O}, their thermal emission dominates Jupiter's spectrum near 5~$\mu$m.\footnote{These dark regions, often referred to as ``5-$\mu$m hot spots'' in the literature, are relatively cloud-free areas that probe deeper, hotter atmospheric layers. This concept of hot spots aligns with the scheme (b) ``patchy clouds'' in our atmospheric inhomogeneity framework (Section~\ref{subsec:inhomo_framework} and Figure~\ref{fig:inhomo}) and differs from the scheme (c) ``cloud-free hot spots'' in our framework; the latter involves two cloud-free columns with different T-P profiles.} Our \texttt{QEQ-1} retrieval suggests that 2MASS~1207~b has thin-cloud patches covering $9\%$ of the atmosphere, with photospheric temperatures of 1600--1700~K, and thick-cloud regions covering $91\%$, with slightly lower photospheric temperatures of 1100--1200~K. These features resemble Jupiter's belts and zones, respectively. As shown in Figure~\ref{fig:patchy_fesifor_evo}, the thin-cloud patches produce much brighter or comparable fluxes than the thick-cloud regions at wavelengths below 2.3~$\mu$m. Consequently, the near-infrared spectral features of 2MASS~1207~b in $Y$, $J$, and $H$ bands predominantly arise from these thin-cloud patches, while this object's unusually red spectral slope is attributed to the thick clouds. 

The thick clouds and potential high-altitude gray cloud content of 2MASS~1207~b, as suggested by our retrievals, may be contributed by this object's ongoing accretion \citep{2023ApJ...949L..36L, 2024ApJ...964...70M, 2024AJ....167..168M} of dust and gas from the circumstellar disk of 2MASS~1207~A \citep[e.g.,][]{2017AJ....154...24R} which has an estimated dust mass of $\sim 0.1$~M$_{\oplus}$ \citep[e.g.,][]{2013ApJ...773..168M, 2017AJ....154...24R}. Additionally, 2MASS~1207~b may host its own disk, though if present, it likely has a low dust mass below 1~M$_{\rm moon}$ \citep[][]{2017AJ....154...24R}. Similar to events that have influenced Jupiter --- such as the impacts of Comet Shoemaker-Levy 9 \citep[e.g.,][]{1995Sci...267.1288H} --- such processes may also influence the atmospheric evolution of 2MASS~1207~b. Follow-up studies similar to \cite{2023AJ....165..238M} would be valuable in assessing whether accretion activity can match the observed cloud properties of 2MASS~1207~b, as well as the present-day atmospheric compositions of both 2MASS~1207~A and b. Furthermore, future retrieval studies that incorporate both NIRSpec and MIRI spectra of 2MASS~1207~b will be crucial for more precisely constraining the type and amount of silicate clouds in its atmosphere and also the presence and properties of this object's circumsecondary disk, which, if present, could contribute to the flux at MIRI wavelengths \citep[][]{2023ApJ...949L..36L}.

\begin{figure*}[t]
\begin{center}
\includegraphics[width=6in]{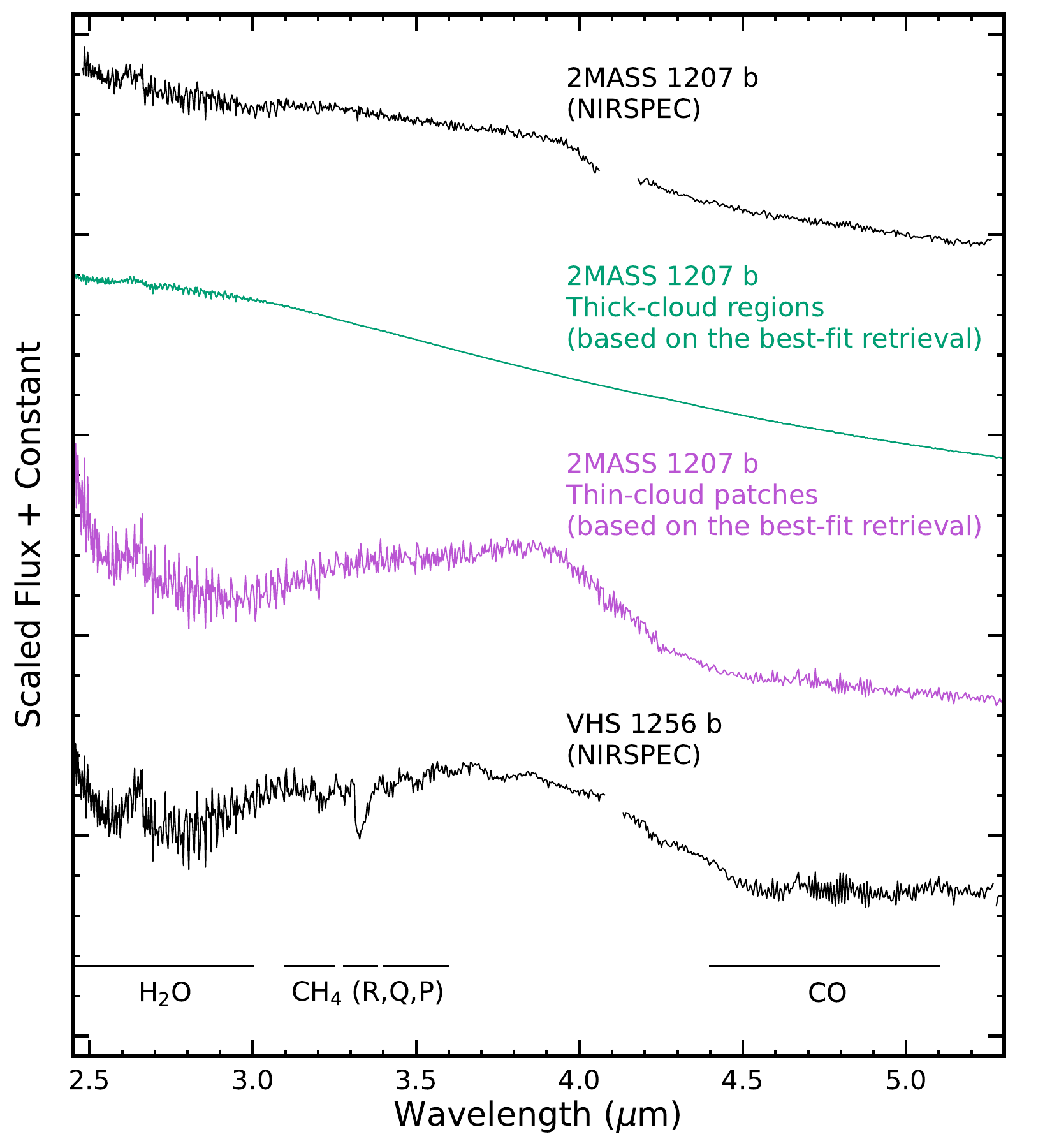}
\caption{The top three spectra show the observed JWST/NIRSpec data of 2MASS~1207~b (black), as well as the best-fit retrieval model spectra emerging from the thick-cloud regions (green) and thin-cloud patches (purple) based on the \texttt{QEQ-1} setup. The JWST/NIRSpec spectrum of VHS~1256~b \citep{2023ApJ...946L...6M} is shown at the bottom. All spectra are plotted at a spectral resolution of $R = 1000$. Key molecular absorption features are labeled. }
\label{fig:2m1207b_vs_vhs1256b}
\end{center}
\end{figure*}

\begin{figure*}[t]
\begin{center}
\includegraphics[width=6.5in]{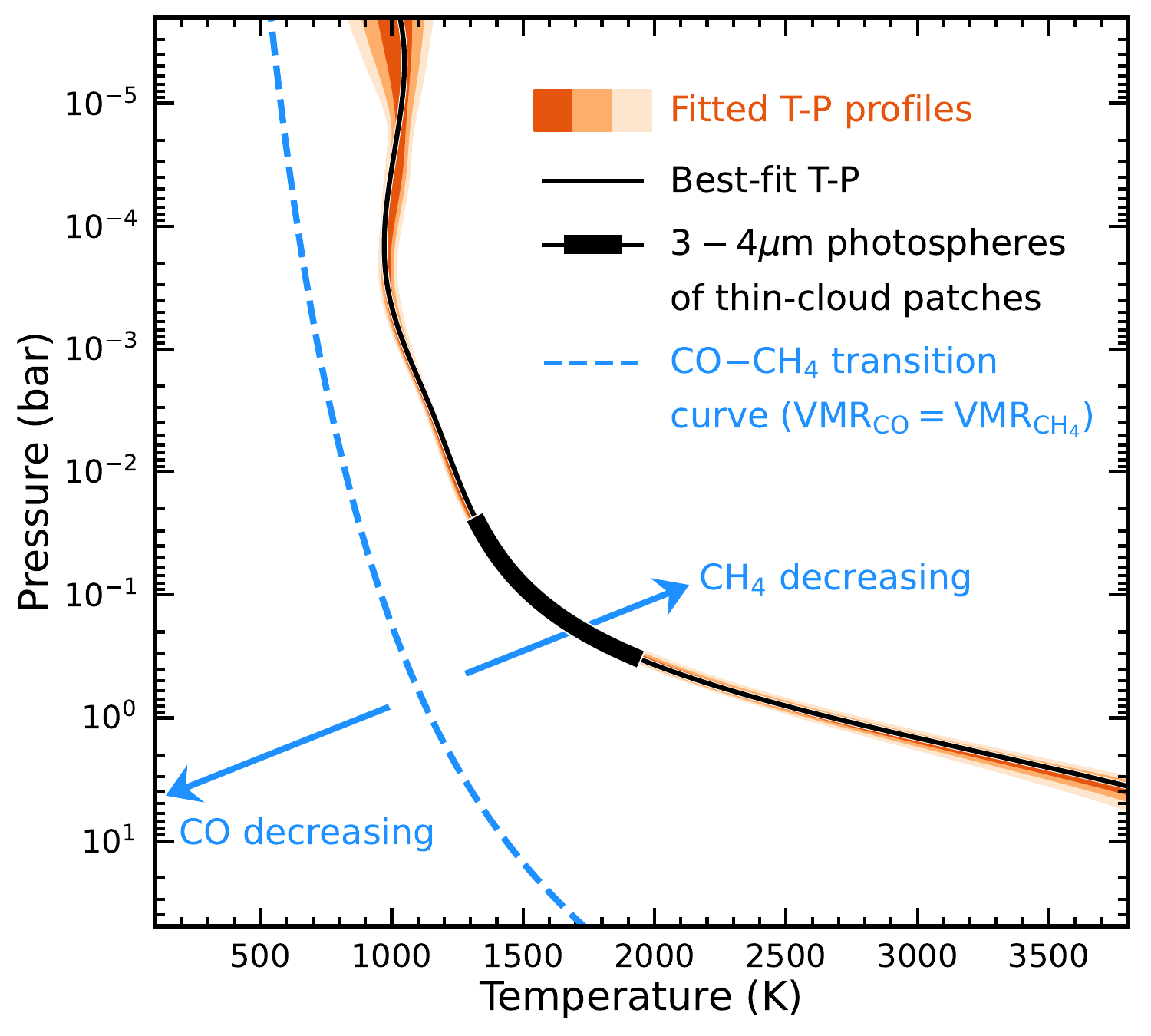}
\caption{Retrieved T-P profiles from the \texttt{QEQ-1} setup, with $1\sigma$, $2\sigma$, and $3\sigma$ confidence intervals shown in orange. The best-fit T-P profile is indicated by the solid black line, with the $3-4$~$\mu$m photosphere of thin-cloud patches highlighted by a thicker line. The CO$-$CH$_{4}$ transition curve for solar metallicity, from \citep{2020AJ....160..288F}, is overlaid as a dashed blue line; along this line, the CO and CH$_{4}$ have equal volume mixing ratios. A thermal structure with hotter temperatures and/or lower pressures than this transition curve would have a decreased CH$_{4}$ abundance, while a cooler and/or higher-pressure thermal structure would have a decreased CO abundance. }
\label{fig:tp_vs_f20}
\end{center}
\end{figure*}

\begin{figure*}[t]
\begin{center}
\includegraphics[width=6.5in]{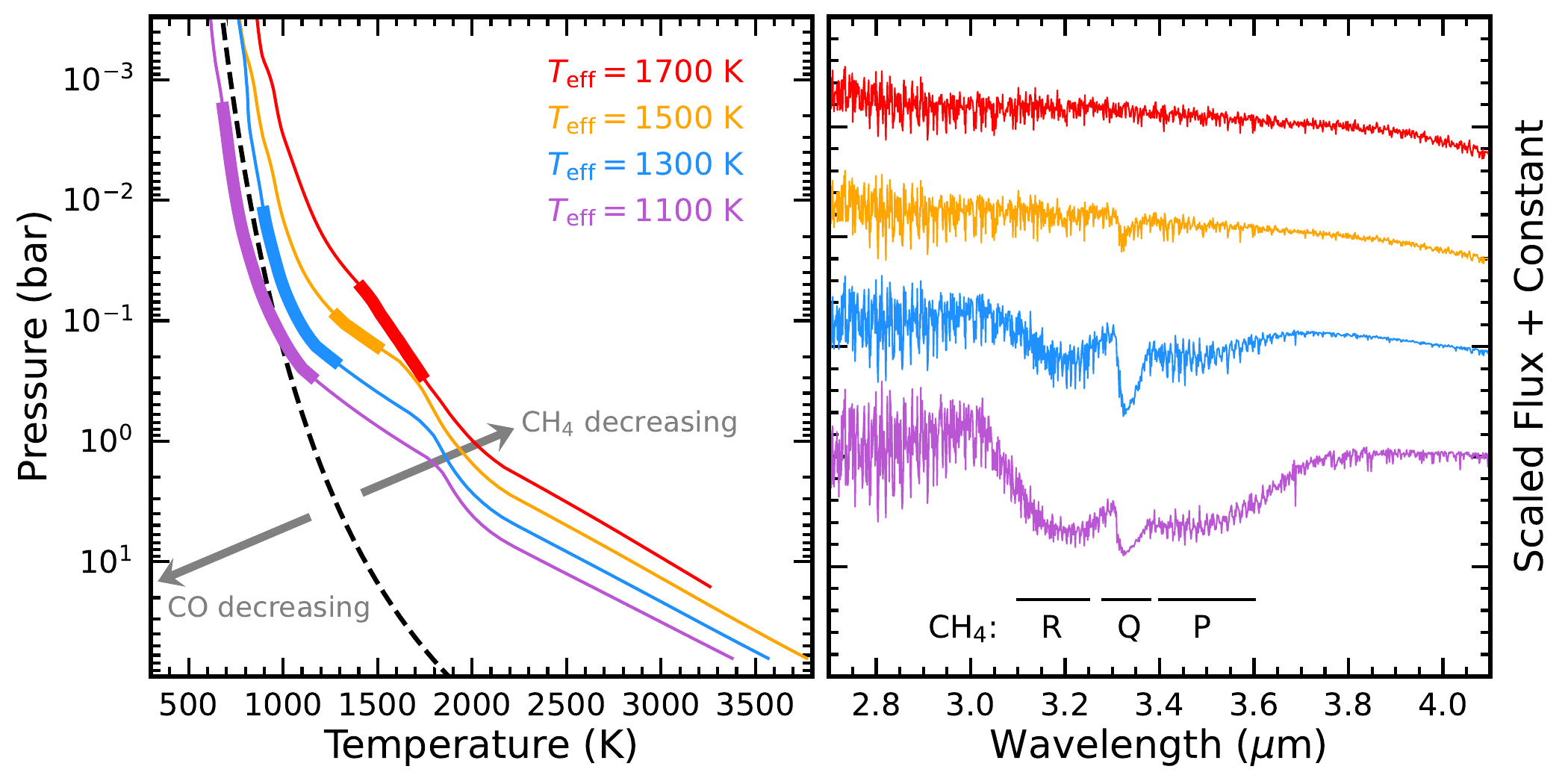}
\includegraphics[width=6.5in]{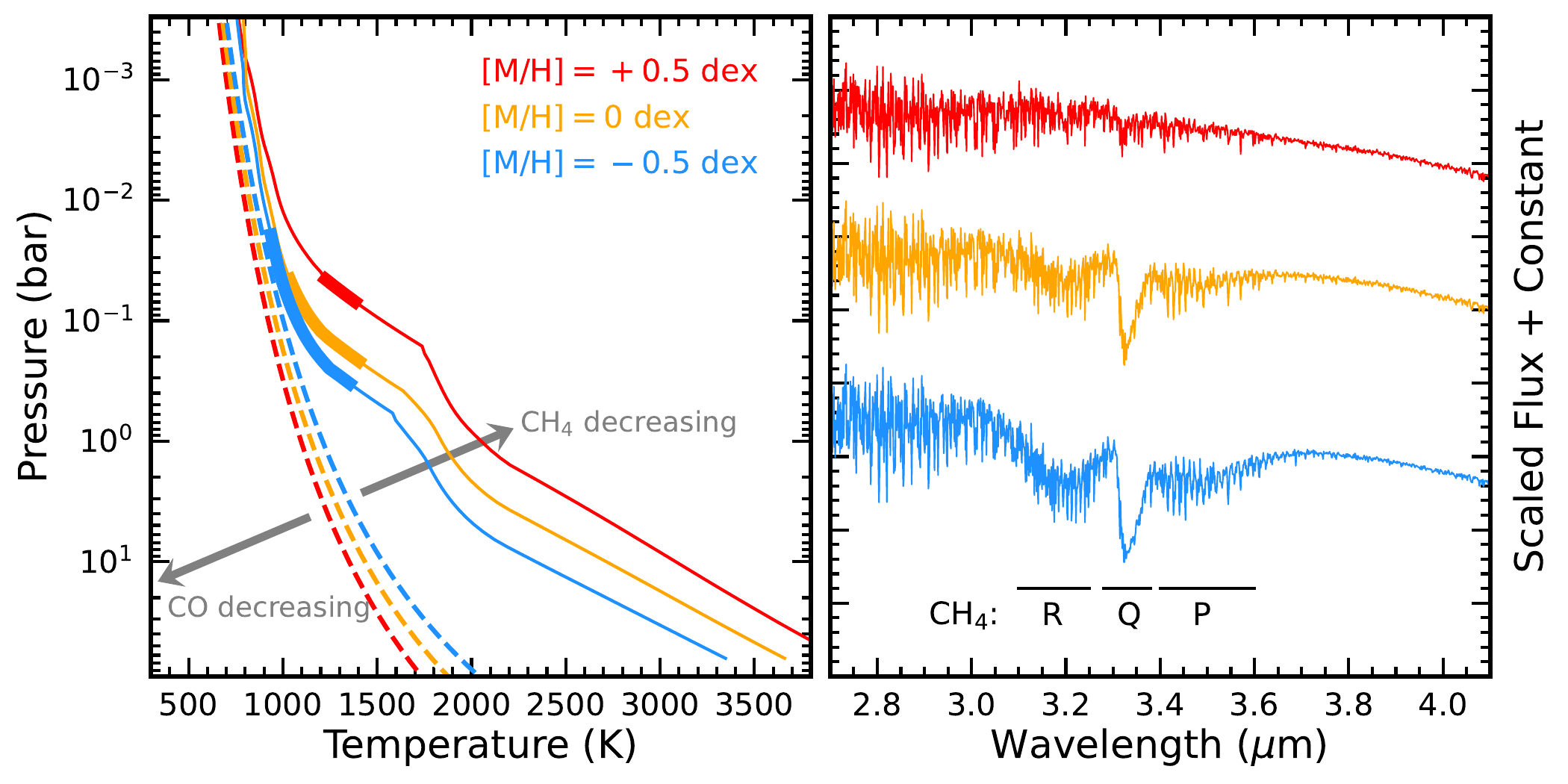}
\caption{{\it Top}: The left panel shows T-P profiles from the \texttt{Sonroa Diamondback} models with varying effective temperatures: $T_{\rm eff} = 1700$~K (red), $1500$~K (orange), 1300~K (blue), and 1100~K (purple). All models have $\log{(g)} = 4.5$~dex, [M/H]$=0$~dex, and $f_{\rm sed} = 2$. The $3-4$~$\mu$m photosphere of each model is highlighted with a thick line. The CO$-$CH$_{4}$ transition curve (see Figure~\ref{fig:tp_vs_f20}) is shown as a dashed line. The right panel presents the corresponding spectra, with CH$_{4}$ absorption features (R, Q, and P branches) labeled.  {\it Bottom}: These two panels follow the same format as the top panels but present the T-P profiles and spectra for models with different [M/H] values: $+0.5$~dex (red), $0$~dex (orange), and $-0.5$~dex (blue). All models have $T_{\rm eff} = 1400$~K, $\log{(g)} = 4.5$~dex, and $f_{\rm sed} = 2$. }
\label{fig:onset_ch4_teffmh}
\end{center}
\end{figure*}

\begin{figure*}[t]
\begin{center}
\includegraphics[width=6.5in]{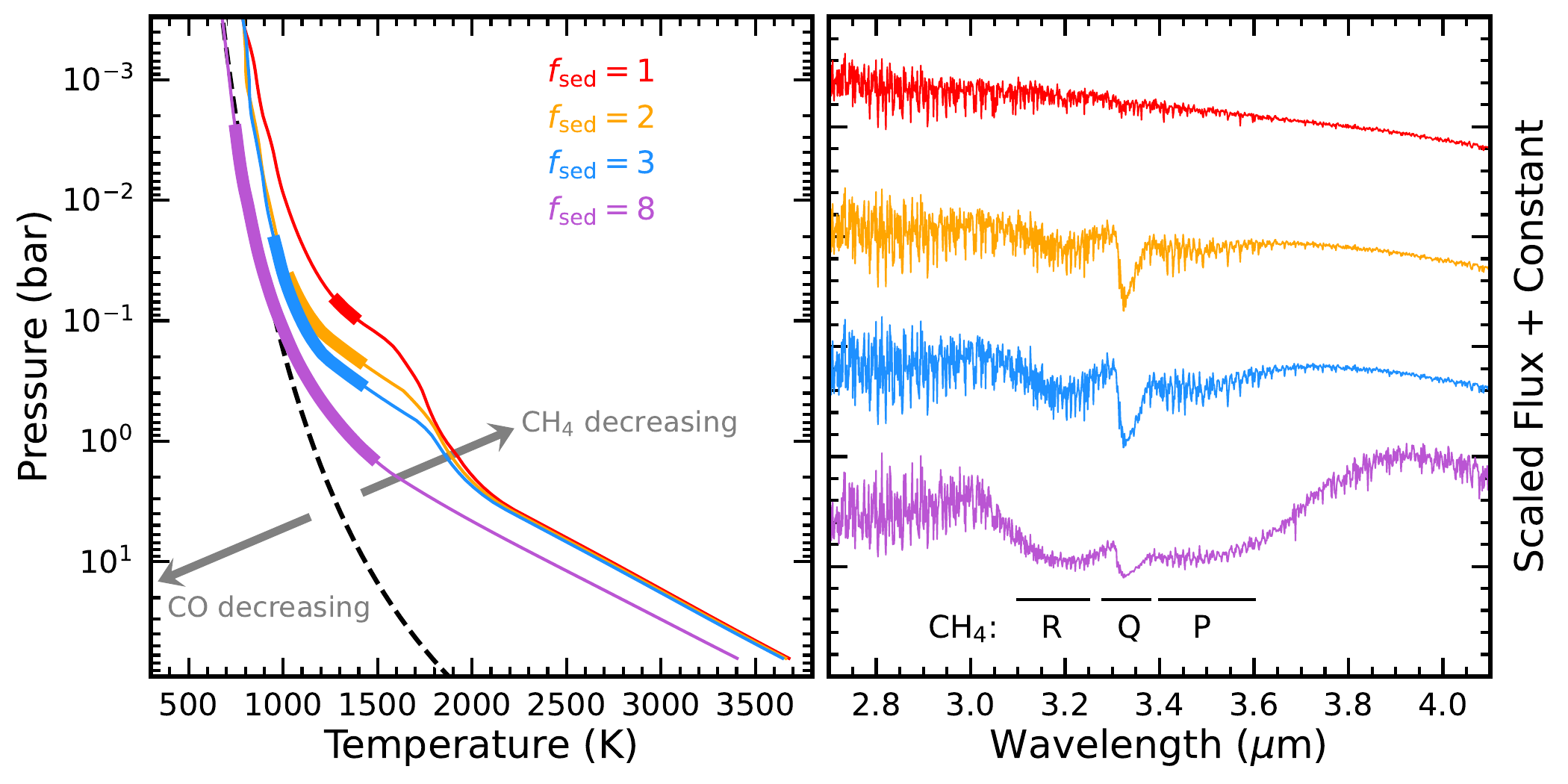}
\includegraphics[width=6.5in]{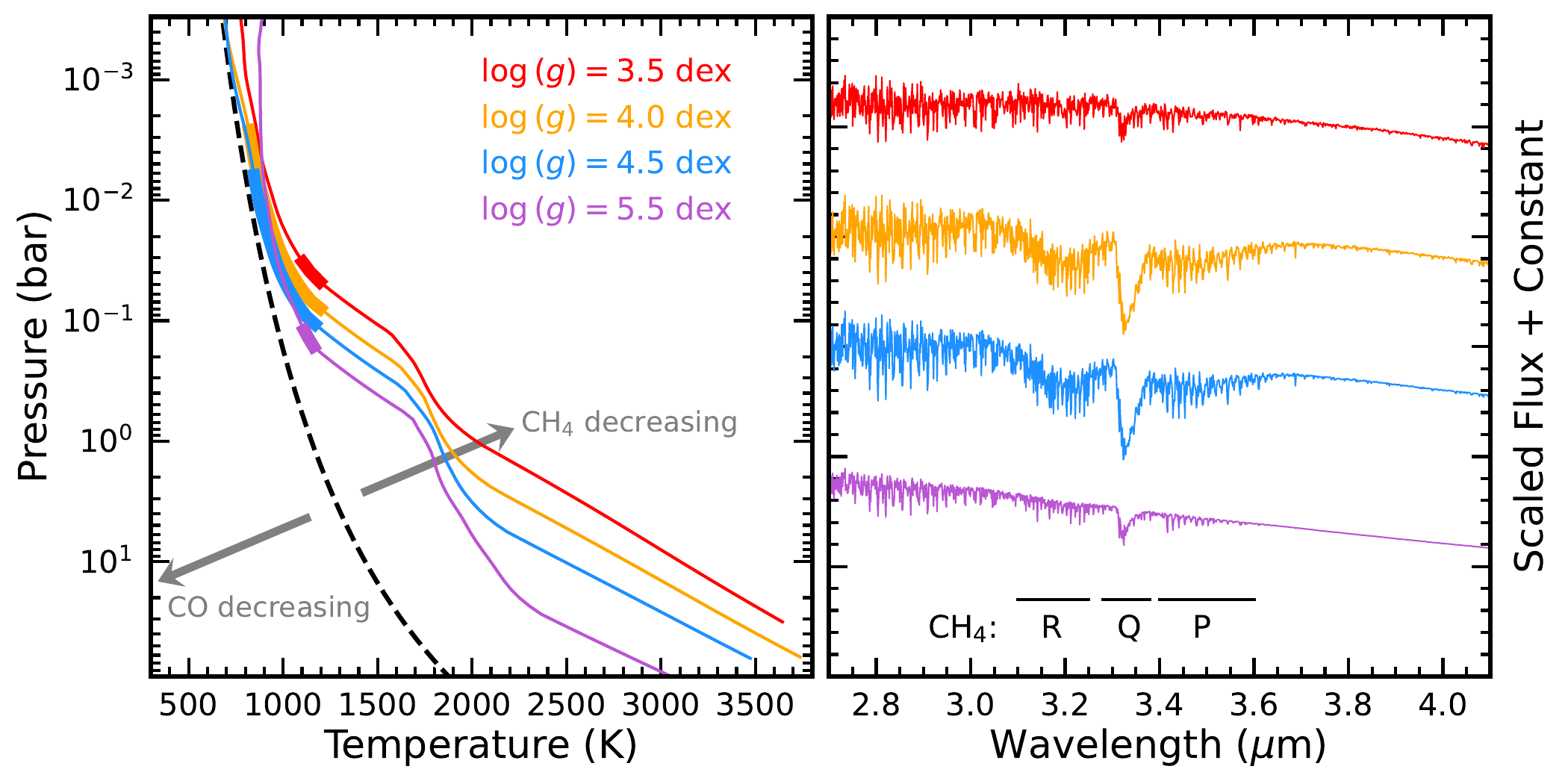}
\caption{The top two panels present T-P profiles and corresponding spectra from the \texttt{Sonroa Diamondback} models with varying  $f_{\rm sed}$ values of $1$ (red), $2$ (orange), $3$ (blue), and $8$ (purple). All models have $T_{\rm eff} = 1400$~K, $\log{(g)} = 4.5$~dex, and [M/H]$=0$~dex. The bottom two panels show models with varying $\log{(g)}$ of $3.5$~dex (red), $4.0$~dex (orange), $4.5$~dex (blue), and $5.5$~dex, all with $T_{\rm eff} = 1200$~K, [M/H]$=0$~dex, and $f_{\rm sed} = 1$. These panels follow the same format as Figure~\ref{fig:onset_ch4_teffmh}. }
\label{fig:onset_ch4_fsedlogg}
\end{center}
\end{figure*}

\subsection{Weak CO and Absent CH$_{4}$ absorption: Net Effects of Patchy Thick Clouds and a Hot Thermal Stucture}
\label{subsec:weakCO_noCH4}

The JWST/NIRSpec spectrum of 2MASS~1207~b offers unprecedented insights into this object's atmosphere, surpassing previous ground-based near-infrared ($1-2.5$~$\mu$m) spectroscopy \citep[e.g.,][]{2004A&A...425L..29C, 2007ApJ...657.1064M, 2010A&A...517A..76P}, and revealing peculiar features that remain unexplained \citep{2023ApJ...949L..36L, 2024AJ....167..168M}. Notably, the $2.5-5.0$~$\mu$m spectrum of 2MASS~1207~b differs significantly from that of VHS~J125601.92$-$125723.9~b \citep[VHS~1256~b;][]{2015ApJ...804...96G, 2023ApJ...946L...6M}, despite similarities in their $1-2.5$~$\mu$m spectra (Figure~\ref{fig:2m1207b_vs_vhs1256b}; see also Figure~6 of \citealt{2023ApJ...949L..36L}). Specifically, 2MASS~1207~b exhibits much weaker CO absorption in the $4.4-5.2$~$\mu$m compared to VHS~1256~b. Also, the prominent 3.3~$\mu$m CH$_{4}$ absorption seen in VHS~1256~b is absent in 2MASS~1207~b's NIRSpec spectrum.

These unique features of 2MASS~1207~b have not been explained by forward modeling efforts. When fitting the \texttt{ATMO} and \texttt{BT-Settl} grid models \citep{2003IAUS..211..325A, 2012RSPTA.370.2765A, 2015ApJ...804L..17T, 2017ApJ...850...46T, 2024ApJ...966L..11P} to 2MASS~1207~b's NIRSpec spectrum, both \cite{2023ApJ...949L..36L} and \cite{2024AJ....167..168M} found that these models overpredict the strengths of CH$_{4}$ and CO absorption compared to the observations. As both studies acknowledged, disequilibrium chemistry can weaken CH$_{4}$ absorption but typically strengthens CO absorption, which conflicts with the observed weak CO features. \cite{2023ApJ...949L..36L} suggested that the weaker CO absorption in 2MASS~1207~b compared to its slightly older counterpart, VHS~1256~b \citep[$\sim 140$~Myr;][]{2023MNRAS.519.1688D}, indicates that certain atmospheric property, such as thermal structure and cloud thickness, may vary with age. Similarly, \cite{2024AJ....167..168M} attributed the unique spectral features of 2MASS~1207~b to its lower surface gravity (or younger age) relative to VHS~1256~b. However, detailed and quantitative explanations for the distinctive CO and CH$_{4}$ features of 2MASS~1207~b remain needed.

As detailed in the following subsections, our retrieval analysis shows that the weak CO absorption observed in 2MASS1207b can be attributed to its patchy clouds (Section~\ref{subsubsec:weakco_explained}). Furthermore, the absence of CH$_{4}$ absorption is consistent with this object's relatively hot atmospheric thermal structure (Section~\ref{subsubsec:noch4_explained}).

\subsubsection{Weak CO absorption due to patchy cloud veiling effects}
\label{subsubsec:weakco_explained}
As discussed in Section~\ref{subsec:inhomo}, the atmosphere of 2MASS~1207~b is characterized by thick-cloud regions and thin-cloud patches. While the thick clouds produce a nearly featureless blackbody-like spectrum, the emergent flux from the thin-cloud patches exhibits prominent absorption features of H$_{2}$O, CO, Na~\textsc{i}, K~\textsc{i}, FeH, and VO (Figure~\ref{fig:patchy_fesifor_evo}). As shown in Figure~\ref{fig:2m1207b_vs_vhs1256b}, the $2.5-5.0$~$\mu$m spectrum from the thin-cloud patches of 2MASS~1207~b closely resembles that of VHS~1256~b, with similar H$_{2}$O and CO absorption bands. However, these absorption features are diluted in the combined emergent spectrum due to (1) the limited $9\%$ coverage of thin clouds and (2) the blackbody-like thermal emission from the thick clouds, which is $2-5$~times brighter than the fluxes from thin-cloud patches across $2.5-5.0$~$\mu$m (Figure~\ref{fig:patchy_fesifor_evo}). As a result, the observed weak CO absorption in 2MASS~1207~b is primarily a result of veiling by the patchy thick clouds that dominate its atmosphere.

\subsubsection{Absent CH$_{4}$ absorption due to the hot atmospheric thermal structure}
\label{subsubsec:noch4_explained}

The absence of 3.3~$\mu$m CH$_{4}$ absorption in 2MASS1207b cannot be explained by cloud veiling effects, as discussed in Section~\ref{subsubsec:weakco_explained}, because this feature is not predicted by our retrieved spectrum from either thin-cloud patches or thick-cloud regions. Instead, we attribute the lack of CH${4}$ absorption to this object's relatively hot atmospheric thermal structure.

In Figure~\ref{fig:tp_vs_f20}, we compare the retrieved T-P profile with a critical curve corresponding to an equilibrium situation where the volume mixing ratios of CH$_{4}$ and CO are equal. This curve, $\log{(P)} = 5.05 - 5807.5/T + 0.5 \times$[M/H], was derived by \cite{2020AJ....160..288F}. Deviations from this curve lead to either a decrease in CH$_{4}$ abundance at higher temperatures and lower pressures, or a decrease in CO abundance at lower temperatures and higher pressures. Objects with thermal structures hotter than this CO--CH$_{4}$ transition curve tend to exhibit weak or absent 3.3~$\mu$m CH$_{4}$ absorption, while cooler objects display stronger CH$_{4}$ absorption. We demonstrate this effect in Figures~\ref{fig:onset_ch4_teffmh}--\ref{fig:onset_ch4_fsedlogg} using the Sonora Diamondback atmospheric models \citep{2024ApJ...975...59M}.

The onset of CH$_{4}$ absorption occurs when an object's $3-4$~$\mu$m photosphere cools sufficiently to approach the CH$_{4}$--CO transition curve. Our analysis reveals that this ``CH$_{4}$ onset'' depends on several physical parameters, including $T_{\rm eff}$, [M/H], $f_{\rm sed}$, and $\log{(g)}$, as summarized below:
\begin{enumerate}
\item[$\bullet$] Effective temperature. As shown in Figure~\ref{fig:onset_ch4_teffmh}, CH$_{4}$ absorption begins at $T_{\rm eff} \approx 1600$~K for objects with $\log{(g)} = 4.5$~dex, [M/H]$=0$~dex, and $f_{\rm sed} = 2$. This critical $T_{\rm eff}$ corresponds to a spectral type of $\sim$L4 \citep[e.g.,][]{2015ApJ...810..158F}, consistent with the findings of \cite{2000ApJ...541L..75N}, who observed CH$_{4}$ absorption in L5 and L7 dwarfs but not in an L3 dwarf with a hotter effective temperature . 

\item[$\bullet$] Metallicity. As shown in Figure~\ref{fig:onset_ch4_teffmh}, higher metallicity shifts the photosphere to lower pressures, moving it farther from the CO--CH$_{4}$ transition curve. For objects with $T_{\rm eff} = 1400$~K, $\log{(g)} = 4.5$~dex, and $f_{\rm sed} = 2$, the strength of the 3.3~$\mu$m CH$_{4}$ absorption increases as [M/H] decreases from $+0.5$~dex to $-0.5$~dex (Figure~\ref{fig:onset_ch4_teffmh}. This trend suggests that CH$_{4}$ onset occurs at a [M/H] slightly above $+0.5$~dex for this particular parameter set.

\item[$\bullet$] Sedimentation efficiency. Figure~\ref{fig:onset_ch4_fsedlogg} shows that more cloudy atmospheres (e.g., $f_{\rm sed}  = 1$) have $3-4$~$\mu$m photospheres farther from the CO--CH$_{4}$ transition curve compared to less cloudy objects (e.g., $f_{\rm sed} = 8$). For $T_{\rm eff} = 1400$~K, $\log{(g)} = 4.5$~dex, and [M/H]$=0$~dex, the CH$_{4}$ onset occurs at $f_{\rm sed} \sim 2$. 

\item[$\bullet$] Surface gravity. The relationship between CH$_{4}$ onset and $\log{(g)}$ is nonlinear. As shown in Figure~\ref{fig:onset_ch4_fsedlogg}, for $T_{\rm eff} = 1200$~K, [M/H]$=0$~dex, and $f_{\rm sed} = 1$, objects with $\log{(g)} = 3.5$~dex show similar CH$_{4}$ absorption strength to those with $\log{(g)} = 5.5$~dex, as their $3-4$~$\mu$m photospheres are similarly distant from the CO--CH$_{4}$ transition curve. However, these models exhibit weaker CH$_{4}$ absorption at 3.3~$\mu$m compared to those with $\log{(g)} = 4.0-4.5$~dex, whose T-P profiles closely approach the CO$-$CH$_{4}$ transition curve.
\end{enumerate}
In summary, the onset of CH$_{4}$ absorption is governed by parameters that significantly alter an object's thermal structure. The presence of vertical mixing \citep[e.g.,][]{2014ApJ...797...41Z} and any variations in the atmospheric adiabatic index \citep[e.g.,][]{2016ApJ...817L..19T}, would also impact the CH$_{4}$ onset.

For 2MASS~1207~b, our retrieved atmospheric thermal structure is hotter than the CO--CH$_{4}$ transition curve at a given pressure layer (Figure~\ref{fig:tp_vs_f20}), explaining the observed absence of CH$_{4}$ absorption features.

\begin{figure*}[t]
\begin{center}
\includegraphics[width=7.in]{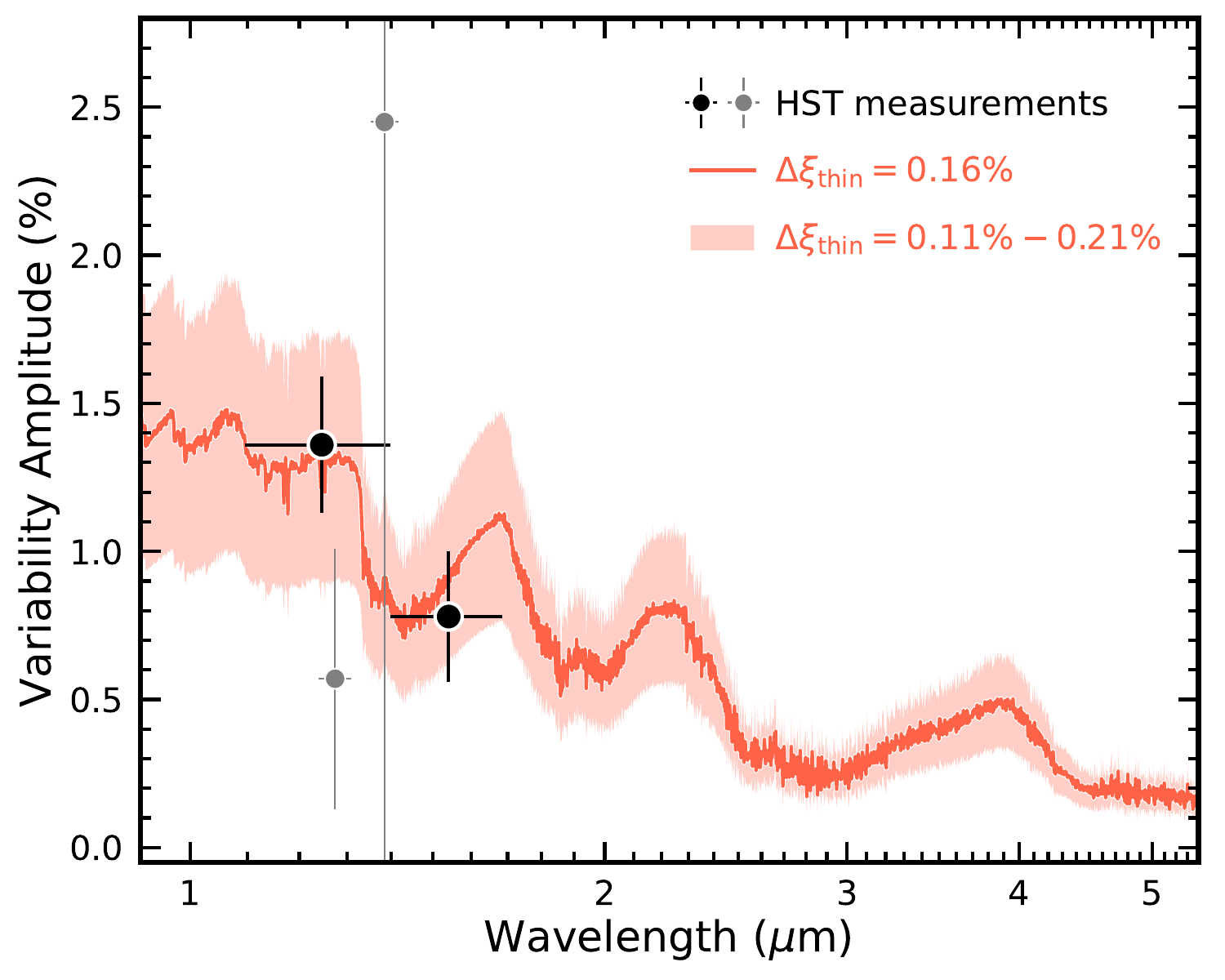}
\caption{The variability amplitude as a function of wavelength predicted by our preferred retrieval setup, \texttt{QEQ-1}. The solid red line represents the variability amplitude when the fraction of thin-cloud patches changes by $\Delta \xi_{\rm col1} = 0.16\%$ (solid red line) from the maximum-likelihood value. The red-shaded region indicates the range of variability amplitudes when $\Delta \xi_{\rm col1} = 0.11\%-0.21\%$. Black and grey circles with error bars represent the measured variability amplitudes from \citet{2016ApJ...818..176Z}, with grey circles corresponding to measurements from \citet{2019AJ....157..128Z}, which are affected by systematics.}
\label{fig:varamp}
\end{center}
\end{figure*}

\subsection{Unconstrained Carbon-Chemistry Quench Pressure}
\label{subsec:unconstrained_quench}

In fifteen out of twenty-four \texttt{QEQ} retrieval runs, including the thirteen with the lowest BIC values (\texttt{QEQ-1} to \texttt{QEQ-12}), the inferred carbon-chemistry quench pressure has a median value around $\log{(P_{\rm quench}/{\rm 1\ bar})} \approx -3.8$~dex with large uncertainties of $\approx 1.0$~dex (Table~\ref{tab:retrieved_qeq_key} and Figure~\ref{fig:corner}). This quench pressure is associated with disequilibrium chemistry (Section~\ref{subsubsec:abund_profiles}), which occurs when the timescale for vertical mixing in the atmosphere is shorter than that of chemical reactions such as CO$\rightarrow$CH$_{4}$ \citep[e.g.,][]{1977Sci...198.1031P, 2014ApJ...797...41Z}. Strong vertical mixing can result in an overabundance of CO relative to CH$_{4}$, often leading to a carbon-chemistry quench pressure that is larger than photospheric pressures \citep[e.g.,][]{2020A&A...640A.131M, 2023AJ....166..198Z, 2024A&A...687A.298N}.

However, for 2MASS1207b, the retrieved low quench pressure of $\log{(P_{\rm quench}/{\rm 1\ bar})} \approx -3.8$~dex should not be interpreted as evidence for the absence of disequilibrium chemistry in its atmosphere. This quench pressure corresponds to atmospheric layers at higher altitudes and lower pressures than the photospheric regions of 2MASS1207b across $1-5$$\mu$m (Figure~\ref{fig:qeq1_properties}). Therefore, the carbon-chemistry quench pressure is unconstrained by the observed JWST/NIRSpec spectrum and holds little physical meaning in this context.

As discussed in Section~\ref{subsubsec:noch4_explained}, the atmospheric thermal structure of 2MASS~1207~b is hotter than the CO--CH$_{4}$ transition curve at a given pressure layer, leading to a naturally CO-dominated and CH$_{4}$-deficient atmosphere, regardless of whether chemical equilibrium or disequilibrium conditions are present. Consequently, the potential effects of disequilibrium chemistry in the particular atmosphere of 2MASS~1207~b cannot be meaningfully constrained using carbon-chemistry quench pressures. This is in contrast to the colder L, T, and Y dwarfs, where the CH$_{4}$ abundances in the upper atmospheres differ significantly between chemical equilibrium and disequilibrium conditions, enabling tighter constraints on the carbon-chemistry quench pressure.

\subsection{Matching the Observed Variability Amplitudes with Retrieved Atmospheric Properties}
\label{subsec:variability}

The atmospheric variability of 2MASS~1207~b has been previously detected. Using the HST Wide Field Camera 3, \cite{2016ApJ...818..176Z} reported rotationally modulated variability amplitudes of $1.36\% \pm 0.23\%$ in the F125W filter (centered around $1.25$~$\mu$m) and $0.78\% \pm 0.22\%$ in the F160W filter (centered around $1.54$~$\mu$m). Additionally, \cite{2019AJ....157..128Z} obtained light curves in the F127M ($\approx 1.27$~$\mu$m) and F139M ($\approx 1.38$~$\mu$m) filters, which have $4-5$ times narrower bandpasses compared to F125W and F160W. They measured variability amplitudes of $0.57\% \pm 0.44\%$ and $2.45\% \pm 2.67\%$ in F127M and F139M, respectively, though these measurements were notably influenced by strong systematics (see their Section~4.3). 

We investigate whether the rotationally modulated variability amplitudes of 2MASS~1207~b are consistent with our most favored retrieval model, \texttt{QEQ-1} (see, e.g., \citealt{2020ApJ...903...15L} and \citealt{2023ApJ...944..138V}, for similar analyses of other objects). In this analysis, we assume the variability signal arises from geometric asymmetries in the coverage of thin-cloud patches and thick-cloud regions. Since the observed spectrum corresponds to the hemisphere of the atmosphere facing the observers, such asymmetries would lead to variations in the fraction of thin-cloud patches, $\xi_{\rm thin}$, as the object rotates. 

Starting with the maximum-likelihood value of $\xi_{\rm thin}^{\star} = 8.65\%$ from \texttt{QEQ-1}, we introduce an increment $\Delta \xi_{\rm thin}$ to simulate the rotational modulations of the cloud coverage and compute a modified emergent spectrum ($F_{\lambda, \rm tuned}^{\star}$), allowing for a comparison with the original, unmodified best-fit spectrum ($F_{\lambda}^{\star}$). 
\begin{equation}
\begin{aligned}
F_{\lambda}^{\star} = & \ \xi_{\rm thin}^{\star} \times F_{\lambda, \rm thin}^{\star} + (1 - \xi_{\rm thin}^{\star}) \times F_{\lambda, \rm thick}^{\star} \\
F_{\lambda, \rm tuned}^{\star} = & \ (\xi_{\rm thin}^{\star} + \Delta \xi_{\rm thin}) \times F_{\lambda, \rm thin}^{\star} \\
& + \left[1 - (\xi_{\rm thin}^{\star} + \Delta \xi_{\rm thin}) \right] \times F_{\lambda, \rm thick}^{\star}
\end{aligned}
\end{equation}
where $F_{\lambda, \rm thin}^{\star}$ and $F_{\lambda, \rm thick}^{\star}$ represent the maximum-likelihood spectra from the thin-cloud and thick-cloud patches, respectively. Since thin-cloud patches probe deeper and hotter layers, they produce higher fluxes. i.e., $F_{\lambda, \rm thin}^{\star} > F_{\lambda, \rm thick}^{\star}$.\footnote{We note that the conclusion of $F_{\lambda, \rm thin}^{\star} > F_{\lambda, \rm thick}^{\star}$ does not conflict with Figure~\ref{fig:patchy_fesifor_evo}. While the figure shows that the emergent spectra from thin-cloud patches, $\xi_{\rm thin}^{\star} \times F_{\lambda, \rm thin}^{\star}$, are not consistently brighter than those from thick-cloud regions, $(1 - \xi_{\rm thin}^{\star}) \times F_{\lambda, \rm thick}^{\star}$, across the 1--5~$\mu$m range, this result arises from the scaling factors ($\xi_{\rm thin}^{\star}$ and $1 - \xi_{\rm thin}^{\star}$) applied to the corresponding contributions, rather than the intrinsic brightness ($F_{\lambda, \rm thin}^{\star}$ and $F_{\lambda, \rm thick}^{\star}$) of the atmospheric components themselves.} Comparing $F_{\lambda}^{\star}$ and $F_{\lambda, \rm tuned}^{\star}$ providers a measure of the variability amplitude as a function of wavelength, $A_{\lambda}$, as follows:
\begin{equation}
\begin{aligned}
A_{\lambda} =& \ \frac{ {\rm abs}\left(F_{\lambda, \rm tuned}^{\star} - F_{\lambda}^{\star} \right) }{ F_{\lambda}^{\star} } \\
=& \ \frac{\Delta \xi_{\rm thin} \times \left(F_{\lambda, \rm thin}^{\star} - F_{\lambda, \rm thick}^{\star} \right) }{\xi_{\rm thin}^{\star} \times F_{\lambda, \rm thin}^{\star} + (1 - \xi_{\rm thin}^{\star}) \times F_{\lambda, \rm thick}^{\star}}
\end{aligned}
\end{equation}
The increment $\Delta \xi_{\rm thin}$ represents the asymmetry in atmospheric brightness distribution as the object rotates.

\begin{figure*}[t]
\begin{center}
\includegraphics[width=7.in]{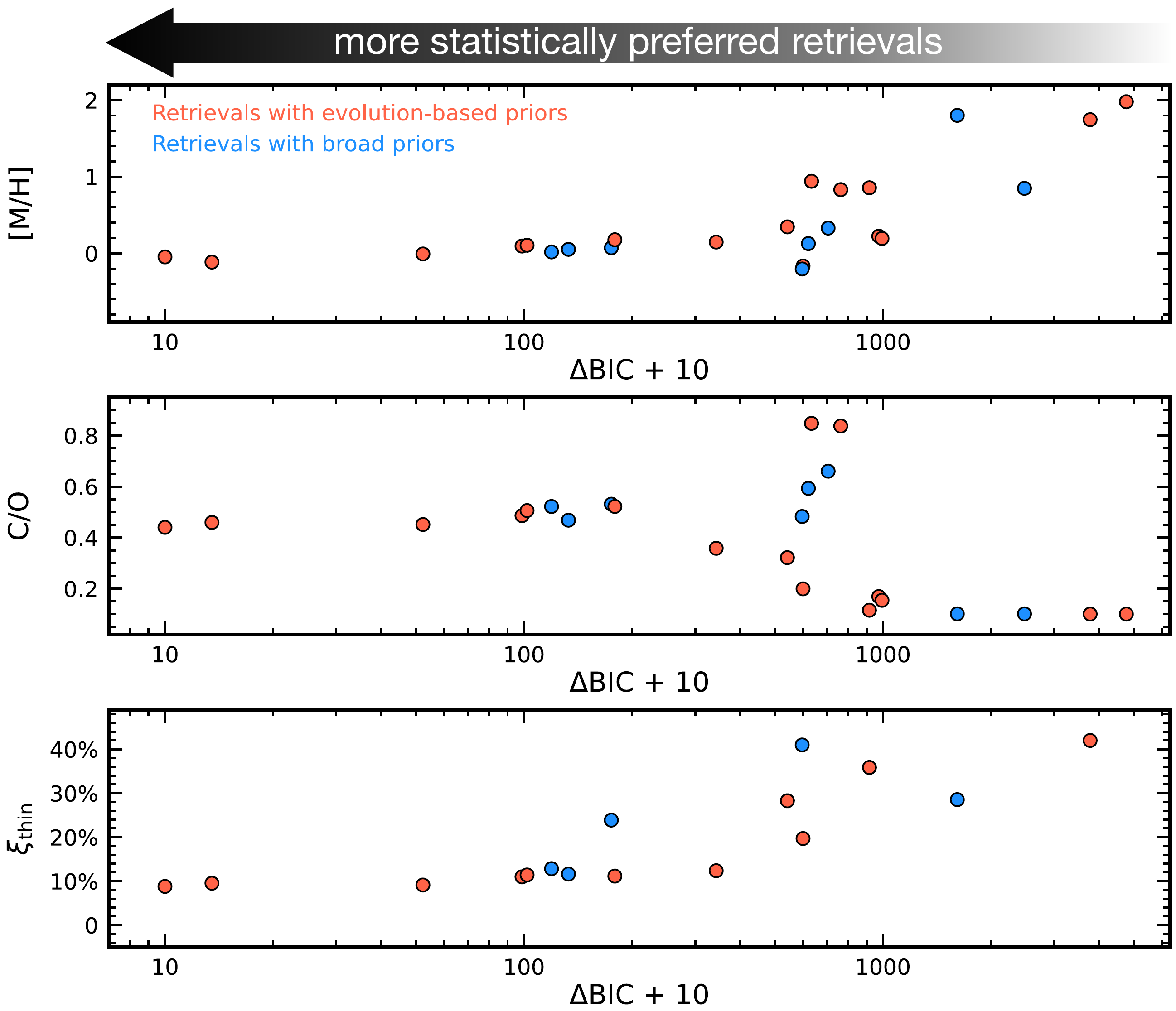}
\caption{The atmospheric metallicity (top), carbon-to-oxygen ratio (middle), and $\xi_{\rm thin}$ (bottom) inferred from all \texttt{QEQ} retrieval setups. The parameter $\xi_{\rm thin}$ is only defined in inhomogeneous atmospheric models. For schemes with patchy clouds or a combination of patchy clouds and hot spots, $\xi_{\rm thin}$ represents the fraction of thin-cloud patches; for the hot-spot scheme, it describes the fraction of the column with a hotter thermal structure (i.e., hot spots; Section~\ref{subsec:inhomo_framework}). Retrievals with evolution-based priors for surface gravity, radius, and mass are shown in orange, while those with broad parameter priors are shown in blue.}
\label{fig:qeq_mhcofraccol}
\end{center}
\end{figure*}

\begin{figure*}[t]
\begin{center}
\includegraphics[width=7.in]{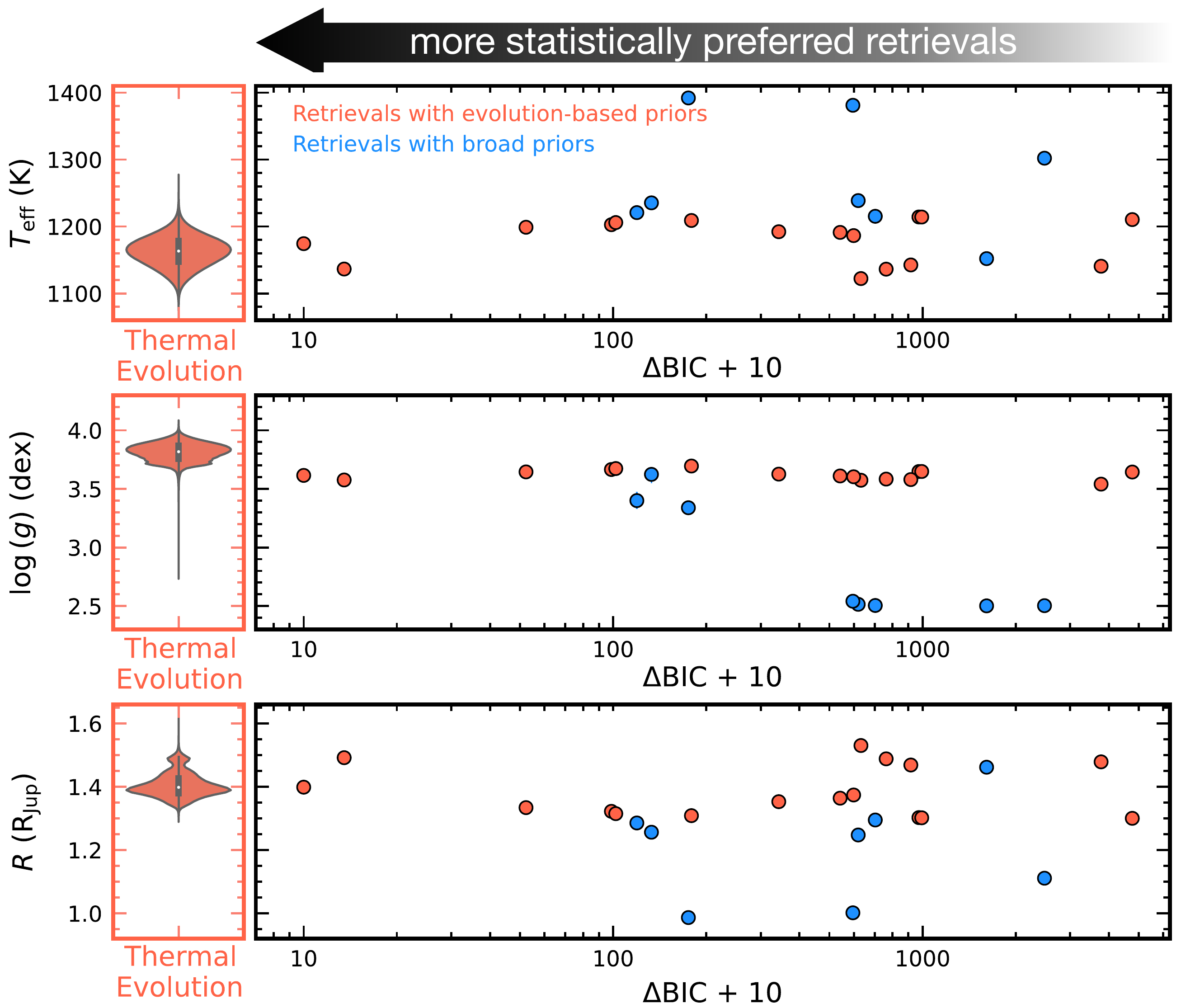}
\caption{The left column presents violin plots for the effective temperature, surface gravity, and radius of 2MASS~1207~b, as determined from thermal evolution models. These plots are based on the concatenated chains of parameters inferred from various evolution models (see Section~\ref{subsec:evo}). The right column presents the $T_{\rm eff}$, $\log{(g)}$, and $R$ values from all \texttt{QEQ} retrieval setups, following the same format as Figure~\ref{fig:qeq_mhcofraccol}. }
\label{fig:qeq_teffloggradmass}
\end{center}
\end{figure*}

In Figure~\ref{fig:varamp}, we show that small changes in the thin-cloud fraction, specifically $\Delta \xi_{\rm thin} = 0.11\% - 0.21\%$, reproduce the near-infrared photometric variability amplitudes observed during the particular epochs studied by \cite{2016ApJ...818..176Z}. These small $\Delta \xi_{\rm thin}$ values also yield variability amplitudes broadly consistent with the measurements by \cite{2019AJ....157..128Z} using narrower filters, despite the large uncertainties and systematics affecting these data. 

Our derived small values of $\Delta \xi_{\rm thin}$ suggest that the line-of-sight distribution of patchy clouds remains nearly symmetric as 2MASS~1207~b rotates during the particular HST monitoring by \cite{2016ApJ...818..176Z}. This can partially result from this object's spin axis inclination, which remains unknown. Observational and theoretical studies have suggested that objects with nearly edge-on viewing angles ($i =90^{\circ}$) tend to exhibit higher variability amplitudes than those with lower inclinations \citep[e.g.,][]{2017ApJ...842...78V, 2021MNRAS.502.2198T, 2024ApJ...975L..32F}. Directly measuring the inclination of 2MASS~1207~b via high-resolution spectroscopy is thus useful to understanding its role in this object's observed variability signals. Additionally, the atmospheric brightness distribution may evolve over multiple rotation periods, as observed in other brown dwarfs and planetary-mass companions \citep[e.g.,][]{2016ApJ...826....8Y, 2017Sci...357..683A, 2022AJ....164..239Z, 2024MNRAS.532.2207B} and as predicted by three-dimensional general circulation models \citep[e.g.,][]{2021MNRAS.502.2198T, 2024MNRAS.529.2686L}. A deeper understanding of 2MASS~1207~b's top-of-atmosphere inhomogeneity would benefit from dedicated variability monitoring observations (e.g., JWST GO program 3181).

\begin{figure*}[t]
\begin{center}
\includegraphics[width=7in]{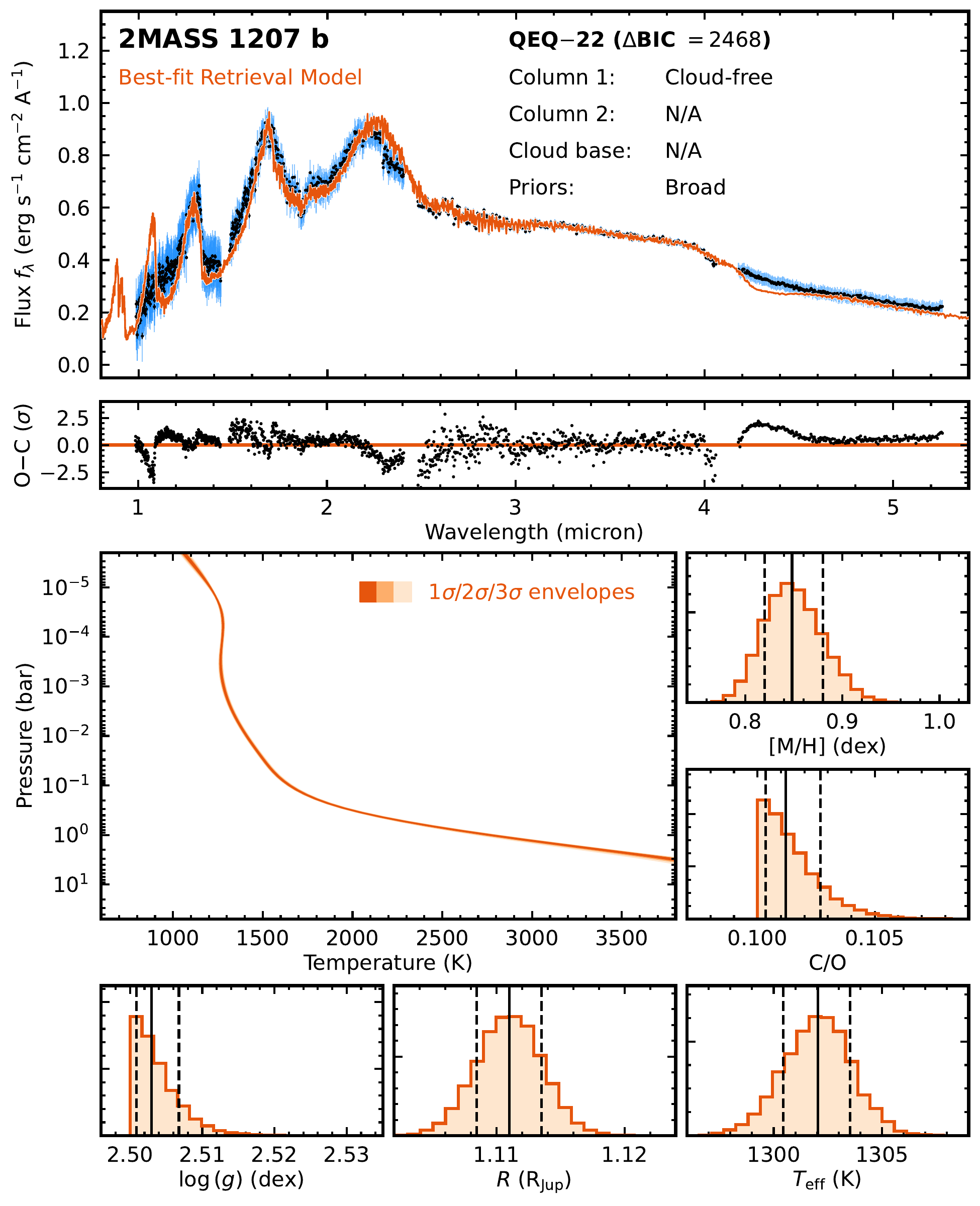}
\caption{Retrieval results for \texttt{QEQ-22}, which assumes a homogeneous cloud-free atmosphere with broad priors for $\log{g}$, $R$, and $M$ (see Section~\ref{subsubsec:comments_nc}). These panels follow the same format as Figure~\ref{fig:patchy_fesifor_evo}.}
\label{fig:homo_nc}
\end{center}
\end{figure*}

\begin{figure*}[t]
\begin{center}
\includegraphics[width=7in]{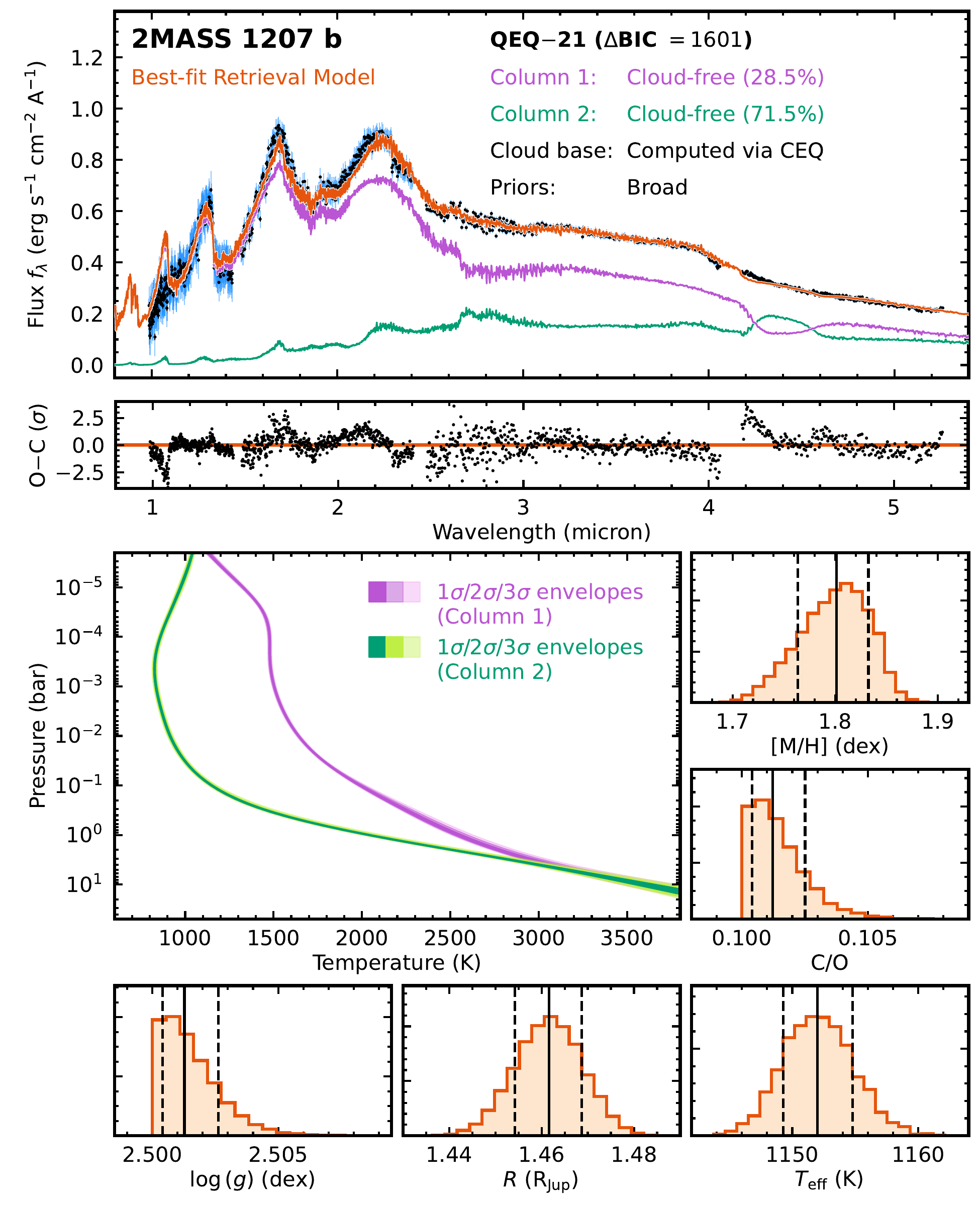}
\caption{Retrieval results for \texttt{QEQ-21}, which assumes an inhomogeneous atmosphere with cloud-free hot/cold spots, along with broad priors for $\log{g}$, $R$, and $M$ (see Section~\ref{subsubsec:comments_nc}). The middle left panel presents the retrieved T-P profiles from individual atmospheric columns. These panels follow the same format as Figure~\ref{fig:patchy_fesifor_evo}.}
\label{fig:inhomo_nc}
\end{center}
\end{figure*}

\subsection{Atmospheric Composition}
\label{subsec:mhco}

Figure~\ref{fig:qeq_mhcofraccol} shows the atmospheric metallicity and carbon-to-oxygen ratio of 2MASS~1207~b, derived from multiple retrieval runs. While these values exhibit significant scatter across retrievals with high $\Delta$BIC, they converge to near-solar compositions in runs with lower $\Delta$BIC values. This trend is also observed in other key retrieved parameters, including the fraction of thin-cloud patches ($\xi_{\rm thin}$), effective temperatures, surface gravities, and radii (Figures~\ref{fig:qeq_mhcofraccol}--\ref{fig:qeq_teffloggradmass}). Our extensive retrieval analysis indicates the sensitivity of inferred atmospheric compositions to retrieval assumptions. For instance, cloud-free retrieval runs (\texttt{QEQ-21} to \texttt{QEQ-24}) yield high [M/H] values of 0.8--2.0~dex and very low C/O values near 0.1. In contrast, the statistically preferred retrievals (e.g., \texttt{QEQ-1} to \texttt{QEQ-7}), which account for patchy clouds, derive solar-like [M/H] snd C/O values for 2MASS~1207~b (see below). These striking differences indicate that relying on a single retrieval setup risks biasing the inferred [M/H], C/O, and other physical parameters, potentially distorting our understanding of the system's formation and evolution history. 

Focusing on retrieval runs with the seven lowest BIC values (\texttt{QEQ-1} to \texttt{QEQ-7}), the inferred [M/H] of 2MASS~1207~b ranges from $-0.12$ to $+0.10$~dex, and C/O from 0.44 to 0.52 (Table\ref{tab:retrieved_qeq_key}). These retrievals incorporate models of atmospheric inhomogeneity, such as patchy clouds or a combination of patchy clouds and hot spots (Figure~\ref{fig:inhomo}), and explore various assumptions about cloud composition and parameter priors. Despite these differences in model assumptions, their consistent results support the robustness of the near-solar [M/H] and C/O values. Specifically, the most favored retrieval setup, \texttt{QEQ-1}, yields a [M/H]$=-0.05 \pm 0.03$~dex and a C/O of $0.440 \pm 0.012$.\footnote{The inferred atmospheric compositions of 2MASS~1207~b --- and more broadly, those of exoplanets and brown dwarfs --- are derived from disk-integrated spectra, which average out potential three-dimensional spatial variations in atmospheric abundances. Studies of Jupiter suggested that certain gas abundances retrieved from disk-integrated spectra can differ from those of localized regions \citep[e.g.,][]{2024A&A...688A..10G, 2025JGRE..13008622I}. Comparing the globally and locally inferred atmospheric properties of solar-system planets, such as Jupiter, provides crucial context for interpreting exoplanet and brown dwarf atmospheres and remains a key topic of ongoing research.} As a member of the TW Hydrae association, 2MASS~1207~b's near-solar [M/H] is consistent with stellar members of other nearby young moving groups \citep[e.g.,][]{2008A&A...480..889S, 2009A&A...508..677A, 2009A&A...501..965V, 2011A&A...526A.103D, 2011A&A...530A..19B, 2012A&A...547A.104B, 2017A&A...605A..66B}.

To further investigate the formation and evolution history of the 2MASS~1207~A+b system, it is essential to compare the atmospheric compositions of 2MASS~1207~b with its primary, 2MASS~1207~A. Previous studies suggest that 2MASS~1207~b likely formed through gravitational instability in a collapsing molecular cloud rather than through core accretion \citep[e.g.,][]{2005MNRAS.364L..91L, 2005ApJ...630L..89G, 2005A&A...438L..25C, 2006ApJ...637L.137B, 2009MNRAS.392..413S}, given that the companion-to-host mass ratio of this system ($\sim 0.2$; see Section~\ref{subsec:2m1207ab}) is much higher than the majority of exoplanet-host systems \citep[e.g., Figure~3 of][]{2021ApJ...916L..11Z}. Gravitational instability could result in similar atmospheric compositions for both components, as observed in members of binary systems and stellar clusters \citep[e.g.,][]{2008A&A...480..889S, 2012MNRAS.427.2905B, 2020MNRAS.492.1164H, 2021ApJ...921..118N}, although interactions between 2MASS~1207~b and its environment, such as planetesimal accretion or orbital migration within the disk surrounding 2MASS~1207~A \citep[e.g.,][]{2006Icar..185...64H, 2010Icar..207..503H, 2009ApJ...697.1256H, 2010ApJ...724..618B, 2021A&A...654A..71S, 2022ApJ...934...74M}, may have altered their relative abundances. Measuring and comparing the present-day atmospheric compositions of 2MASS~1207~A and b will provide insights into the presence and impacts of these post-formation processes.

\begin{figure*}[t]
\begin{center}
\includegraphics[width=7in]{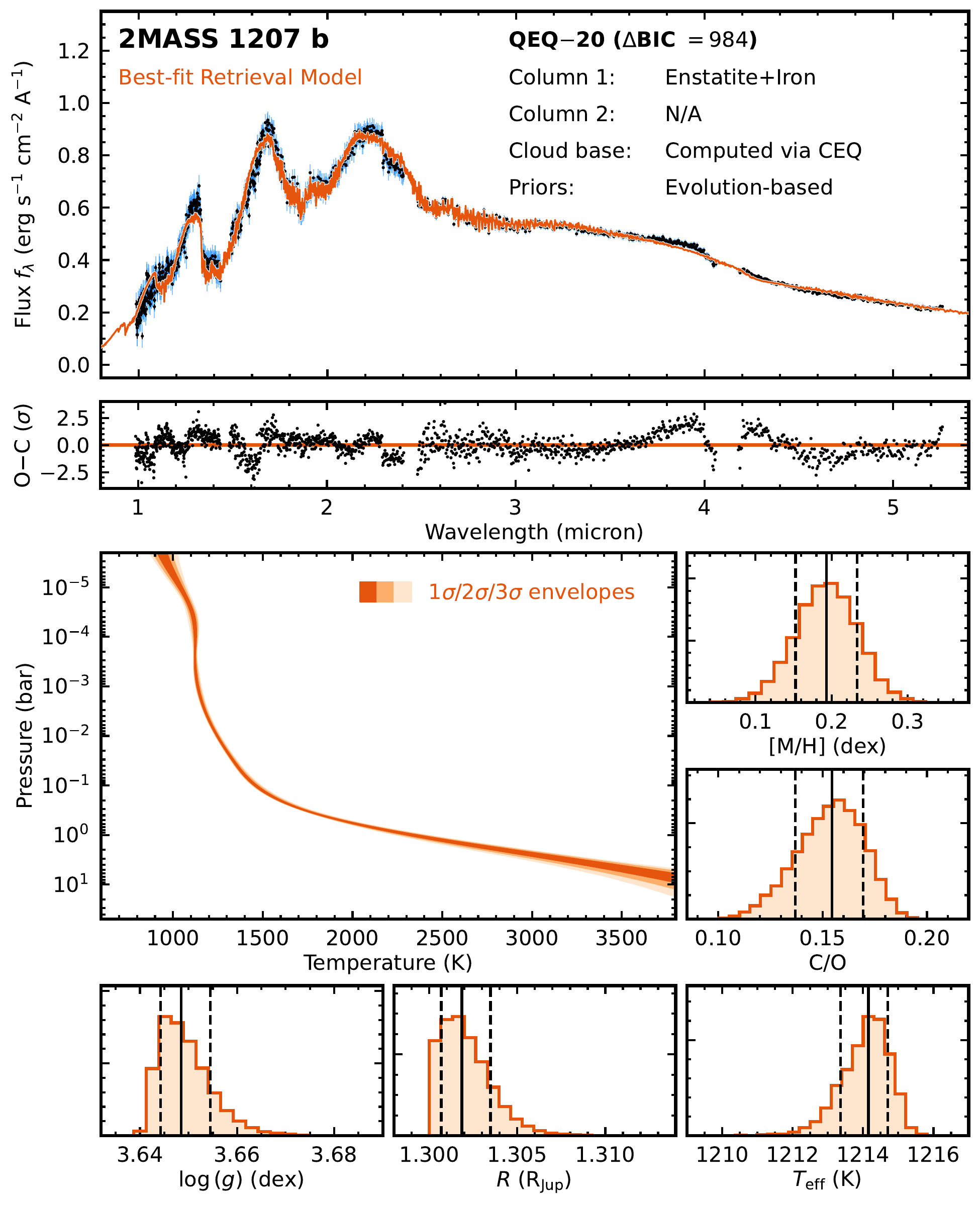}
\caption{Retrieval results for \texttt{QEQ-20}, which assumes a homogeneous atmosphere with enstatite and iron clouds, with their base pressured determined from equilibrium chemistry (see Section~\ref{subsubsec:comments_combo}). Evolution-based priors are adopted for $\log{g}$, $R$, and $M$. These panels follow the same format as Figure~\ref{fig:patchy_fesifor_evo}.}
\label{fig:homo_cloud}
\end{center}
\end{figure*}

\begin{figure*}[t]
\begin{center}
\includegraphics[width=7in]{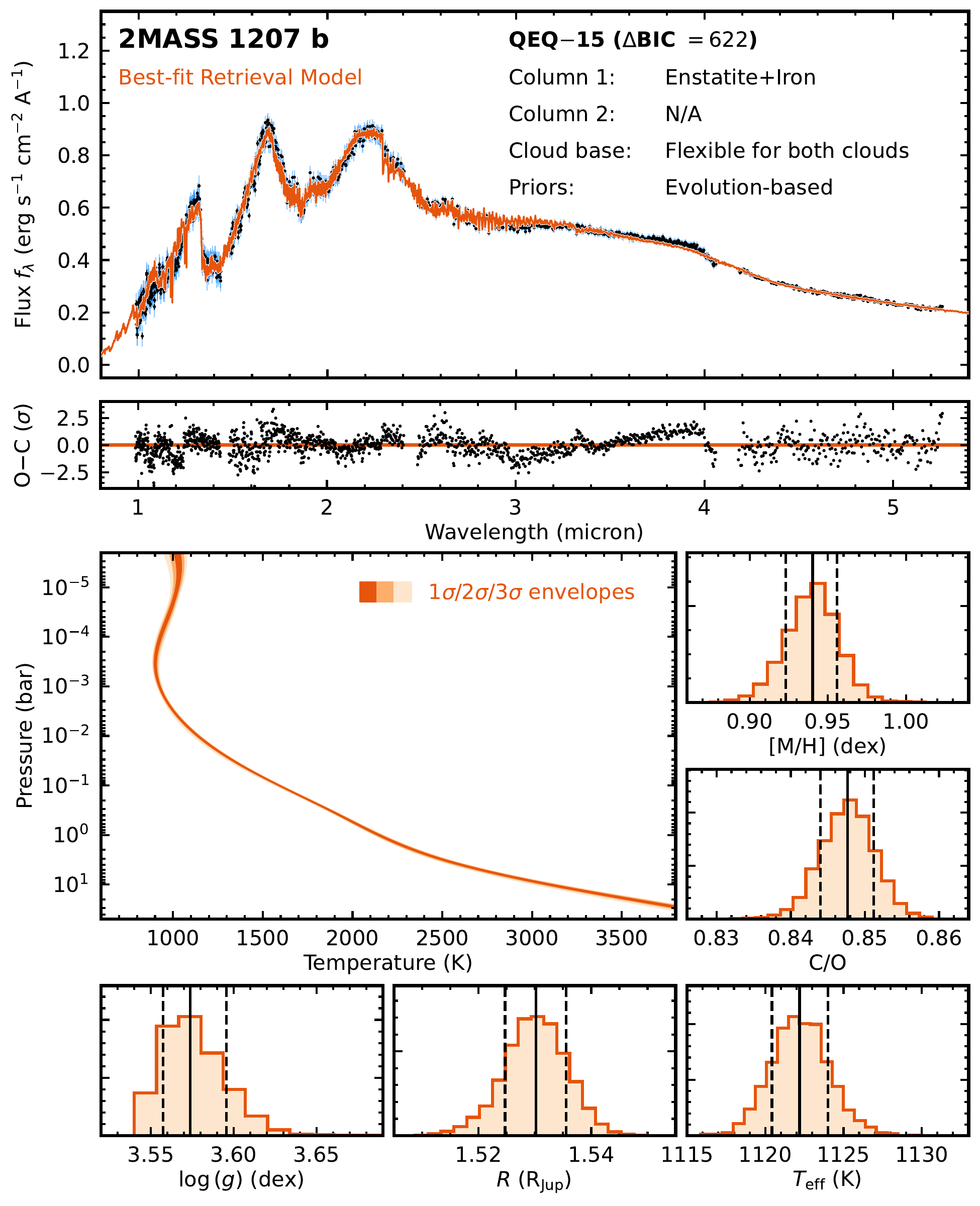}
\caption{Retrieval results for \texttt{QEQ-15}, which resembles the \texttt{QEQ-20} setup (Figure~\ref{fig:homo_cloud}) but allows the cloud base pressures to vary as free parameters (see discussions in Section~\ref{subsubsec:comments_combo}). Evolution-based priors are adopted for $\log{g}$, $R$, and $M$. These panels follow the same format as Figure~\ref{fig:patchy_fesifor_evo}.}
\label{fig:homo_cloud_flex}
\end{center}
\end{figure*}

\begin{figure*}[t]
\begin{center}
\includegraphics[width=7in]{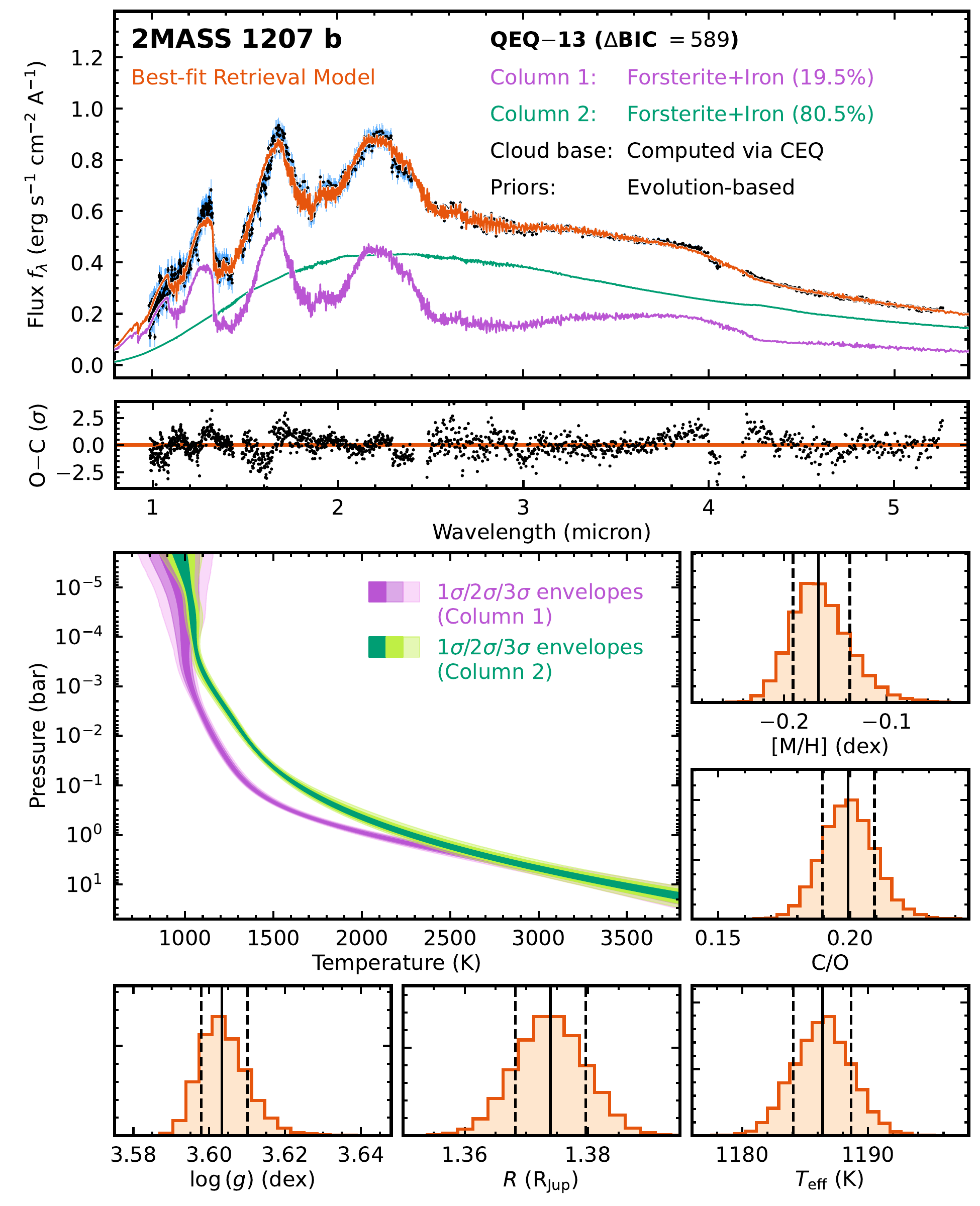}
\caption{Retrieval results for \texttt{QEQ-13}, which assumes an inhomogeneous atmosphere with patchy forsterite and iron clouds with hot spots (see discussions in Section~\ref{subsubsec:comments_combo}). Evolution-based priors are adopted for $\log{g}$, $R$, and $M$. These panels follow the same format as Figure~\ref{fig:patchy_fesifor_evo}.}
\label{fig:patchyhot}
\end{center}
\end{figure*}

\begin{figure*}[t]
\begin{center}
\includegraphics[width=7in]{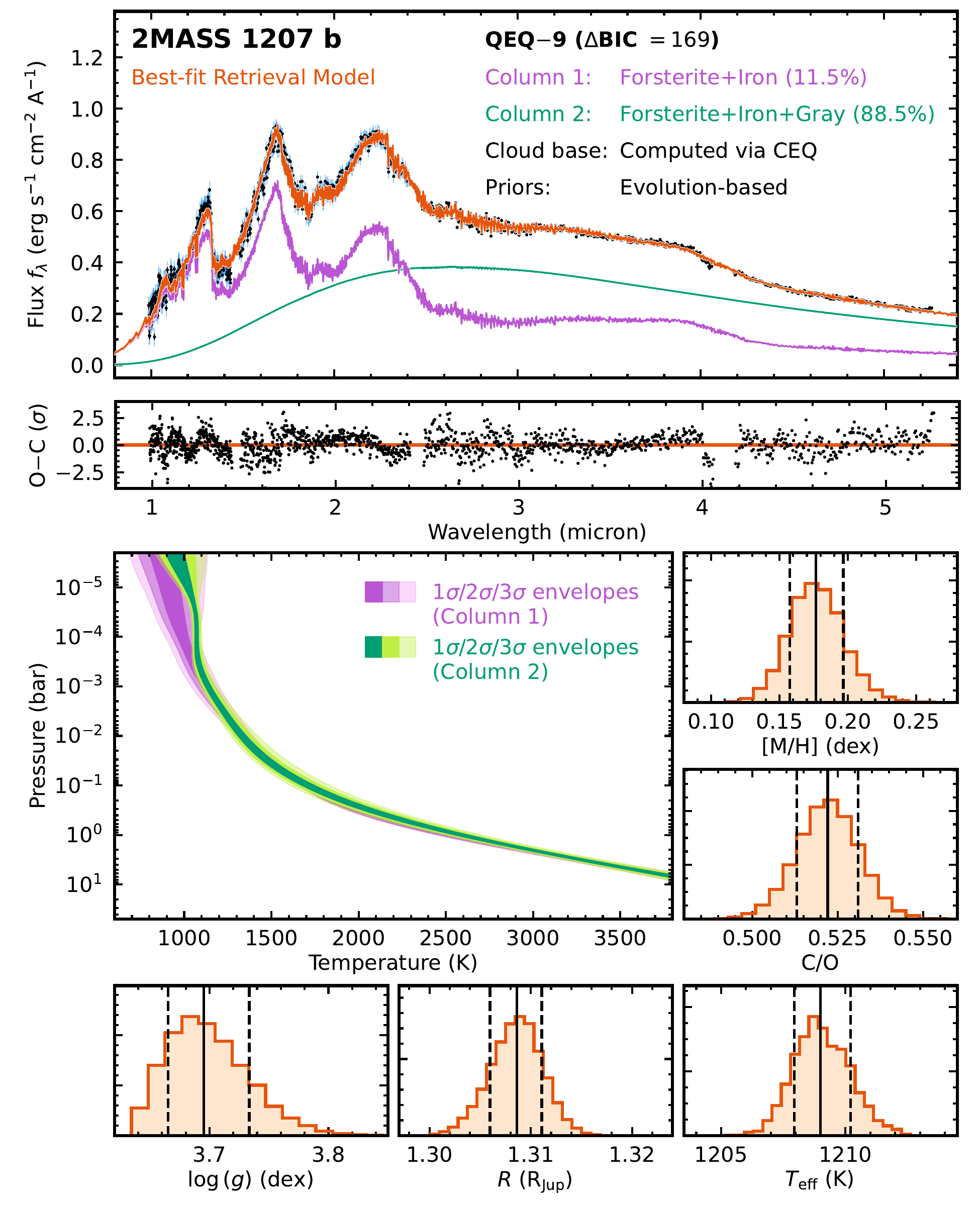}
\caption{Retrieval results for \texttt{QEQ-9}, which resembles the \texttt{QEQ-13} setup (Figure~\ref{fig:patchyhot}) but adds a gray cloud component to the second atmospheric column (see discussions in Section~\ref{subsubsec:comments_combo}). Evolution-based priors are adopted for $\log{g}$, $R$, and $M$. These panels follow the same format as Figure~\ref{fig:patchy_fesifor_evo}.}
\label{fig:patchyhot_gray}
\end{center}
\end{figure*}

\subsection{Comments on the Variations in Retrieval Setups}
\label{subsec:impact}

In this section, we discuss the results of retrievals that use different assumptions from the \texttt{QEQ-1} setup to provide additional insights into the atmospheric properties of 2MASS1207b.

\subsubsection{Retrievals with Cloud-free Atmospheres}
\label{subsubsec:comments_nc}

Figure~\ref{fig:homo_nc} presents the retrieval results of \texttt{QEQ-22}, which assumes a homogeneous, cloud-free atmosphere with broad priors for $\log{(g)}$, $R$, and $M$. This setup corresponds to a simplified scenario under scheme (a) in our atmospheric inhomogeneity framework (Section~\ref{subsec:inhomo_framework}). Significant discrepancies are evident between data and fitted models in the $Y$ band, the blue wing of the $H$ band, the flux peak of the $K$ band, and around 4.3~$\mu$m. Moreover, the retrieved values for $T_{\rm eff}$, $\log{(g)}$, and $R$ (Table~\ref{tab:retrieved_qeq_key}) deviate significantly from the inferred parameters based on thermal evolution models (Table~\ref{tab:evoparams}). 

A more sophisticated cloud-free model that incorporates hot and cold spots, such as \texttt{QEQ-21}, provides a better, though still inadequate, fit to the observations (Figure~\ref{fig:inhomo_nc}). This model corresponds to the scheme (c) in our atmospheric inhomogeneity framework (Section~\ref{subsec:inhomo_framework}) and combines two T-P profiles with consistent temperatures and temperature gradients at $P \geqslant 10$~bar (Section~\ref{subsubsec:scheme_c}). However, as shown in Figure~\ref{fig:inhomo_nc}, the retrieved T-P profiles show temperature contrasts of up to $\sim 800$~K at pressures between $\sim 10^{-5}$~bar and $\sim 1$~bar, along with an upper-atmosphere thermal inversion in cold spots. It is challenging to justify how such large temperature differences between the two atmospheric columns could remain stable over time, especially given that 2MASS~1207~b is not tidally locked nor exposed to intense external radiation from 2MASS~1207~A.

Thus, within our retrieval frameworks, the JWST/NIRSpec spectrum of 2MASS~1207~b cannot be fully explained by cloud-free models. This conclusion aligns with the findings of \cite{2023ApJ...949L..36L}, who showed that cloud-free \texttt{ATMO2020} models --- assuming one-dimensional homogeneous atmosphere with adjusted adiabatic indices --- cannot reproduce all the observed spectral features of 2MASS~1207~b.

\subsubsection{Retrievals with Homogeneous Cloudy Atmospheres}
\label{subsubsec:comments_homocloud}

Figure~\ref{fig:homo_cloud} presents the retrieval results of \texttt{QEQ-20}, which assumes a homogeneous atmosphere with MgSiO$_{3}$ and Fe clouds with base pressures determined from equilibrium chemistry. Evolution-based priors are used for $\log{(g)}$, $R$, and $M$. While this cloudy model significantly improves the fit to the JWST/NISPEC spectrum of 2MASS~1207~b compared to cloud-free retrievals (Section~\ref{subsubsec:comments_nc}), notable systematic discrepancies between the data and models remain. Specifically, the retrieved model spectrum shows (1) fainter fluxes near the peaks of $J$, $H$, and $K$ bands, (2) inconsistent depths of the K~\textsc{i} doublet near 1.25~$\mu$m, (3) mismatched shapes in the blue wing of $H$ band and the red wing of the $K$ band, and (4) overall fainter fluxes between $3.7$ and $4.0$~$\mu$m. 

The retrieved iron cloud in this model has an unusually high mass fraction at its base pressure, with $\log{(X_{\rm 0, Fe})} \approx -0.11$. The enstatite cloud is vertically extended, with a very low sedimentation efficiency of $f_{\rm sed, MgSiO_{3}} \approx 0.04$ and a large mass fraction of $\log{(X_{\rm 0, MgSiO_{3}})} \sim -2.90$ at its base pressure. These parameters suggest that both cloud species contribute significantly to the atmospheric opacity, helping to explain the red spectral slope of 2MASS~1207~b but failing to reproduce detailed spectral features.

To refine the fit, we adjust the \texttt{QEQ-20} setup by treating the base pressures of the enstatite and iron clouds as free parameters, resulting in the \texttt{QEQ-15} retrieval (Figure~\ref{fig:homo_cloud_flex}). This modification yields improved agreement with the observations but still shows discrepancies. Specifically, this model predicts slightly brighter fluxes in $1.1-1.3$~$\mu$m, fainter fluxes in $3.7-4.0$~$\mu$m, and a weak 3.3~$\mu$m CH$_{4}$ feature that is not observed. The retrieved base pressures for iron and enstatite clouds are $\log{(P_{\rm base, Fe}/{\rm 1\ bar})} \approx -1.48$ and $\log{(P_{\rm base, Fe}/{\rm 1\ bar})} \approx -4.7$, respectively. These pressures indicate that the cloud layers are located at higher altitudes and lower pressures than predicted by chemical equilibrium models 

The systematic mismatches between data and models observed in \texttt{QEQ-20} and \texttt{QEQ-13} are consistent across other retrievals assuming homogeneous cloudy atmospheres. These findings highlight the value of inhomogeneous atmospheric models for explaining the spectrum of 2MASS1207b.

\subsubsection{Retrievals with a Combination of Patchy Clouds and Hot Spots}
\label{subsubsec:comments_combo}

Figure~\ref{fig:patchyhot} presents the retrieval results for \texttt{QEQ-13}, which assumes a combination of patchy forsterite and iron clouds along with hot spots. Evolution-based priors are used for $\log{(g)}$, $R$, and $M$. This model corresponds to scheme (d) in our atmospheric inhomogeneity framework (Section~\ref{subsec:inhomo_framework}) and significantly improves the spectral fits compared to cloud-free models (Section~\ref{subsubsec:comments_nc}) and homogeneous cloudy models (Section~\ref{subsubsec:comments_homocloud}). 

\texttt{QEQ-13} suggests that thin-cloud patches cover $\sim 20\%$ of 2MASS~1207~b's atmosphere. These regions contain vertically extended Mg$_{2}$SiO$_{4}$ clouds with $f_{\rm sed, Mg_{2}SiO_{4}} \approx 1$, combined with low Fe cloud content, resulting in an emergent spectrum resembling that of an L dwarf. The remaining $80\%$ of the atmosphere are thick-cloud regions with vertically compact ($f_{\rm sed} > 9$) but significantly abundant Mg$_{2}$SiO$_{4}$ and Fe clouds, producing a blackbody-like spectrum.  These properties align with the results from the most favored retrieval model, \texttt{QEQ-1} (Section~\ref{subsec:inhomo}). 

We further adjust the \texttt{QEQ-13} setup by incorporating a gray cloud layer into the second atmospheric column, leading to the \texttt{QEQ-9} model (Figure~\ref{fig:patchyhot_gray}). This adjustment improves the data-model consistencies further. \texttt{QEQ-9} suggests that thin-cloud patches cover $\sim 11\%$ of 2MASS~1207~b's atmosphere and produce an L-dwarf-like spectrum, while the remaining $89\%$ thick-cloud regions have low Mg$_{2}$SiO$_{4}$ and Fe cloud content, combined with a high-opacity gray cloud layer located at $\sim 10^{-5}$~bar. The gray cloud dominates the opacity, resulting in a blackbody-like emergent spectrum. The T-P profiles between the two atmospheric columns are consistent in this model, justifying the adequacy of a single T-P profile, as used in \texttt{QEQ-1}, to describe the NIRSpec spectrum of 2MASS~1207~b.

The combination of thin-cloud patches with L-dwarf-like spectra and thick-cloud regions with blackbody-like spectra, as seen in \texttt{QEQ-13} and \texttt{QEQ-9}, is consistently supported by other retrievals that incorporate patchy clouds and hot/cold spots. These findings strongly corroborate the atmospheric scenario inferred by \texttt{QEQ-1}, emphasizing the critical role of inhomogeneous atmospheres in explaining the observations of 2MASS1207b.

\section{Summary} 
\label{sec:summary}

2MASS~1207~b is the first planetary-mass companion discovered by direct imaging \citep{2004A&A...425L..29C, 2005A&A...438L..25C}. We have studied the atmospheric properties of this remarkable object based on its JWST/NIRSpec spectrum and atmospheric retrievals, with the main findings summarized below. 
\begin{enumerate}
\item[$\bullet$] We contextualized the physical parameters of 2MASS~1207~b using several thermal evolution models based on its bolometric luminosity and age. These evolution-based properties --- including $T_{\rm eff} = 1163^{+23}_{-23}$~K, $\log{(g)} = 3.82^{+0.09}_{-0.07}$~dex, $R = 1.40^{+0.03}_{-0.04}$~R$_{\rm Jup}$, and $M = 5.2^{+0.6}_{-0.7}$~M$_{\rm Jup}$ --- provide key physics-driven priors for our atmospheric retrieval analyses. 

\item[$\bullet$] We have performed extensive atmospheric retrieval analyses of 2MASS~1207~b's JWST NIRSpec spectrum using a newly developed atmospheric inhomogeneity framework. This framework enables the characterization of self-luminous exoplanets, brown dwarfs, and low-mass stars under four atmospheric schemes: (1) homogeneous atmospheres, (2) patchy clouds, (3) cloud-free hot spots, and (4) a combination of patchy clouds and hot spots. The latter three schemes are particularly suited for objects exhibiting atmospheric variabilities, such as 2MASS~1207~b. 

\item[$\bullet$] We have conducted 24 retrieval runs with variations in atmospheric inhomogeneity, parameter priors, and cloud properties. The most statistically preferred retrieval model, \texttt{QEQ-1}, corresponds to the patchy cloud scheme. This retrieval model incorporates evolution-based priors for key physical parameters and yields an $T_{\rm eff} = 1174^{+4}_{-3}$~K, a surface gravity of $\log{(g)} = 3.62^{+0.03}_{-0.02}$~dex, and a radius of $R = 1.399^{+0.008}_{-0.010}$~R$_{\rm Jup}$. 

\item[$\bullet$] The \texttt{QEQ-1} retrieval model suggests that $\sim 9\%$ of 2MASS~1207~b's atmosphere is covered by thin-cloud patches with Fe clouds and a lower Mg$_{2}$SiO$_{4}$ cloud content, producing L-dwarf-like emergent spectra. The remaining $91\%$ of the atmosphere consists of thick-cloud regions with vertically extended Mg$_{2}$SiO$_{4}$ and Fe clouds, resulting in blackbody-like spectra with fluxes peaking at $\sim 2.5$~$\mu$m. This scenario is consistently supported by other retrieval runs that incorporate clouds and inhomogeneous atmospheres. Furthermore, when a gray cloud layer is incorporated into one of the atmospheric columns, the retrieved models indicate that the thick-cloud regions are characterized by a high-altitude gray cloud layer at low pressures around $10^{-5}$~bar. 

\item[$\bullet$] The inhomogeneous atmosphere of 2MASS~1207~b is analogous to that of Jupiter, which exhibits a banded structure of belts and zones. Our retrievals indicate that 2MASS~1207~b has thin-cloud patches, with photospheric temperatures of 1600--1700~K and thick-cloud regions with slightly cooler photospheric temperatures of 1100--1200~K, resembling Jupiter's belts and zones, respectively. 

\item[$\bullet$] The thick clouds and potentially high-altitude gray cloud content of 2MASS~1207~b, as suggested by our retrieval analyses, may be contributed by this object's ongoing accretion of dust and gas. Extreme events such as the impact of Comet Shoemaker-Levy 9 on Jupiter may also influence the evolution of 2MASS~1207~b. Follow-up studies are needed to investigate the potential occurrence and impact of such processes. 

\item[$\bullet$] The previously noted but unexplained unique spectral features of 2MASS~1207~b can now be understood through our retrieved atmospheric properties. The weak CO absorption over 4.4--5.2~$\mu$m likely arises from veiling effects caused by patchy thick clouds. The blackbody-like featureless spectra from the thick-cloud regions dilute the contributions from the thin-cloud patches, which exhibit prominent CO absorption and share a spectral morphology similar to VHS~1256~b. 

\item[$\bullet$] The absence of 3.3~$\mu$m CH$_{4}$ absorption in 2MASS~1207~b can be explained by this object's relatively hot atmospheric thermal structure, which leads to a CO-dominant, CH$_{4}$-deficient atmosphere. As a result, the potential effects of disequilibrium chemistry in the particular atmosphere of 2MASS~1207~b cannot be meaningfully constrained using carbon-chemistry quench pressures. This contrasts with colder L, T, and Y dwarfs, where CH$_{4}$ abundances in the upper atmospheres differ significantly between chemical equilibrium and disequilibrium conditions, enabling tighter constraints on the carbon-chemistry quench pressure.

\item[$\bullet$] The retrieved atmospheric models can match the observed rotationally modulated variabilities of 2MASS~1207~b. Based on \texttt{QEQ-1}, small perturbations in the thin-cloud coverage ($\xi_{\rm thin}$) from its maximum-likelihood value by increments of $\Delta \xi_{\rm thin} = 0.11\% - 0.21\%$ reproduce the variability amplitudes observed in broadband HST/WFC3 filters. 

\item[$\bullet$] Our analysis reveals that the inferred atmospheric properties of 2MASS1207b are sensitive to assumptions in retrieval models. Atmospheric properties such as [M/H] and C/O show significant scatter across retrieval runs with high BICs but converge to consistent values among the statically preferred models. Cloud-free retrieval runs yield high [M/H] values of 0.8--2.0~dex and very low C/O values near 0.1. In contrast, statistically preferred retrieval runs consistently suggest that 2MASS~1207~b has near-solar compositions that line up with those of stellar members in nearby young moving groups. The most preferred retrieval model (\texttt{QEQ-1}) derives [M/H]$= -0.05 \pm 0.03$~dex and C/O$=0.440 \pm 0.012$ for 2MASS~1207~b. These findings underscore the importance of exploring diverse assumptions in retrievals to avoid biased interpretations of atmospheres, formation pathways, and evolution histories. 
\end{enumerate}

Beyond these findings specific to 2MASS~1207~b, we have developed several new methodologies for atmospheric studies of self-luminous exoplanets, brown dwarfs, and low-mass stars:
\begin{enumerate}
\item[$\bullet$] A new spectroscopy-based bolometric correction approach designed to mitigate modeling systematics in bolometric luminosity estimates (Section~\ref{subsec:lbol} and Appendix~\ref{app:bc}).

\item[$\bullet$] An expanded grid of chemical equilibrium abundances\footnote{\url{https://doi.org/10.5281/zenodo.15654694}} for \texttt{petitRADTRANS}, which extends the temperature range and incorporates additional atomic and molecular species (Section~\ref{subsubsec:ceq_grid}).

\item[$\bullet$] An atmospheric retrieval framework for characterizing both homogeneous and inhomogeneous atmospheres (Section~\ref{subsec:inhomo_framework}), as summarized above.
\end{enumerate}

Future observational studies of 2MASS~1207~b, such as JWST MIRI spectroscopy (GTO program 1270) and time-series NIRSpec spectroscopy (GO program 3181), will provide new insights into this remarkable planetary-mass companion. In particular, follow-up atmospheric retrieval analyses of combined NIRSpec and MIRI data for 2MASS~1207~b should account for both its inhomogeneous atmosphere and the possible presence of its circumsecondary disk, which, if present, could contribute to the flux at MIRI wavelengths. Additionally, detailed atmospheric studies of its host, 2MASS~1207~A, will offer a crucial context for understanding formation pathways and the evolutionary history of this planetary system.

\begin{acknowledgments}
We thank the anonymous referee for suggestions that improved the manuscript. Z.Z. thanks Ga\"{e}l Chauvin, Gabriel-Dominique Marleau, and Elena Manjavacas for helpful comments and thanks Andy Skemer and Xi Zhang for discussions during the early stages of this work. Z.Z. acknowledges the support of the NASA Hubble Fellowship grant HST-HF2-51522.001-A awarded by the Space Telescope Science Institute, which is operated by the Association of Universities for Research in Astronomy, Inc., for NASA, under contract NAS5-26555. We acknowledge the use of the lux supercomputer at UC Santa Cruz, funded by NSF MRI grant AST 1828315. This work has made use of data from the European Space Agency (ESA) mission Gaia (\url{https://www.cosmos.esa.int/gaia}), processed by the Gaia Data Processing and Analysis Consortium (DPAC, \url{https://www.cosmos.esa.int/web/gaia/dpac/consortium}). Funding for the DPAC has been provided by national institutions, in particular, the institutions participating in the Gaia Multilateral Agreement. The data presented in this article were obtained from the Mikulski Archive for Space Telescopes (MAST) at the Space Telescope Science Institute. The specific observations analyzed can be accessed via \dataset[DOI: 10.17909/jzze-5611]{https://doi.org/10.17909/jzze-5611}.

Z.Z. warmly acknowledges the love and support of his wife, Yanxia Li, and the arrival of our daughter, Allison Muzi Zhang, whose presence has brought immense joy and inspiration during the completion of this work.
\end{acknowledgments}

\facilities{JWST (NIRSpec)}

\software{\texttt{petitRADTRANS} \citep[][]{2019A&A...627A..67M, 2024JOSS....9.5875N}, \texttt{easyCHEM} \citep[][]{2017A&A...600A..10M, 2024arXiv241021364L}, \texttt{MultiNest} \citep[][]{2008MNRAS.384..449F, 2009MNRAS.398.1601F, 2019OJAp....2E..10F}, \texttt{PyMultiNest} \citep[][]{2014A&A...564A.125B}, \texttt{corner.py} \citep[][]{corner}, \texttt{astropy} \citep{2013A&A...558A..33A, 2018AJ....156..123A}, \texttt{ipython} \citep{PER-GRA:2007}, \texttt{numpy} \citep{numpy}, \texttt{scipy} \citep{scipy}, \texttt{matplotlib} \citep{Hunter:2007}.}

\appendix

\section{A New Spectroscopy-based Bolometric Correction with Mitigated Modeling Systematics}
\label{app:bc}

\subsection{An Overview of Bolometric Luminosity Measurements}
\label{app:sub:overview}

Bolometric luminosity is a critical physical property for understanding the thermal evolution and atmospheric energy budget of exoplanets, brown dwarfs, and stars. However, directly measuring $L_{\rm bol}$ is challenging due to the limited wavelength coverage of typical spectroscopic observations. To address this, several methods have been commonly adopted to estimate the bolometric luminosities of self-luminous exoplanets and brown dwarfs by extrapolating their observed spectrophotometry to cover the full wavelength range. 

For objects with spectrophotometry spanning wavelengths typically broader than 1~$\mu$m, atmospheric model spectra that best match the observations are often used for extrapolation \citep[e.g.,][]{2013ApJ...779..188M, 2015ApJ...810..158F, 2021ApJ...916...53Z, 2021ApJ...921...95Z, 2023ApJ...949L..36L}. However, this spectral-fitting approach may be excessive if the primary goal is only to derive $L_{\rm bol}$ or if an $L_{\rm bol}$ value is required beforehand to inform spectral fits (e.g., \citealt{2023AJ....166..198Z}; see also Section~\ref{sec:evo}). Moreover, rigorous spectral fitting typically involves accounting for model grid interpolation uncertainties \citep[e.g.,][]{2015ApJ...812..128C, 2017ApJ...836..200G, 2021ApJ...916...53Z, 2021ApJ...921...95Z} and employing complex covariance matrices for likelihood evaluations, which are unnecessary when the objective is limited to $L_{\rm bol}$ estimation.

These studies also often rely on a single grid of model atmospheres, which can introduce systematics arising from the specific assumptions inherent in the chosen models, potentially leading to model-dependent biases and underestimated uncertainties in the inferred $L_{\rm bol}$ values. For example, \cite{2025AJ....169....9Z} used fourteen different grids of atmospheric models to estimate the bolometric luminosities of the T/Y planetary-mass companion COCONUTS-2b. While their uncertainties in individual $L_{\rm bol}$ estimates were small ($0.004-0.01$~dex), \cite{2025AJ....169....9Z} found a significant scatter of $\approx 0.5$~dex among these estimates, driven by the vastly different assumptions within the models. Eliminating such modeling systematics is nearly impossible in exoplanet and brown dwarf studies. Directly observing a substantial portion of a given object's SED across broad wavelengths --- enabled by the combination of JWST NIRSpec and MIRI --- provides a robust approach to reducing the modeling systematics to the level of or below measurement uncertainties \citep[e.g.,][]{2024ApJ...973..107B}. However, for the majority of objects lacking such data, it is essential to combine multiple grids of atmospheric models to mitigate systematics and derive $L_{\rm bol}$ estimates with more conservative uncertainties (see Appendix~\ref{app:sub:new}). 

As an alternative approach for estimating $L_{\rm bol}$, when observed spectrophotometry extends to longer wavelengths near or beyond the $L$ and $M$ bands, a Rayleigh-Jeans tail is often appended to the longest-wavelength flux to approximate the full SED \citep[e.g.,][]{2001ApJ...548..908L, 2004AJ....127.3516G, 2006ApJ...648..614C, 2020ApJ...891..171Z, 2022ApJ...935...15Z, 2024ApJ...973..107B}. This method is significantly less complex than spectral fitting. However, when the observed spectrophotometry does not extend beyond the $M$ band, these objects' mid-infrared features such as CH$_{4}$, CO, and silicate cloud absorption features can introduce significant deviations from the Rayleigh-Jeans approximation \citep[e.g.,][]{2004AJ....127.3516G}. 

Additionally, for objects lacking spectroscopic data, bolometric corrections (BCs) are traditionally employed to estimate $L_{\rm bol}$ from broadband photometry, provided the objects' distances are known. BCs are generally derived using empirical polynomials based on spectral types or single-band absolute magnitudes, calibrated using spectroscopic samples with bolometric luminosities estimated via the methods described above \citep[e.g.,][]{2015ApJ...810..158F, 2017ApJS..231...15D, 2023ApJ...959...63S}. This approach is technically straightforward to implement. While the bolometric luminosities estimated using BCs share some limitations with spectral fitting and the Rayleigh-Jeans approximation, the intrinsic scatter within large samples tends to increase the uncertainties of the estimated $L_{\rm bol}$, helping to mitigate systematics associated with atmospheric models and Rayleigh-Jeans tails. However, these empirical polynomials are not well-established for objects with very young ages ($\lesssim 100$~Myr) or extreme metallicities (e.g., $\big|$[M/H]$\big| > 0.5$~dex) due to the limited number of known objects in these parameter spaces. 

Motivated by the challenges associated with existing methods for bolometric luminosity estimation, we propose a new approach with mitigated modeling systematics, tailored for objects with observed spectrophotometry. This approach is presented in the next section.

{ 
\begin{deluxetable*}{lccccccccc} 
\renewcommand{\arraystretch}{1.2} 
\setlength{\tabcolsep}{4pt} 
\tablecaption{Atmospheric Models and Parameter Space Investigated in Appendix~\ref{app:sub:new}} \label{tab:atm_grids} 
\tablehead{ \multicolumn{1}{l}{Atmospheric Models} &  \multicolumn{6}{c}{Parameters} &  \multicolumn{1}{c}{} &  \multicolumn{2}{c}{Assumptions} \\ 
\cline{2-7} \cline{9-10} 
\multicolumn{1}{l}{} &  \multicolumn{1}{c}{$T_{\rm eff}$} &  \multicolumn{1}{c}{$\log{(g)}$} &  \multicolumn{1}{c}{[M/H]} &  \multicolumn{1}{c}{C/O} &  \multicolumn{1}{c}{Cloud Param.} &  \multicolumn{1}{c}{$\log{(K_{\rm zz})}$} &  \multicolumn{1}{c}{} &  \multicolumn{1}{c}{Clouds?} &  \multicolumn{1}{c}{Chem. Eq?} \\ 
\multicolumn{1}{l}{} &  \multicolumn{1}{c}{(K)} &  \multicolumn{1}{c}{(dex)} &  \multicolumn{1}{c}{(dex)} &  \multicolumn{1}{c}{} &  \multicolumn{1}{c}{} &  \multicolumn{1}{c}{(dex)} &  \multicolumn{1}{c}{} &  \multicolumn{1}{c}{} &  \multicolumn{1}{c}{} } 
\startdata 
\texttt{ATMO2020}  &  $[400, 3000]$  &  $[2.5, 5.5]$  &  $0$  &  $0.55$   &  --  &  multiple\tablenotemark{\scriptsize a}   &  &   Cloudless  &  CEQ + NEQ  \\  
\texttt{Sonora Diamondback}  &  $[900, 2400]$  &  $[3.5, 5.5]$  &  $[-0.5, +0.5]$  &  $0.458$   &  $f_{\rm sed} \in [1,2,3,4,8,{\rm nc}]$  &  --   &  &   Cloudy  &  CEQ  \\  
\texttt{Exo-REM}  &  $[400, 2000]$  &  $[3.0, 5.0]$  &  $[-0.5, +1.0]$  &  $[0.1, 0.8]$   &  microphysics\tablenotemark{\scriptsize b}  &  profiles\tablenotemark{\scriptsize b}   &  &   Cloudy  &  CEQ + NEQ  \\  
\texttt{SPHINX}  &  $[2000, 4000]$  &  $[4.0, 5.5]$  &  $[-1.0, +1.0]$  &  $[0.3, 0.9]$   &  --  &  --   &  &   Cloudless  &  CEQ  \\  
\texttt{BT-Settl}  &  $[1800, 4000]$  &  $[3.0, 5.5]$  &  $[-1.0, +0.5]$  &  0.55   &  microphysics  &  --   &  &   Cloudy  &  CEQ  \\  
\enddata 
\tablenotetext{a}{The $K_{\rm zz}$ varies with the surface gravity in the ATMO2020 chemical disequilibrium models \citep[see Figure~1 of][]{2020AandA...637A..38P}.  }  
\tablenotetext{b}{The Exo-REM models incorporate iron and silicate clouds. Also, the $K_{zz}$ values are computed as a function of pressures \citep[see Section~2.2.2 of][]{2018ApJ...854..172C}. }  
\end{deluxetable*} 
}

\onecolumngrid

\begin{figure*}[t]
\begin{center}
\includegraphics[width=6.in]{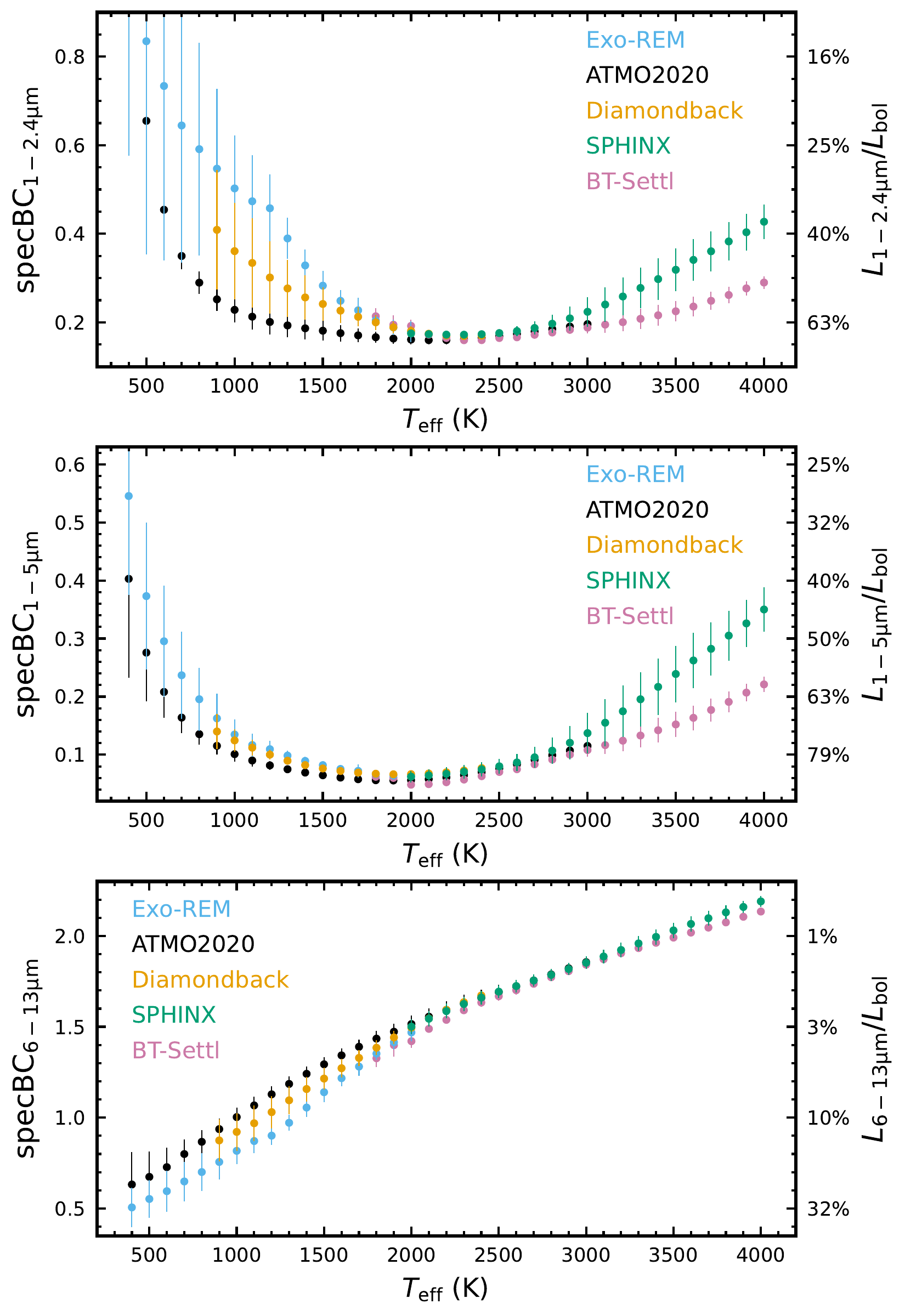}
\caption{The bolometric correction (\texttt{specBC}$_{\lambda\lambda}$; defined by Equation~\ref{eq:bc}) for wavelength ranges of $\lambda\lambda=1-2.4$~$\mu$m (top), $1-5$~$\mu$m (middle), and $6-13$~$\mu$m (bottom), calculated using multiple grids of atmospheric models. The y-axis on the right presents the ratio between the integrated spectral luminosity over $\lambda\lambda$ and the bolometric luminosity, $L_{\lambda\lambda}/L_{\rm bol} = 10^{-{\rm \texttt{specBC}_{\lambda\lambda}}}$. These plotted values are summarized in Tables~\ref{tab:bc_1to2p4}, \ref{tab:bc_1to5}, and \ref{tab:bc_6to13}.  }
\label{fig:bc}
\end{center}
\end{figure*}

\subsection{A New Bolometric Correction Approach with Mitigated Modeling Systematics}
\label{app:sub:new}

Here, we present a new approach to determining the bolometric luminosities using the observed spectra of self-luminous exoplanets, brown dwarfs, and low-mass stars. Specifically, we define \texttt{specBC$_{\lambda\lambda}$} as the difference between the bolometric luminosity, $\log{(L_{\rm bol}/L_{\odot})}$, and the integrated spectral luminosity over a narrower wavelength range ($\lambda\lambda$), $\log{(L_{\lambda\lambda}/L_{\odot})}$, as follows,
\begin{equation} \label{eq:bc}
\begin{aligned} 
{\rm \texttt{specBC}}_{\lambda\lambda} \equiv & \log{(L_{\rm bol}/L_{\odot})} - \log{(L_{\lambda\lambda}/L_{\odot})} \\
\equiv & \log{(L_{\rm bol}/L_{\odot})} - \log{\left(4\pi d^{2} \int_{\lambda\lambda} f_{\lambda} d\lambda\ \big/\ L_{\odot} \right)}
\end{aligned} 
\end{equation}
For a given object with an observed spectrum ($f_{\lambda}$) and a known distance ($d$), once its corresponding \texttt{specBC$_{\lambda\lambda}$} is estimated, the bolometric luminosity can be derived as:
\begin{equation} \label{eq:lbol}
\log{(L_{\rm bol}/L_{\odot})} = \log{\left(4\pi d^{2} \int_{\lambda\lambda} f_{\lambda} d\lambda\ \big/\ L_{\odot} \right)} + {\rm \texttt{specBC}}_{\lambda\lambda}
\end{equation}

For the rest of this section, we explain how to estimate \texttt{specBC$_{\lambda\lambda}$} for a given object. First, we present pre-calculated \texttt{specBC$_{\lambda\lambda}$} values over a wide parameter space. These \texttt{specBC$_{\lambda\lambda}$} values are computed using multiple grids of atmospheric models, including 
\begin{enumerate}
\item[(1)] \texttt{ATMO2020} \citep{2020AandA...637A..38P},
\item[(2)] \texttt{Sonora Diamondback} \citep{2024ApJ...975...59M},
\item[(3)] \texttt{Exo-REM} \citep{2015A&A...582A..83B, 2018ApJ...854..172C, 2021A&A...646A..15B},
\item[(4)] \texttt{SPHINX} \citep{2023ApJ...944...41I}, and
\item[(5)] \texttt{BT-Settl} \citep{2003IAUS..211..325A, 2011ASPC..448...91A, 2012RSPTA.370.2765A, 2013MSAIS..24..128A}
\end{enumerate}
The parameter space and key assumptions of these model grids are summarized in Table~\ref{tab:atm_grids}. 

For each synthetic spectrum ($F_{\lambda}$) in each of the above atmospheric model grids, we calculate the \texttt{specBC$_{\lambda\lambda}$} $= \log{(L_{\rm bol} / L_{\lambda\lambda})} = \log{\left[ \left( \int_{\rm bol} F_{\lambda} d\lambda \right) / \left( \int_{\lambda\lambda} F_{\lambda} d\lambda \right) \right]}$, where $\left(\int_{\rm bol} F_{\lambda} d\lambda \right)$ is the bolometric model flux integrated over the full wavelength range (up to 250~$\mu$m). For \texttt{SPHINX} models, which only extend to 20~$\mu$m, we approximate the bolometric flux using $\sigma_{\rm SB} T_{\rm eff}^{4}$, where $\sigma_{\rm SB}$ is the Stefan-Boltzmann constant and $T_{\rm eff}$ is the effective temperature of the corresponding model spectrum. The term, $(\int_{\lambda\lambda} F_{\lambda} d\lambda)$, represents the integrated model flux over a pre-defined wavelength range. In this work, we explore three ranges based on the wavelength coverage of modern spectrographs:
\begin{enumerate}
\item[$\bullet$] $\lambda\lambda =$1--2.4~$\mu$m, covered by ground-based near-infrared spectrographs, e.g., IRTF/SpeX \citep{2003PASP..115..362R}, Gemini-N/GNIRS \citep{2006SPIE.6269E..4CE}, Gemini-S/FLAMINGOS-2 \citep{2004SPIE.5492.1196E, 2008SPIE.7014E..0VE}, Magellan/FIRE \citep{2013PASP..125..270S}, VLT/X-shooter \citep{2011A&A...536A.105V}.
\item[$\bullet$] $\lambda\lambda =$1--5~$\mu$m, covered by, e.g., JWST/NIRSpec \citep{2022A&A...661A..80J, 2023PASP..135c8001B}..
\item[$\bullet$] $\lambda\lambda =$6--13~$\mu$m, covered by, e.g., JWST/MIRI \citep{2015PASP..127..584R, 2023PASP..135d8003W}.
\end{enumerate}
For JWST NIRSpec (0.6--5.3~$\mu$m) and MIRI (4.9--27.9~$\mu$m) spectrographs, we adopt slightly narrower wavelength ranges than their full wavelength sensitivities. This approach enhances the applicability of our methods across various observing modes while minimizing potential calibration uncertainties near detector edges. 

Figure~\ref{fig:bc} summarizes our computed \texttt{specBC$_{\lambda\lambda}$} as a function of effective temperature for each atmospheric model grid. At a given $T_{\rm eff}$, the value and errorbar of \texttt{specBC$_{\lambda\lambda}$}$(T_{\rm eff})$ represent the mean and the standard deviation of the computed bolometric correction for all model spectra with the same $T_{\rm eff}$ but varying $\log{(g)}$, [M/H], C/O, and cloud properties. The pre-computed \texttt{specBC$_{\lambda\lambda}$} for each $\lambda\lambda$ and model grid are provided in Tables~\ref{tab:bc_1to2p4}, \ref{tab:bc_1to5}, and \ref{tab:bc_6to13}. 

With these pre-computed \texttt{specBC$_{\lambda\lambda}$} values, one can estimate the bolometric correction term for a given object by computing the mean and standard deviation of the \texttt{specBC$_{\lambda\lambda}$} values over a customized $\{T_{\rm eff}$, $\log{(g)}$, [M/H], CO$\}$ parameter space relevant to the object (see Section~\ref{subsec:lbol}) and then compute the bolometric luminosity via Equation~\ref{eq:lbol}.

The bolometric correction approach presented in this work is applicable for objects with $T_{\rm eff}$ from 400~K to 4000~K, $\log{(g)}$. Our analysis also shows that the bolometric correction is highly model-dependent, except for objects with specific $T_{\rm eff}$ ranges. For objects with 1--2.4~$\mu$m spectra, various model grids yield consistent \texttt{specBC}$_{\rm 1-2.4\mu m}$ over $T_{\rm eff} = 1900-3200$~K; for objects with 1--5~$\mu$m spectra, various model grids yield consistent \texttt{specBC}$_{\rm 1-5\mu m}$ for $T_{\rm eff} = 900-3200$~K. Outside these $T_{\rm eff}$ ranges, the systematic differences bewteen model grids are much larger than the scatter within each model grid. In this situation, we recommend incorporating a large uncertainty in \texttt{specBC}$_{\lambda\lambda}$ to account for the modeling systematics.

{ 
\begin{deluxetable*}{ccccccccccccccc}
\setlength{\tabcolsep}{4pt} 
\tablecaption{Spectroscopy-based bolometric correction for $\lambda\lambda=1-2.4$~$\mu$m, i.e., \texttt{specBC$_{\rm 1-2.4\mu m}$}$=\log{(L_{\rm bol}/L_{\odot})} - \log{(L_{\rm 1-2.4\mu m}/L_{\odot})}$} \label{tab:bc_1to2p4} 
\tablehead{ \multicolumn{1}{c}{$T_{\rm eff}$}  & \multicolumn{2}{c}{\texttt{Exo-REM}}  & \multicolumn{1}{c}{}  & \multicolumn{2}{c}{\texttt{Diamondback}}  & \multicolumn{1}{c}{}   & \multicolumn{2}{c}{\texttt{ATMO2020}}  & \multicolumn{1}{c}{}   & \multicolumn{2}{c}{\texttt{BT-Settl}}  & \multicolumn{1}{c}{}   & \multicolumn{2}{c}{\texttt{SPHINX}}   \\ 
\cline{2-3} \cline{5-6} \cline{8-9} \cline{11-12} \cline{14-15}  
\multicolumn{1}{c}{}  & \multicolumn{1}{c}{Value}  & \multicolumn{1}{c}{$N_{\rm model}$}  & \multicolumn{1}{c}{}  & \multicolumn{1}{c}{Value}  & \multicolumn{1}{c}{$N_{\rm model}$}  & \multicolumn{1}{c}{}  & \multicolumn{1}{c}{Value}  & \multicolumn{1}{c}{$N_{\rm model}$}  & \multicolumn{1}{c}{}  & \multicolumn{1}{c}{Value}  & \multicolumn{1}{c}{$N_{\rm model}$}  & \multicolumn{1}{c}{}  & \multicolumn{1}{c}{Value}  & \multicolumn{1}{c}{$N_{\rm model}$}   \\ 
\multicolumn{1}{c}{(K)} & \multicolumn{1}{c}{(dex)} & \multicolumn{1}{c}{} & \multicolumn{1}{c}{} & \multicolumn{1}{c}{(dex)} & \multicolumn{1}{c}{} & \multicolumn{1}{c}{} & \multicolumn{1}{c}{(dex)} & \multicolumn{1}{c}{} & \multicolumn{1}{c}{}   & \multicolumn{1}{c}{(dex)} & \multicolumn{1}{c}{} & \multicolumn{1}{c}{}   & \multicolumn{1}{c}{(dex)} & \multicolumn{1}{c}{}   } 
\startdata 
 400  &  $1.169 \pm 0.593$ & 293  &  &  ... & ...  &  &  $1.132 \pm 0.085$ & 21  &  &  ... & ...  &  &  ... & ... \\ 
 500  &  $0.835 \pm 0.481$ & 283  &  &  ... & ...  &  &  $0.655 \pm 0.053$ & 20  &  &  ... & ...  &  &  ... & ... \\ 
 600  &  $0.734 \pm 0.394$ & 294  &  &  ... & ...  &  &  $0.454 \pm 0.038$ & 21  &  &  ... & ...  &  &  ... & ... \\ 
 700  &  $0.645 \pm 0.310$ & 289  &  &  ... & ...  &  &  $0.350 \pm 0.030$ & 20  &  &  ... & ...  &  &  ... & ... \\ 
 800  &  $0.591 \pm 0.240$ & 289  &  &  ... & ...  &  &  $0.290 \pm 0.025$ & 21  &  &  ... & ...  &  &  ... & ... \\ 
 900  &  $0.547 \pm 0.180$ & 294  &  &  $0.409 \pm 0.135$ & 90  &  &  $0.252 \pm 0.027$ & 21  &  &  ... & ...  &  &  ... & ... \\ 
 1000  &  $0.502 \pm 0.120$ & 292  &  &  $0.361 \pm 0.109$ & 90  &  &  $0.228 \pm 0.028$ & 21  &  &  ... & ...  &  &  ... & ... \\ 
 1100  &  $0.473 \pm 0.104$ & 287  &  &  $0.334 \pm 0.101$ & 90  &  &  $0.213 \pm 0.029$ & 20  &  &  ... & ...  &  &  ... & ... \\ 
 1200  &  $0.458 \pm 0.077$ & 290  &  &  $0.301 \pm 0.081$ & 90  &  &  $0.201 \pm 0.029$ & 21  &  &  ... & ...  &  &  ... & ... \\ 
 1300  &  $0.389 \pm 0.046$ & 290  &  &  $0.277 \pm 0.064$ & 90  &  &  $0.193 \pm 0.027$ & 21  &  &  ... & ...  &  &  ... & ... \\ 
 1400  &  $0.329 \pm 0.036$ & 285  &  &  $0.256 \pm 0.050$ & 90  &  &  $0.187 \pm 0.025$ & 21  &  &  ... & ...  &  &  ... & ... \\ 
 1500  &  $0.283 \pm 0.033$ & 284  &  &  $0.241 \pm 0.038$ & 90  &  &  $0.181 \pm 0.022$ & 21  &  &  ... & ...  &  &  ... & ... \\ 
 1600  &  $0.249 \pm 0.024$ & 293  &  &  $0.226 \pm 0.028$ & 90  &  &  $0.176 \pm 0.019$ & 21  &  &  ... & ...  &  &  ... & ... \\ 
 1700  &  $0.227 \pm 0.029$ & 290  &  &  $0.212 \pm 0.022$ & 90  &  &  $0.171 \pm 0.015$ & 21  &  &  ... & ...  &  &  ... & ... \\ 
 1800  &  $0.204 \pm 0.014$ & 291  &  &  $0.200 \pm 0.018$ & 90  &  &  $0.166 \pm 0.013$ & 21  &  &  $0.214 \pm 0.018$ & 25  &  &  ... & ... \\ 
 1900  &  $0.188 \pm 0.017$ & 294  &  &  $0.189 \pm 0.014$ & 90  &  &  $0.163 \pm 0.012$ & 7  &  &  $0.195 \pm 0.021$ & 25  &  &  ... & ... \\ 
 2000  &  $0.179 \pm 0.024$ & 294  &  &  $0.180 \pm 0.010$ & 90  &  &  $0.161 \pm 0.010$ & 7  &  &  $0.192 \pm 0.014$ & 30  &  &  $0.175 \pm 0.006$ & 252 \\ 
 2100  &  ... & ...  &  &  $0.174 \pm 0.008$ & 90  &  &  $0.160 \pm 0.010$ & 7  &  &  $0.172 \pm 0.003$ & 30  &  &  $0.173 \pm 0.005$ & 252 \\ 
 2200  &  ... & ...  &  &  $0.171 \pm 0.006$ & 90  &  &  $0.160 \pm 0.009$ & 7  &  &  $0.165 \pm 0.003$ & 30  &  &  $0.172 \pm 0.005$ & 252 \\ 
 2300  &  ... & ...  &  &  $0.169 \pm 0.004$ & 90  &  &  $0.162 \pm 0.009$ & 7  &  &  $0.160 \pm 0.004$ & 30  &  &  $0.172 \pm 0.006$ & 252 \\ 
 2400  &  ... & ...  &  &  $0.170 \pm 0.003$ & 90  &  &  $0.165 \pm 0.009$ & 7  &  &  $0.160 \pm 0.005$ & 30  &  &  $0.173 \pm 0.007$ & 252 \\ 
 2500  &  ... & ...  &  &  ... & ...  &  &  $0.169 \pm 0.008$ & 7  &  &  $0.165 \pm 0.006$ & 30  &  &  $0.176 \pm 0.009$ & 252 \\ 
 2600  &  ... & ...  &  &  ... & ...  &  &  $0.174 \pm 0.007$ & 7  &  &  $0.166 \pm 0.006$ & 18  &  &  $0.180 \pm 0.011$ & 252 \\ 
 2700  &  ... & ...  &  &  ... & ...  &  &  $0.179 \pm 0.005$ & 7  &  &  $0.172 \pm 0.005$ & 18  &  &  $0.187 \pm 0.015$ & 252 \\ 
 2800  &  ... & ...  &  &  ... & ...  &  &  $0.185 \pm 0.005$ & 7  &  &  $0.177 \pm 0.006$ & 18  &  &  $0.197 \pm 0.021$ & 251 \\ 
 2900  &  ... & ...  &  &  ... & ...  &  &  $0.190 \pm 0.008$ & 7  &  &  $0.183 \pm 0.009$ & 18  &  &  $0.209 \pm 0.027$ & 252 \\ 
 3000  &  ... & ...  &  &  ... & ...  &  &  $0.196 \pm 0.012$ & 7  &  &  $0.188 \pm 0.013$ & 18  &  &  $0.224 \pm 0.033$ & 250 \\ 
 3100  &  ... & ...  &  &  ... & ...  &  &  ... & ...  &  &  $0.195 \pm 0.018$ & 18  &  &  $0.240 \pm 0.039$ & 252 \\ 
 3200  &  ... & ...  &  &  ... & ...  &  &  ... & ...  &  &  $0.200 \pm 0.021$ & 18  &  &  $0.258 \pm 0.043$ & 252 \\ 
 3300  &  ... & ...  &  &  ... & ...  &  &  ... & ...  &  &  $0.208 \pm 0.022$ & 18  &  &  $0.278 \pm 0.046$ & 252 \\ 
 3400  &  ... & ...  &  &  ... & ...  &  &  ... & ...  &  &  $0.216 \pm 0.023$ & 18  &  &  $0.298 \pm 0.047$ & 251 \\ 
 3500  &  ... & ...  &  &  ... & ...  &  &  ... & ...  &  &  $0.225 \pm 0.023$ & 18  &  &  $0.319 \pm 0.048$ & 246 \\ 
 3600  &  ... & ...  &  &  ... & ...  &  &  ... & ...  &  &  $0.236 \pm 0.022$ & 18  &  &  $0.341 \pm 0.047$ & 237 \\ 
 3700  &  ... & ...  &  &  ... & ...  &  &  ... & ...  &  &  $0.249 \pm 0.021$ & 18  &  &  $0.360 \pm 0.046$ & 245 \\ 
 3800  &  ... & ...  &  &  ... & ...  &  &  ... & ...  &  &  $0.262 \pm 0.019$ & 18  &  &  $0.383 \pm 0.043$ & 241 \\ 
 3900  &  ... & ...  &  &  ... & ...  &  &  ... & ...  &  &  $0.277 \pm 0.016$ & 18  &  &  $0.403 \pm 0.041$ & 247 \\ 
 4000  &  ... & ...  &  &  ... & ...  &  &  ... & ...  &  &  $0.290 \pm 0.014$ & 18  &  &  $0.427 \pm 0.039$ & 248 \\ 
\enddata 
\end{deluxetable*} 
}

{ 
\begin{deluxetable*}{ccccccccccccccc}
\setlength{\tabcolsep}{4pt} 
\tablecaption{Spectroscopy-based bolometric correction for $\lambda\lambda=1-5$~$\mu$m, i.e., \texttt{specBC$_{\rm 1-5\mu m}$}$=\log{(L_{\rm bol}/L_{\odot})} - \log{(L_{\rm 1-5\mu m}/L_{\odot})}$} \label{tab:bc_1to5} 
\tablehead{ \multicolumn{1}{c}{$T_{\rm eff}$}  & \multicolumn{2}{c}{\texttt{Exo-REM}}  & \multicolumn{1}{c}{}  & \multicolumn{2}{c}{\texttt{Diamondback}}  & \multicolumn{1}{c}{}   & \multicolumn{2}{c}{\texttt{ATMO2020}}  & \multicolumn{1}{c}{}   & \multicolumn{2}{c}{\texttt{BT-Settl}}  & \multicolumn{1}{c}{}   & \multicolumn{2}{c}{\texttt{SPHINX}}   \\ 
\cline{2-3} \cline{5-6} \cline{8-9} \cline{11-12} \cline{14-15}  
\multicolumn{1}{c}{}  & \multicolumn{1}{c}{Value}  & \multicolumn{1}{c}{$N_{\rm model}$}  & \multicolumn{1}{c}{}  & \multicolumn{1}{c}{Value}  & \multicolumn{1}{c}{$N_{\rm model}$}  & \multicolumn{1}{c}{}  & \multicolumn{1}{c}{Value}  & \multicolumn{1}{c}{$N_{\rm model}$}  & \multicolumn{1}{c}{}  & \multicolumn{1}{c}{Value}  & \multicolumn{1}{c}{$N_{\rm model}$}  & \multicolumn{1}{c}{}  & \multicolumn{1}{c}{Value}  & \multicolumn{1}{c}{$N_{\rm model}$}   \\ 
\multicolumn{1}{c}{(K)} & \multicolumn{1}{c}{(dex)} & \multicolumn{1}{c}{} & \multicolumn{1}{c}{} & \multicolumn{1}{c}{(dex)} & \multicolumn{1}{c}{} & \multicolumn{1}{c}{} & \multicolumn{1}{c}{(dex)} & \multicolumn{1}{c}{} & \multicolumn{1}{c}{}   & \multicolumn{1}{c}{(dex)} & \multicolumn{1}{c}{} & \multicolumn{1}{c}{}   & \multicolumn{1}{c}{(dex)} & \multicolumn{1}{c}{}   } 
\startdata 
 400  &  $0.545 \pm 0.170$ & 293  &  &  ... & ...  &  &  $0.403 \pm 0.171$ & 21  &  &  ... & ...  &  &  ... & ... \\ 
 500  &  $0.373 \pm 0.127$ & 283  &  &  ... & ...  &  &  $0.276 \pm 0.084$ & 20  &  &  ... & ...  &  &  ... & ... \\ 
 600  &  $0.295 \pm 0.096$ & 294  &  &  ... & ...  &  &  $0.208 \pm 0.045$ & 21  &  &  ... & ...  &  &  ... & ... \\ 
 700  &  $0.237 \pm 0.075$ & 289  &  &  ... & ...  &  &  $0.164 \pm 0.026$ & 20  &  &  ... & ...  &  &  ... & ... \\ 
 800  &  $0.196 \pm 0.054$ & 289  &  &  ... & ...  &  &  $0.135 \pm 0.018$ & 21  &  &  ... & ...  &  &  ... & ... \\ 
 900  &  $0.163 \pm 0.042$ & 294  &  &  $0.140 \pm 0.030$ & 90  &  &  $0.115 \pm 0.015$ & 21  &  &  ... & ...  &  &  ... & ... \\ 
 1000  &  $0.135 \pm 0.026$ & 292  &  &  $0.125 \pm 0.019$ & 90  &  &  $0.101 \pm 0.013$ & 21  &  &  ... & ...  &  &  ... & ... \\ 
 1100  &  $0.117 \pm 0.020$ & 287  &  &  $0.112 \pm 0.014$ & 90  &  &  $0.090 \pm 0.011$ & 20  &  &  ... & ...  &  &  ... & ... \\ 
 1200  &  $0.110 \pm 0.014$ & 290  &  &  $0.100 \pm 0.009$ & 90  &  &  $0.082 \pm 0.008$ & 21  &  &  ... & ...  &  &  ... & ... \\ 
 1300  &  $0.098 \pm 0.009$ & 290  &  &  $0.090 \pm 0.006$ & 90  &  &  $0.075 \pm 0.005$ & 21  &  &  ... & ...  &  &  ... & ... \\ 
 1400  &  $0.089 \pm 0.007$ & 285  &  &  $0.082 \pm 0.004$ & 90  &  &  $0.069 \pm 0.003$ & 21  &  &  ... & ...  &  &  ... & ... \\ 
 1500  &  $0.082 \pm 0.007$ & 284  &  &  $0.076 \pm 0.004$ & 90  &  &  $0.065 \pm 0.002$ & 21  &  &  ... & ...  &  &  ... & ... \\ 
 1600  &  $0.076 \pm 0.006$ & 293  &  &  $0.072 \pm 0.003$ & 90  &  &  $0.061 \pm 0.001$ & 21  &  &  ... & ...  &  &  ... & ... \\ 
 1700  &  $0.072 \pm 0.011$ & 290  &  &  $0.069 \pm 0.003$ & 90  &  &  $0.058 \pm 0.001$ & 21  &  &  ... & ...  &  &  ... & ... \\ 
 1800  &  $0.067 \pm 0.007$ & 291  &  &  $0.067 \pm 0.003$ & 90  &  &  $0.056 \pm 0.001$ & 21  &  &  $0.062 \pm 0.003$ & 25  &  &  ... & ... \\ 
 1900  &  $0.064 \pm 0.007$ & 294  &  &  $0.066 \pm 0.003$ & 90  &  &  $0.056 \pm 0.001$ & 7  &  &  $0.060 \pm 0.003$ & 25  &  &  ... & ... \\ 
 2000  &  $0.063 \pm 0.008$ & 294  &  &  $0.067 \pm 0.003$ & 90  &  &  $0.056 \pm 0.000$ & 7  &  &  $0.048 \pm 0.001$ & 30  &  &  $0.062 \pm 0.010$ & 252 \\ 
 2100  &  ... & ...  &  &  $0.068 \pm 0.003$ & 90  &  &  $0.058 \pm 0.000$ & 7  &  &  $0.049 \pm 0.002$ & 30  &  &  $0.064 \pm 0.010$ & 252 \\ 
 2200  &  ... & ...  &  &  $0.070 \pm 0.003$ & 90  &  &  $0.061 \pm 0.000$ & 7  &  &  $0.053 \pm 0.002$ & 30  &  &  $0.067 \pm 0.011$ & 252 \\ 
 2300  &  ... & ...  &  &  $0.073 \pm 0.004$ & 90  &  &  $0.065 \pm 0.000$ & 7  &  &  $0.057 \pm 0.001$ & 30  &  &  $0.071 \pm 0.012$ & 252 \\ 
 2400  &  ... & ...  &  &  $0.077 \pm 0.004$ & 90  &  &  $0.070 \pm 0.001$ & 7  &  &  $0.063 \pm 0.001$ & 30  &  &  $0.075 \pm 0.012$ & 252 \\ 
 2500  &  ... & ...  &  &  ... & ...  &  &  $0.076 \pm 0.001$ & 7  &  &  $0.070 \pm 0.001$ & 30  &  &  $0.080 \pm 0.013$ & 252 \\ 
 2600  &  ... & ...  &  &  ... & ...  &  &  $0.083 \pm 0.001$ & 7  &  &  $0.075 \pm 0.002$ & 18  &  &  $0.087 \pm 0.015$ & 252 \\ 
 2700  &  ... & ...  &  &  ... & ...  &  &  $0.091 \pm 0.002$ & 7  &  &  $0.083 \pm 0.003$ & 18  &  &  $0.096 \pm 0.018$ & 252 \\ 
 2800  &  ... & ...  &  &  ... & ...  &  &  $0.099 \pm 0.004$ & 7  &  &  $0.092 \pm 0.005$ & 18  &  &  $0.107 \pm 0.023$ & 251 \\ 
 2900  &  ... & ...  &  &  ... & ...  &  &  $0.107 \pm 0.007$ & 7  &  &  $0.100 \pm 0.007$ & 18  &  &  $0.121 \pm 0.029$ & 252 \\ 
 3000  &  ... & ...  &  &  ... & ...  &  &  $0.115 \pm 0.010$ & 7  &  &  $0.108 \pm 0.011$ & 18  &  &  $0.137 \pm 0.035$ & 250 \\ 
 3100  &  ... & ...  &  &  ... & ...  &  &  ... & ...  &  &  $0.117 \pm 0.015$ & 18  &  &  $0.155 \pm 0.040$ & 252 \\ 
 3200  &  ... & ...  &  &  ... & ...  &  &  ... & ...  &  &  $0.124 \pm 0.018$ & 18  &  &  $0.175 \pm 0.044$ & 252 \\ 
 3300  &  ... & ...  &  &  ... & ...  &  &  ... & ...  &  &  $0.133 \pm 0.020$ & 18  &  &  $0.195 \pm 0.047$ & 252 \\ 
 3400  &  ... & ...  &  &  ... & ...  &  &  ... & ...  &  &  $0.142 \pm 0.021$ & 18  &  &  $0.217 \pm 0.048$ & 251 \\ 
 3500  &  ... & ...  &  &  ... & ...  &  &  ... & ...  &  &  $0.152 \pm 0.022$ & 18  &  &  $0.239 \pm 0.049$ & 246 \\ 
 3600  &  ... & ...  &  &  ... & ...  &  &  ... & ...  &  &  $0.164 \pm 0.021$ & 18  &  &  $0.262 \pm 0.048$ & 237 \\ 
 3700  &  ... & ...  &  &  ... & ...  &  &  ... & ...  &  &  $0.177 \pm 0.020$ & 18  &  &  $0.282 \pm 0.046$ & 245 \\ 
 3800  &  ... & ...  &  &  ... & ...  &  &  ... & ...  &  &  $0.191 \pm 0.018$ & 18  &  &  $0.305 \pm 0.043$ & 241 \\ 
 3900  &  ... & ...  &  &  ... & ...  &  &  ... & ...  &  &  $0.207 \pm 0.015$ & 18  &  &  $0.326 \pm 0.040$ & 247 \\ 
 4000  &  ... & ...  &  &  ... & ...  &  &  ... & ...  &  &  $0.221 \pm 0.013$ & 18  &  &  $0.350 \pm 0.038$ & 248 \\ 
\enddata 
\end{deluxetable*} 
}

{ 
\begin{deluxetable*}{ccccccccccccccc}
\setlength{\tabcolsep}{4pt} 
\tablecaption{Spectroscopy-based bolometric correction for $\lambda\lambda=6-13$~$\mu$m, i.e., \texttt{specBC$_{\rm 6-13\mu m}$}$=\log{(L_{\rm bol}/L_{\odot})} - \log{(L_{\rm 6-13\mu m}/L_{\odot})}$} \label{tab:bc_6to13} 
\tablehead{ \multicolumn{1}{c}{$T_{\rm eff}$}  & \multicolumn{2}{c}{\texttt{Exo-REM}}  & \multicolumn{1}{c}{}  & \multicolumn{2}{c}{\texttt{Diamondback}}  & \multicolumn{1}{c}{}   & \multicolumn{2}{c}{\texttt{ATMO2020}}  & \multicolumn{1}{c}{}   & \multicolumn{2}{c}{\texttt{BT-Settl}}  & \multicolumn{1}{c}{}   & \multicolumn{2}{c}{\texttt{SPHINX}}   \\ 
\cline{2-3} \cline{5-6} \cline{8-9} \cline{11-12} \cline{14-15}  
\multicolumn{1}{c}{}  & \multicolumn{1}{c}{Value}  & \multicolumn{1}{c}{$N_{\rm model}$}  & \multicolumn{1}{c}{}  & \multicolumn{1}{c}{Value}  & \multicolumn{1}{c}{$N_{\rm model}$}  & \multicolumn{1}{c}{}  & \multicolumn{1}{c}{Value}  & \multicolumn{1}{c}{$N_{\rm model}$}  & \multicolumn{1}{c}{}  & \multicolumn{1}{c}{Value}  & \multicolumn{1}{c}{$N_{\rm model}$}  & \multicolumn{1}{c}{}  & \multicolumn{1}{c}{Value}  & \multicolumn{1}{c}{$N_{\rm model}$}   \\ 
\multicolumn{1}{c}{(K)} & \multicolumn{1}{c}{(dex)} & \multicolumn{1}{c}{} & \multicolumn{1}{c}{} & \multicolumn{1}{c}{(dex)} & \multicolumn{1}{c}{} & \multicolumn{1}{c}{} & \multicolumn{1}{c}{(dex)} & \multicolumn{1}{c}{} & \multicolumn{1}{c}{}   & \multicolumn{1}{c}{(dex)} & \multicolumn{1}{c}{} & \multicolumn{1}{c}{}   & \multicolumn{1}{c}{(dex)} & \multicolumn{1}{c}{}   } 
\startdata 
 400  &  $0.506 \pm 0.108$ & 293  &  &  ... & ...  &  &  $0.632 \pm 0.180$ & 21  &  &  ... & ...  &  &  ... & ... \\ 
 500  &  $0.552 \pm 0.104$ & 283  &  &  ... & ...  &  &  $0.674 \pm 0.141$ & 20  &  &  ... & ...  &  &  ... & ... \\ 
 600  &  $0.595 \pm 0.112$ & 294  &  &  ... & ...  &  &  $0.728 \pm 0.108$ & 21  &  &  ... & ...  &  &  ... & ... \\ 
 700  &  $0.649 \pm 0.108$ & 289  &  &  ... & ...  &  &  $0.800 \pm 0.081$ & 20  &  &  ... & ...  &  &  ... & ... \\ 
 800  &  $0.701 \pm 0.104$ & 289  &  &  ... & ...  &  &  $0.867 \pm 0.065$ & 21  &  &  ... & ...  &  &  ... & ... \\ 
 900  &  $0.756 \pm 0.095$ & 294  &  &  $0.875 \pm 0.116$ & 90  &  &  $0.937 \pm 0.056$ & 21  &  &  ... & ...  &  &  ... & ... \\ 
 1000  &  $0.817 \pm 0.074$ & 292  &  &  $0.921 \pm 0.105$ & 90  &  &  $1.003 \pm 0.052$ & 21  &  &  ... & ...  &  &  ... & ... \\ 
 1100  &  $0.871 \pm 0.066$ & 287  &  &  $0.969 \pm 0.099$ & 90  &  &  $1.067 \pm 0.049$ & 20  &  &  ... & ...  &  &  ... & ... \\ 
 1200  &  $0.901 \pm 0.052$ & 290  &  &  $1.030 \pm 0.090$ & 90  &  &  $1.128 \pm 0.044$ & 21  &  &  ... & ...  &  &  ... & ... \\ 
 1300  &  $0.972 \pm 0.044$ & 290  &  &  $1.095 \pm 0.078$ & 90  &  &  $1.186 \pm 0.041$ & 21  &  &  ... & ...  &  &  ... & ... \\ 
 1400  &  $1.056 \pm 0.051$ & 285  &  &  $1.157 \pm 0.068$ & 90  &  &  $1.241 \pm 0.040$ & 21  &  &  ... & ...  &  &  ... & ... \\ 
 1500  &  $1.140 \pm 0.055$ & 284  &  &  $1.215 \pm 0.060$ & 90  &  &  $1.294 \pm 0.039$ & 21  &  &  ... & ...  &  &  ... & ... \\ 
 1600  &  $1.217 \pm 0.045$ & 293  &  &  $1.272 \pm 0.050$ & 90  &  &  $1.343 \pm 0.040$ & 21  &  &  ... & ...  &  &  ... & ... \\ 
 1700  &  $1.281 \pm 0.052$ & 290  &  &  $1.329 \pm 0.044$ & 90  &  &  $1.390 \pm 0.041$ & 21  &  &  ... & ...  &  &  ... & ... \\ 
 1800  &  $1.352 \pm 0.033$ & 291  &  &  $1.385 \pm 0.040$ & 90  &  &  $1.435 \pm 0.042$ & 21  &  &  $1.327 \pm 0.049$ & 25  &  &  ... & ... \\ 
 1900  &  $1.415 \pm 0.039$ & 294  &  &  $1.441 \pm 0.037$ & 90  &  &  $1.473 \pm 0.045$ & 7  &  &  $1.399 \pm 0.062$ & 25  &  &  ... & ... \\ 
 2000  &  $1.469 \pm 0.047$ & 294  &  &  $1.495 \pm 0.031$ & 90  &  &  $1.515 \pm 0.046$ & 7  &  &  $1.421 \pm 0.036$ & 30  &  &  $1.501 \pm 0.041$ & 252 \\ 
 2100  &  ... & ...  &  &  $1.546 \pm 0.028$ & 90  &  &  $1.556 \pm 0.047$ & 7  &  &  $1.489 \pm 0.017$ & 30  &  &  $1.545 \pm 0.041$ & 252 \\ 
 2200  &  ... & ...  &  &  $1.593 \pm 0.027$ & 90  &  &  $1.593 \pm 0.048$ & 7  &  &  $1.538 \pm 0.010$ & 30  &  &  $1.586 \pm 0.041$ & 252 \\ 
 2300  &  ... & ...  &  &  $1.635 \pm 0.026$ & 90  &  &  $1.628 \pm 0.048$ & 7  &  &  $1.591 \pm 0.018$ & 30  &  &  $1.626 \pm 0.039$ & 252 \\ 
 2400  &  ... & ...  &  &  $1.673 \pm 0.027$ & 90  &  &  $1.660 \pm 0.045$ & 7  &  &  $1.634 \pm 0.022$ & 30  &  &  $1.661 \pm 0.038$ & 252 \\ 
 2500  &  ... & ...  &  &  ... & ...  &  &  $1.691 \pm 0.039$ & 7  &  &  $1.668 \pm 0.024$ & 30  &  &  $1.693 \pm 0.036$ & 252 \\ 
 2600  &  ... & ...  &  &  ... & ...  &  &  $1.722 \pm 0.033$ & 7  &  &  $1.702 \pm 0.023$ & 18  &  &  $1.724 \pm 0.034$ & 252 \\ 
 2700  &  ... & ...  &  &  ... & ...  &  &  $1.755 \pm 0.026$ & 7  &  &  $1.737 \pm 0.019$ & 18  &  &  $1.754 \pm 0.033$ & 252 \\ 
 2800  &  ... & ...  &  &  ... & ...  &  &  $1.788 \pm 0.019$ & 7  &  &  $1.773 \pm 0.015$ & 18  &  &  $1.786 \pm 0.032$ & 251 \\ 
 2900  &  ... & ...  &  &  ... & ...  &  &  $1.822 \pm 0.014$ & 7  &  &  $1.806 \pm 0.011$ & 18  &  &  $1.818 \pm 0.033$ & 252 \\ 
 3000  &  ... & ...  &  &  ... & ...  &  &  $1.856 \pm 0.010$ & 7  &  &  $1.842 \pm 0.007$ & 18  &  &  $1.852 \pm 0.035$ & 250 \\ 
 3100  &  ... & ...  &  &  ... & ...  &  &  ... & ...  &  &  $1.872 \pm 0.007$ & 18  &  &  $1.887 \pm 0.037$ & 252 \\ 
 3200  &  ... & ...  &  &  ... & ...  &  &  ... & ...  &  &  $1.905 \pm 0.005$ & 18  &  &  $1.923 \pm 0.040$ & 252 \\ 
 3300  &  ... & ...  &  &  ... & ...  &  &  ... & ...  &  &  $1.933 \pm 0.006$ & 18  &  &  $1.959 \pm 0.041$ & 252 \\ 
 3400  &  ... & ...  &  &  ... & ...  &  &  ... & ...  &  &  $1.962 \pm 0.006$ & 18  &  &  $1.995 \pm 0.042$ & 251 \\ 
 3500  &  ... & ...  &  &  ... & ...  &  &  ... & ...  &  &  $1.991 \pm 0.006$ & 18  &  &  $2.031 \pm 0.043$ & 246 \\ 
 3600  &  ... & ...  &  &  ... & ...  &  &  ... & ...  &  &  $2.019 \pm 0.006$ & 18  &  &  $2.066 \pm 0.042$ & 237 \\ 
 3700  &  ... & ...  &  &  ... & ...  &  &  ... & ...  &  &  $2.046 \pm 0.006$ & 18  &  &  $2.097 \pm 0.040$ & 245 \\ 
 3800  &  ... & ...  &  &  ... & ...  &  &  ... & ...  &  &  $2.075 \pm 0.005$ & 18  &  &  $2.130 \pm 0.038$ & 241 \\ 
 3900  &  ... & ...  &  &  ... & ...  &  &  ... & ...  &  &  $2.106 \pm 0.006$ & 18  &  &  $2.160 \pm 0.034$ & 247 \\ 
 4000  &  ... & ...  &  &  ... & ...  &  &  ... & ...  &  &  $2.134 \pm 0.005$ & 18  &  &  $2.190 \pm 0.030$ & 248 \\ 
\enddata 
\end{deluxetable*} 
}

\end{CJK*}

%%%%%%%%%%%%%%%
%%% BIBLIOGRAPHY %%%
%%%%%%%%%%%%%%%
\clearpage
\bibliographystyle{aasjournal}
\bibliography{ms}{}

%%%%%%%%%%%%
%%% FIGURES %%%
%%%%%%%%%%%%

\end{document}